\documentclass[11pt,preprint]{article}
\pdfoutput=1
\usepackage{jheppub}
\usepackage[usenames,dvipsnames]{xcolor}
\usepackage{hyperref}
\usepackage{tikz}
\usepackage{subcaption}
\usepackage{diagbox}
\usepackage{multirow}
\usepackage{amssymb}
\usepackage{bbding}
\usepackage{shuffle}
\usepackage{mathrsfs}

\usetikzlibrary{decorations.markings,arrows.meta}
\usetikzlibrary{decorations.pathmorphing}
\usetikzlibrary{calc}
\newcommand{\circleleft}{
\begin{tikzpicture}
 \draw[
        decoration={markings, mark=at position 0.3 with   {\arrow[scale=1.4]{>}}},
        postaction={decorate}
        ] (-0.01,0.1) -- (-0.011,0.1);
\draw[] (0.05,0) circle (0.1);
\end{tikzpicture}
}

\hypersetup{
    pdfencoding=unicode,
	colorlinks=true,
	urlcolor=Maroon,
	linkcolor=RoyalBlue,
	citecolor=Maroon,
	pdftitle={},
	pdfauthor={Ruth Britto and Holmfridur S. Hannesdottir},
	pdfdisplaydoctitle=true,
	pdfstartview=FitH,
	linktocpage=true
}

\def\d{\mathrm{d}}
\def\be{\begin{equation}}
\def\ee{\end{equation}}

\def\F{\mathcal{F}}
\def\U{\mathcal{U}}
\def\I{\mathcal{I}}

\def\triopm{\mathrm{op-}m}
\def\triadjm{\mathrm{adj-}m}
\def\trithree{\mathrm{hook}}
\def\threem{3\mathrm{m}}
\def\icecream{\mathrm{ice-sym}}
\def\icecreamA{\mathrm{ice-asym}}
\def\Li{\mathrm{Li}}
\def \D {D}
\def \E {E}
\def\rd{\mathrm{d}}

\def\GL{\mathrm{GL}}
\def\leq{\leqslant}
\def\geq{\geqslant}

\def\eps{\epsilon}
\def\bsp#1\esp{\begin{split}#1\end{split}}
\newcommand{\hypgeo}[4]{\,_2F_1\left(#1,#2;#3;#4\right)}
\def\tri{\mathrm{tri}}
\def\zb{\bar{z}}

\def \disc{\mathrm{Disc}}
\def \Cut{\mathrm{Cut}}
\def \LS{\mathrm{LS}}

\newcommand{\cF}{\mathcal{F}}
\newcommand{\cU}{\mathcal{U}}

\newcommand{\cI}{\mathcal{I}}
\newcommand{\cC}{\mathcal{C}}
\newcommand{\cA}{\mathcal{A}}
\newcommand{\cJ}{\mathcal{J}}

\definecolor{darkorange}{HTML}{E28413}
\definecolor{choral}{HTML}{E09891}
\definecolor{darkred}{HTML}{6B0F1A}
\definecolor{darkgreen}{rgb}{0.0, 0.4, 0.0}
\definecolor{darkmagenta}{rgb}{0.55, 0.0, 0.55}
\def\HSH#1{}
\def\rbnote#1{}

\tikzset{
    vector/.style={
        decoration={snake, aspect=0.75, mirror, segment length=2mm},
        decorate
    },
	photon/.style={decorate, decoration={snake, amplitude=1pt, segment length=6pt}
	}
}

\definecolor{charcoal}{HTML}{343837}

\title{Tracking discontinuities in parameter space}

\usepackage{orcidlink}
\author[\!a \orcidlink{0000-0003-2462-6481}]{Ruth Britto,}\emailAdd{britto@maths.tcd.ie}

\author[\!b \orcidlink{0000-0002-5440-2086}]{Holmfridur S. Hannesdottir}\emailAdd{hofie@ias.edu}

\affiliation{$^a$School of Mathematics and Hamilton Mathematics Institute, Trinity College, Dublin 2, Ireland}

\affiliation{$^b$Institute for Advanced Study, Einstein Drive, Princeton, NJ 08540, USA}

\abstract{We develop a geometric framework in Feynman-parameter space to determine constraints on the sequential discontinuities of Feynman integrals. Our method is based on tracking  the deformation of the integration contour as external kinematics are analytically continued. This procedure imposes powerful constraints on the analytic structure of Feynman integrals, providing crucial inputs for their bootstrap. 
We demonstrate the usefulness of this framework by applying it to integrals in dimensional regularization, with higher propagator powers, and to examples with non-uniform transcendental weight. The method is illustrated with several 
one- and two-loop calculations.
}

\begin{document} 

\maketitle

\setcounter{page}{2}

\section{Introduction}

Computations of scattering amplitudes for collider experiments usually involve computing their expansion in perturbation theory via Feynman diagrams. A more ambitious philosophy rooted in the S-matrix bootstrap program of the 1960s involved trying to compute amplitudes directly from physical principles such as unitarity, causality and locality. While this goal may have been too ambitious for the time, the core philosophy has been revived in recent years, leading to a host of different bootstrap programs. In planar $\mathcal{N}=4$ Super Yang-Mills theory, for instance, amplitudes are now computed by constraining a general ansatz with symmetries and analytic data~\cite{Arkani-Hamed:2022rwr}.
Analogous ideas underpin the bootstrap of Feynman integrals themselves (the Landau bootstrap~\cite{Hannesdottir:2024hke}),
the soft bootstrap ~\cite{Elvang:2018dco} which has been used to constrain the Wilson coefficients in effective field theories, and the modern S-matrix bootstrap program~\cite{Paulos:2016fap} (see~\cite{Kruczenski:2022lot} for a review). In these programs, first principles like unitarity, causality and locality are necessary but not sufficient to pin down the amplitudes. The crucial ingredient, and the focus of this paper, is a detailed knowledge of the amplitude's analytic structure---its branch points, discontinuities, and monodromies---which provides the ultimate constraints needed to determine the result.  Practitioners of the 1960s S-matrix bootstrap pioneered the probe of this information, paving the way for modern approaches that employ a precise understanding of the specific type of functions that are needed and the accompanying technology for treating divergences.

Understanding the analytic structure begins with locating the potential singularities in the complex planes of the kinematic variables. This is the role of the Landau equations~\cite{nakanishi1959,landau1959,Bjorken:1959fd,Fevola:2023kaw,Fevola:2023fzn,Helmer:2024wax,Caron-Huot:2024brh,Correia:2025yao}, whose solutions, called Landau singularities, pinpoint the kinematic configurations where an amplitude can become non-analytic. Once a singularity is located, we can deduce further information e.g.~from a series expansion in its neighborhood~\cite{landau1959, Hannesdottir:2021kpd}. It is further characterized by its \emph{discontinuities}, i.e.,\ the changes across its branch cuts. The Cutkosky rules~\cite{Cutkosky:1960sp,Bloch:2015efx,Muhlbauer:2022ylo} provide a recipe in loop-momentum space for computing these discontinuities using an integration contour that puts some internal particles on shell. More abstractly, the set of all possible discontinuities of an amplitude is constrained by its monodromy group, which governs how the function transforms as its kinematic arguments are analytically continued around its branch points~\cite{Ponzano:1969tk, Ponzano:1970ch,Regge:1972ns,Lee:2025rik}. 
This deep understanding of discontinuities now sets the stage for tackling the next level of complexity.

While discontinuities and their relation to cut integrals are fairly well understood (see~\cite{Britto:2024mna} for a review), 
deeper analytic structure can be probed if we learn to understand \emph{sequential discontinuities}: discontinuities of a previously computed discontinuity function of a Feynman integral or scattering amplitude. An early application of sequential discontinuities was the double dispersion relation of Mandelstam \cite{Mandelstam:1958xc}. They have more recently been applied in the context of the modern unitarity method initiated in \cite{Bern:1994zx,Bern:1994cg}, where classic unitarity cuts expose the first discontinuities \cite{Gaiotto:2011dt} and generalized cuts can be related to sequential discontinuities \cite{Abreu:2014cla,Abreu:2015zaa}. However, reconstructing the uncut integrals by this method relies on the ability to compute the values of the discontinuities, which is simple in some cases but can be complicated for general multiloop integrals.

Here, we would like to follow a more general approach based primarily on analyzing the existence of sequential discontinuities. This program started with the seminal papers on {Steinmann relations} from the 1960s, which forbid the presence of a sequential discontinuity in certain partially overlapping kinematic Mandelstam invariants as a consequence of causality in axomatic S-matrix theory~\cite{Steinmann,Steinmann2,araki:1961,Cahill:1973qp,Stapp:1976mx}. Later, a thesis by Pham~\cite{pham_thesis} described singularities of Feynman integrals geometrically 
in the on-shell space of loop momenta, also leading to constraints on sequential discontinuities~\cite{Hannesdottir:2022xki}.  Originally, these relations were only applicable to mass-gapped theories, but have more recently been extended to other cases~\cite{Drummond:2017ssj,Caron-Huot:2019bsq,Bourjaily:2020wvq,Benincasa:2020aoj}. The Steinmann relations have been a cornerstone in the perturbative bootstrap program in planar $\mathcal{N}=4$ supersymmetric Yang-Mills theory, providing valuable constraints on an ansatz such that the scattering amplitude can be computed to high loop order, and the Pham relations were recently used to determine scalar Feynman integrals in the Landau Bootstrap~\cite{Hannesdottir:2024hke}. This suggests that the mastery of sequential discontinuities can allow us to make progress in other quantum field theories as well.

Sequential-discontinuity constraints coming from \emph{genealogical constraints} and minimal cuts were formulated in Ref.~\cite{Hannesdottir:2024cnn}. The idea was to use the parametric-space formulation of Feynman integrals to constrain the existence of further discontinuities based on which boundaries $\alpha_i=0$ are not used to compute a certain discontinuity integral. If we start with a discontinuity integral $\disc_s \I$ whose integration contour is \emph{not} bounded by some $\alpha_{i}=0$, then all Landau singularities that require $\alpha_{i}=0$ are absent from $\disc_s \I$. Thus, all further discontinuities around those singularities vanish. Ultimately, the genealogical constraints are of the form $\disc_t \cdots \disc_s \cdots \I = 0$, where $t=0$ and $s=0$ correspond to two different Landau singularities, and the dots represent discontinuities around any branch points of $\I$. In other words, these formulas are simultaneously constraints on the sequential discontinuities of $\I$ and any analytic continuations of $\I$.

In this paper, we take a more detailed approach to analyzing sequential discontinuities. We will start with the original integration contour for $\I$ over the positive orthant of parameter space, and track what happens to the integral formulation as the external variables are analytically continued. This gives us more fine-grained information: while the genealogical constraints might allow $\disc_t \cdots \disc_s \cdots \I \neq 0$, the tracking of integration contours could still show that $\disc_t \, \disc_s  \, \I = 0$. We will also discuss sequential discontinuities around the same branch point, providing constraints such as $\disc_s \, \disc_s  \, \I = 0$.

Our method relies on being able to rewrite discontinuities as modified integration contours. To be more concrete, let us consider arguably the simplest example of this phenomenon: the logarithm, $\log(z) = \int_1^z \frac{\d t}{t}$. If we plug $z = -1$ into this integral representation, it naively looks like it diverges, since we integrate straight through the value $t=0$, where the denominator blows up. However, as is well known from elementary complex analysis, we simply have to deform the integration contour to either go slightly above the singularity at $t=0$ or slightly below $t=0$. The difference of the two choices is the discontinuity across the branch cut of the logarithm, or the monodromy around its branch point at $z=0$ and is given by a circular integration contour around the singularity, $(1-\mathscr{M}_{z=0}) \log(z) = \int_{\rm{\circleleft}} \frac{\d t}{t}$. The lesson we draw from this simple example is that monodromies can be represented by using the \emph{same} integrand integrated over a \emph{different} integration contour.

Feynman integrals in parametric space generally have an analogous branch-cut ambiguity as in the example of the logarithm. The ambiguity is resolved by the $i \varepsilon$ prescription, placing the physical value on the causal sides of all branch cuts. Analogous to the logarithmic example, we can relate monodromies around branch points to modified integration contours for the parametric integrand. In this formulation, a key role is played by the~\emph{boundaries of integration}: while the original integration contour is the positive orthant in the Feynman parameters $\alpha_i$, the discontinuity contours are instead bounded by some of the $\alpha_i=0$, and potentially the singular surfaces of the Symanzik polynomials, i.e.~$\cF=0$ and $\cU=0$, as discussed in~\cite{Britto:2023rig}, with a close connection to the ideas of~\cite{boyling1968homological,Berghoff:2022mqu}. Our first step will be to identify the homology class of the integral and what contours correspond to first discontinuities. 
These modified integration contours will be the starting point for studying sequential discontinuities: the next discontinuity we take can now be analyzed in the same way as first discontinuities, except that we start with a different integration contour than the positive orthant.

This paper is organized as follows. We propose that sequential discontinuities are visible in parameter space as singularities in the shape of integration regions. In Sec.~\ref{sec:monodromy_theory}, we explain how to obtain integration contours for sequential discontinuities. We discuss the dimensional and region-dependence of these integration contours. To illustrate this proposal in concrete examples at one and two loops in Secs.~\ref{sec:one-loop} and~\ref{sec:two-loop} respectively, we work at the level of the symbol of multiple polylogarithms.\footnote{See \cite{Arkani-Hamed:2017ahv,Gong:2022erh,Gong:2025qry} for strategies for obtaining symbols of 1-loop integrals in Feynman parameters.}  The first-entry letters are known to be mass invariants and physical thresholds. Given a first-entry letter, we are able to list the possible second-entry letters that can follow it. Combined with general considerations such as integrability of the symbol and symmetry of the integral, we find that these sequential constraints can be strong enough to fix the symbol uniquely at a given weight, up to overall normalization. We conclude and discuss future directions in~\ref{sec:conclusions}.

\section{Monodromies in Feynman parameter space}
\label{sec:monodromy_theory}

The integration contours corresponding to discontinuities and cuts of Feynman integrals in parameter space in dimensional regularization were argued in \cite{Britto:2023rig} to be bounded by specific subsets of the coordinate hyperplanes, along with the zero-locus of the second Symanzik polynomial, $\F = 0$. We now review this result and show how to generalize it to the case where we take monodromies around individual branch points, instead of total discontinuities. The former is needed for the generalization to sequential discontinuities, as discussed in more detail below.

\subsection{Preliminaries: discontinuities and monodromies in parameter space}

We start with the parametric representation of an $L$-loop Feynman integral with $E$ edges in $\D$ spacetime dimensions,  \begin{equation}
\label{eq:FP-gen}
\cI = \frac{e^{L\gamma_E \epsilon} \, 
\mu^{2 \lambda} \Gamma\left(\lambda\right)}{\prod_{i=1}^E \Gamma(\nu_i)}
\int_{\alpha_i \geq 0}
\frac{\rd^\E \alpha}{\text{GL}(1)}
\left(\prod_{i=1}^E \alpha_i^{\nu_i-1} \right)
\,\frac{\cU^{\lambda-\D/2}}
{(\F- i \varepsilon)^{\lambda}}\,,
\end{equation}
where $\cU$ and $\F$ are 
the first and second Symanzik polynomials, $\lambda = \nu- L \D/2$, $\nu = \sum_{i=1}^E \nu_i$, and $\epsilon$ is set so that $D+2\epsilon$ is the integer dimension of spacetime. The factor $e^{L\gamma_E \epsilon} \, 
\mu^{2 \lambda}$ is a convention \cite{Weinzierl:2022eaz}, where $\mu$ is a parameter of mass dimension 1, so that the overall integral is dimensionless. It is often set equal to 1 unless we need to keep track of dimensions and we will do so in some of the examples below.
In loop-momentum space, $\nu_i$ corresponds to the power of the propagator labeled with $i$, so setting $\nu_i=1$ corresponds to unit propagators. The $\text{GL}(1)$ in the denominator removes the redundancy of the integrand under rescalings ($\alpha_i \to \lambda \alpha_i$ for all $i$ and any $\lambda>0$). A common choice is to set  $\sum_{i=1}^\E \alpha_i=1$. 

Focusing on the kinematic dependence contained in the polynomial $\F$,  we use following identity obtained by flipping the sign of the $i \varepsilon$,
\begin{equation} \label{eq:discF}
    (\F - i\varepsilon)^{-\lambda} - (\F + i\varepsilon)^{-\lambda} =
    \begin{cases}
         -\theta(-\F) (-\F)^{-\lambda} 2 i \sin(\pi \lambda), & \text{if } \lambda \not\in \mathbb{Z}_+
        \\ 
        2 i \pi \delta^{\lambda-1} (-\F) / \Gamma(\lambda)
         & \text{if } \lambda \in \mathbb{Z}_+
    \end{cases}    
\end{equation} 
to find that the total discontinuity of $\I$ in the case where $\lambda \not\in \mathbb{Z}_+$ is given by\footnote{If the $i \varepsilon$ can be transferred to be associated to the Mandelstam invariants, e.g.\ taking $s \to s + i\varepsilon$ has the same effect as taking $\F \to \F + i \varepsilon$, then the formula in~\eqref{eq:simp_disc} becomes the discontinuity of $\I$ across the branch cut in $s$, $\disc_s \I$.}
\begin{equation}
    \I - \I \vert_{i \varepsilon \to - i \varepsilon} = - \frac{e^{L\gamma_E\epsilon} \mu^{2 \lambda} \, \Gamma\left(\lambda\right)}{\prod_{i=1}^E \Gamma(\nu_i)} 2 i \sin(\pi \lambda) \int_{\alpha_i \geq 0, \, \F <0}
    \frac{\rd^E \alpha}{\text{GL}(1)}
    \,\frac{\cU^{\kappa}}
    {(-\F)^{\lambda}} \,.
    \label{eq:simp_disc}
\end{equation} 
The most significant feature of the discontinuity integral (\ref{eq:simp_disc}) compared to the original integral (\ref{eq:FP-gen}) is the effect of the unit step function $\theta(-\F)$ identifying where the discontinuity is present, which  changes the integration region from the positive orthant bounded by all hyperplanes $\alpha_i=0$, to one bounded as well by $\F=0$. In addition, the discontinuity integrand contains a phase factor proportional to $2 i \sin(\pi \lambda)$. For example, if $\lambda$ is a half-integer, corresponding to a square-root branch cut, this phase factor is equal to $\pm 2 i$, indicating that the square root branch cut separates values of different signs, and leading to a difference proportional to a factor of 2. For other values of $\lambda$, the phase factor represents the value obtained after analytic continuation around the branch cut $\F^\lambda$. Note that the multivalued function $\F^\lambda$ that was responsible for the discontinuity has been replaced by $(-\F)^\lambda$, avoiding all branch-cut ambiguity.

The formula of equation (\ref{eq:simp_disc}), where we replace the $i \varepsilon$ in the denominator with a $-i \varepsilon$, gives the total discontinuity of the Feynman integral, following the terminology of~\cite{Britto:2024mna}. An alternative prescription is to compute the {monodromy}, which is obtained by analytically continuing around a specific branch point in the external kinematic space and tracking what happens to the integration contour along the way. Note that total discontinuities and monodromies are two different prescriptions for manipulations in the external variables: one is obtained by flipping the signs of all $i \varepsilon$, while the other involves an analytic continuation in the space of external variables. When analyzing sequential discontinuities, the latter is the only option, since the expression in~\ref{eq:simp_disc} does not have any $i \varepsilon$ and thus the prescription cannot be iterated. Computing the monodromy instead of the total discontinuity, however, allows for an iteration. Since we here aim to explore what happens to sequential discontinuities, we start by classifying the different contours over which the integral can be evaluated (i.e., determine the homology classes), and the relation of different integration contours to monodromies.

Since total discontinuities and monodromies are different operations, the resulting integration contours that compute them may also differ. The brute-force way to find the monodromy contour is to analytically continue the integral in the kinematic parameters, and manually track what happens to the integration contour as we do so. A powerful result from Picard-Lefschetz theory gives us a shortcut for computing the monodromy as follows:
\begin{equation}
    (1 - \mathscr{M}_s) \I = \frac{e^{L\gamma_E\epsilon} \, \Gamma\left(\lambda\right)}{\prod_{i=1}^E \Gamma(\nu_i)} N_{\gamma} \int_{\gamma}
    \frac{\rd^E \alpha}{\text{GL}(1)}
    \,\frac{\cU^{\kappa}}
    {(-\F)^{\lambda}} \,.
    \label{eq:simp_mon}
\end{equation} 
Here, the operator $\mathscr{M}_{s}$ denotes an analytic continuation in the kinematic space around a branch point at $s=0$,\footnote{Here, we use $s$ for clarity to represent any kinematic variety of $\I$. Note that $s$ need not be a Mandelstam invariant.} and $N_{\gamma}$ is an integer-valued number called the {intersection number}.
The monodromy contour $\gamma$ can be deduced from the so-called \emph{vanishing cell}, which is a cycle in the homology that continuously shrinks to a point when the external variables are analytically continued towards the branch point. Since we will only need to know the vanishing cell for our purposes, we will not go into details of Picard-Lefschetz theory here (see~\cite{hannesdottir:2022bmo,Hannesdottir:2022xki,Berghoff:2022mqu} for references), but rather illustrate the basic idea on a simple example.

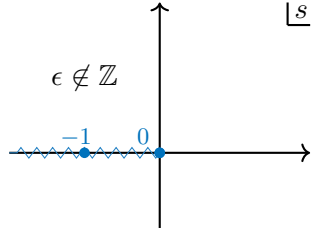
\begin{figure}
    \centering
    \begin{tikzpicture}[scale=1]
    \draw[->,thick] (-2,0) -- (2,0);
    \draw[->,thick] (0,-1) -- (0,2);
    \draw[thick] (1.7,1.7) -- (2,1.7);
    \draw[thick] (1.7,1.7) -- (1.7,2);
    \draw[thick] (1.65,1.65) node[above right] {$s$};
    \draw[decorate, decoration={zigzag, segment length=6, amplitude=2}, RoyalBlue!80] (0,0) -- (-2,0);
    \fill[RoyalBlue] (0,0) circle[radius=0.07] node[below left, yshift=1.2em] {\footnotesize$0$};
    \fill[RoyalBlue] (-1,0) circle[radius=0.07] node[below left, yshift=1.2em,xshift=0.6em] {\footnotesize$-1$};
    \fill[RoyalBlue] (0,0) circle[radius=0.07] node[above, yshift=-20] {};
    \node[] at (-1,1) {$\epsilon \not\in \mathbb{Z}$};
    \coordinate (start2) at (0,0); 
    \coordinate (end2) at (1,0); 
    \coordinate (p1) at (0.25,0.2);      
    \coordinate (p2) at (0.5,0.3);     
    \coordinate (p3) at (0.75,0.2);     
    \end{tikzpicture}
    \caption{The analytic structure of $\cI$ in the complex $s$ plane.}
    \label{fig:complex_s}
\end{figure}

Consider the following integral.
\begin{equation}
    \I = \int_0^1 \frac{\d x}{(x+s + i \varepsilon)^{1-\epsilon}} = \frac{(s + i\varepsilon+1)^{\epsilon }-(s+i\varepsilon)^{\epsilon }}{\epsilon } \,.
    \label{eq:toy_I_monodromy}
\end{equation}
Note that the integral formulation of $\cI$ has a pole at $x = - s$. If $s<0$, we need to use an integration contour that avoids the pole, as determined by the $i \varepsilon$. This prescription works everywhere except at the end points of the integral, at $x=0$ and $x=1$, which cannot be shifted. Thus, the Landau singularities of this integral can be found as the points where the pole at $x=-s$ hits one of the endpoints, i.e., at
\begin{equation}
    s = 0\,, \qquad s = -1 \,.
\end{equation}
When compactifying the $x$-space, we also find a Landau singularity at $s=\infty$, although it is not a singularity on the principal sheet of $\cI$ (note that the singularity cancels between the two terms on the r.h.s.~of Eq.~\eqref{eq:toy_I_monodromy}).
Thus, in the complex space of the external variable $s$, this integral has a branch cut on the negative real axis with branch points at $s=0$, $s=-1$ and $s=\infty$. See Figure~\ref{fig:complex_s}. The $i\varepsilon$ tells us which side of the branch cut in $x$ the integral is evaluated on (i.e., deforms the contour into the upper half-plane, assuming  $\varepsilon>0$).

\begin{figure}
\centering

    \begin{subfigure}[]{0.45\textwidth}
    \centering
    
    \begin{tikzpicture}[scale=1]
    \draw[decorate, decoration={zigzag, segment length=6, amplitude=2}, RoyalBlue!80] (-2,0) -- (0.6,0);
    \draw[->,thick] (-2,0) -- (2,0);
    \draw[->,thick] (0,-1) -- (0,2);
    \draw[thick] (1.7,1.7) -- (2,1.7);
    \draw[thick] (1.7,1.7) -- (1.7,2);
    \draw[thick] (1.65,1.65) node[above right] {$x$};
    \fill[Maroon] (0,0) circle[radius=0.07] node[below left, yshift=1.2em] {\footnotesize$0$};
    \fill[Maroon] (1,0) circle[radius=0.07] node[below right, yshift=1.2em] {\footnotesize$1$};
    \draw[->, line width=0.9, Maroon] (0.8,0.6) -- (0.7,0.4);
    \node[Maroon, scale=0.7, align=center] at (1,0.8) {original contour}; 
    \fill[RoyalBlue] (0.6,0) circle[radius=0.07] node[above, yshift=-20] {\footnotesize$x=-s$};
    \coordinate (start2) at (0,0); 
    \coordinate (end2) at (1,0); 
    \coordinate (p1) at (0.25,0.2);      
    \coordinate (p2) at (0.5,0.3);     
    \coordinate (p3) at (0.75,0.2);     
    \draw[line width=1.2, Maroon] 
        (start2) to[out=45, in=-135] 
        (p1) to[out=45, in=180] 
        (p2) to[out=0, in=135] 
        (p3) to[out=-45, in=135] 
        (end2);
    \draw[Maroon, line width=1.2, -{>[scale=0.6]}] ($(0.2,0.15)$) -- ($(0.2,0.15)+(0.05,0.025)$);
    \end{tikzpicture}
    \end{subfigure}
    \begin{subfigure}[]{0.45\textwidth}
    \centering
    
    \begin{tikzpicture}[scale=1]
    \draw[decorate, decoration={zigzag, segment length=6, amplitude=2}, RoyalBlue!80] (-2,0) -- (0.2,0);
    \draw[->,thick] (-2,0) -- (2,0);
    \draw[->,thick] (0,-1) -- (0,2);
    \draw[thick] (1.7,1.7) -- (2,1.7);
    \draw[thick] (1.7,1.7) -- (1.7,2);
    \draw[thick] (1.65,1.65) node[above right] {$x$};
    \coordinate (start) at (0.2,0); 
    \coordinate (end) at (-0.3,0.5); 
    \draw[decorate, decoration={zigzag, segment length=6, amplitude=1}, RoyalBlue!80]
    (start) to [out=110, in=-45] 
    (end);
    \fill[Maroon] (0,0) circle[radius=0.07] node[below left, yshift=1.2em,xshift=-0.8em] {\footnotesize$0$};
    \fill[Maroon] (1,0) circle[radius=0.07] node[below right, yshift=1.2em] {\footnotesize$1$};
    \draw[->, line width=0.9, Maroon, yshift=10] (0.8,0.6) -- (0.7,0.4);
    \node[Maroon, scale=0.7, align=center, yshift=20] at (1,0.8) {during analytic\\continuation}; 
    \fill[RoyalBlue] (-0.3,0.5) circle[radius=0.07] node[above, yshift=0] {\footnotesize$x=-s$};
    \coordinate (start2) at (0,0); 
    \coordinate (end2) at (1,0); 
    \coordinate (p1) at (-0.4,0.3);  
    \coordinate (p2) at (-0.3,0.8);   
    \coordinate (p3) at (-0.1,0.7);
    \coordinate (p4) at (0.1,0.5);
    \draw[line width=1.2, Maroon] 
        (start2) to[out=135, in=-45] 
        (p1) to[out=135, in=180] 
        (p2) to[out=0, in=135] 
        (p3) to[out=-45, in=135] 
        (p4) to[out=-45, in=135] 
        (end2);
    \draw[Maroon, line width=1.2, -{>[scale=0.6]}] ($(0.25,0.39)$) -- ($(0.25,0.39)+(0.05,-0.025)$);
    \end{tikzpicture}
    \end{subfigure}
    \begin{subfigure}[]{0.45\textwidth}
    \centering
    
    \begin{tikzpicture}[scale=1]
    \draw[decorate, decoration={zigzag, segment length=6, amplitude=2}, RoyalBlue!80] (-2,0) -- (-0.4,0);
    \draw[->,thick] (-2,0) -- (2,0);
    \draw[->,thick] (0,-1) -- (0,2);
    \draw[thick] (1.7,1.7) -- (2,1.7);
    \draw[thick] (1.7,1.7) -- (1.7,2);
    \draw[thick] (1.65,1.65) node[above right] {$x$};
    \fill[Maroon] (0,0) circle[radius=0.07] node[below left, yshift=1.2em,xshift=-0.3em] {\footnotesize$0$};
    \fill[Maroon] (1,0) circle[radius=0.07] node[below right, yshift=1.2em] {\footnotesize$1$};
    \draw[->, line width=0.9, Maroon, yshift=10] (0.8,0.6) -- (0.7,0.4);
    \node[Maroon, scale=0.7, align=center, yshift=20] at (1,0.8) {after analytic\\continuation}; 
    \fill[RoyalBlue] (0.6,0) circle[radius=0.07] node[above, yshift=-25] {\footnotesize$x=-s$};
    \coordinate (start2) at (0,0); 
    \coordinate (end2) at (1,0); 
    \coordinate (p3) at (0.8,0);
    \coordinate (p5) at (0,-0.3);
    \coordinate (p55) at (-0.2,0);
    \coordinate (p6) at (0,0.5);
    \coordinate (p7) at (0.5,0.7);
    \draw[line width=1.2, Maroon] 
        (start2) to[out=45, in=100] 
        (p3) to[out=-90, in=-30] 
       (p5) to[out=150, in=-90] 
       (p55) to[out=90, in=-135] 
        (p6) to[out=45, in=180] 
        (p7) to[out=0, in=90] 
        (end2);
    \draw[Maroon, line width=1.2, -{>[scale=0.6]}] ($(0.35,0.18)$) -- ($(0.35,0.18)+(0.05,0.025)$);
    \coordinate (start) at (-0.4,0); 
    \coordinate (end) at (0.6,0); 
    \coordinate (q1) at (0,-0.5);  
    \coordinate (q2) at (0.87,0);   
    \coordinate (q3) at (0.4,0.45);   
    \coordinate (q4) at (-0.1,0);   
    \draw[decorate, decoration={zigzag, segment length=6, amplitude=0.8}, RoyalBlue!80]
    (start) to [out=-45, in=180] 
    (q1) to [out=0, in=-90]
    (q2) to [out=90, in=0]
    (q3) to [out=180, in=90]
    (q4) to [out=-90, in=-125]
    (end);
    \end{tikzpicture}
    \end{subfigure}
    \begin{subfigure}[]{0.45\textwidth}
    \centering
    
    \begin{tikzpicture}[scale=1]
    \draw[decorate, decoration={zigzag, segment length=6, amplitude=2}, RoyalBlue!80] (-2,0) -- (0.7,0);
    \draw[->,thick] (-2,0) -- (2,0);
    \draw[->,thick] (0,-1) -- (0,2);
    \draw[thick] (1.7,1.7) -- (2,1.7);
    \draw[thick] (1.7,1.7) -- (1.7,2);
    \draw[thick] (1.65,1.65) node[above right] {$x$};
    \fill[Maroon] (0,0) circle[radius=0.07] node[below left, yshift=1.2em,xshift=-0.3em] {\footnotesize$0$};
    \fill[Maroon] (1,0) circle[radius=0.07] node[below right, yshift=1.2em] {\footnotesize$1$};
    \draw[->, line width=0.9, Maroon, yshift=10] (0.9,0.6) -- (0.76,0.2);
    \node[Maroon, scale=0.7, align=center, yshift=20] at (1,0.8) {monodromy\\contour}; 
    \fill[RoyalBlue] (0.6,0) circle[radius=0.07] node[above, yshift=-25] {\footnotesize$x=-s$};
    \coordinate (start2) at (0,0); 
    \coordinate (p1) at (0.5,0.2);  
    \coordinate (p2) at (0.85,0);   
    \coordinate (p3) at (0.5,-0.2);
    \draw[line width=1.2, Maroon] 
        (start2) to[out=45, in=180] 
        (p1) to[out=0, in=90] 
        (p2) to[out=-90, in=0] 
        (p3) to[out=180, in=-45]
        (start2);
    \draw[Maroon, line width=1.2, -{>[scale=0.6]}] ($(0.25,0.15)$) -- ($(0.25,0.15)+(0.05,0.025)$);
    \end{tikzpicture}
    \end{subfigure}
    
    \caption{\textbf{Top left.} The original integration contour for $\I$ avoids the singularity of the integrand by a deformation in the upper half-plane. \textbf{Top right.} During analytic continuation of $s$ in the complex plane in a counterclockwise circle around $s=0$, the root at $x=-s$ traverses around $x=0$, dragging the integration contour with it. \textbf{Bottom left.} After the analytic continuation in $s$, the integration contour has been suitably modified. \textbf{Bottom right.} After subtracting the original contour, we see that the contour representing the monodromy of $\I$ around $s=0$ is given as a contour around the branch cut.}
    \label{fig:monodromies_tracking}
\end{figure}
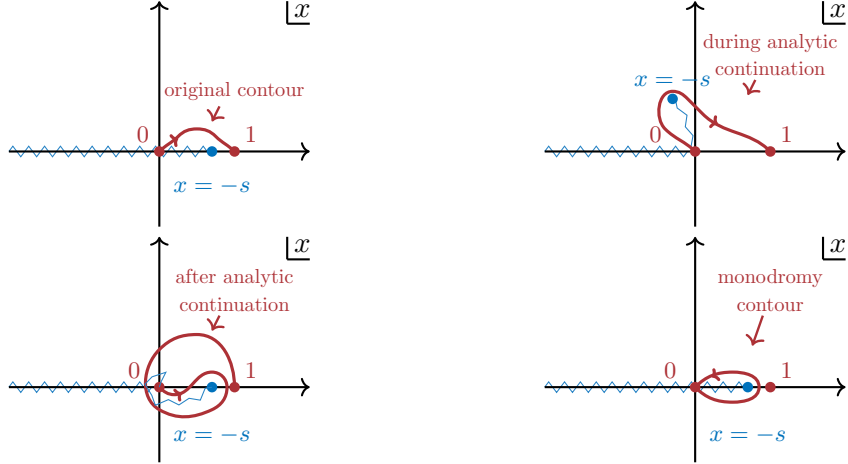

In the space of the integration variable $x$,  the integral formulation of $\cI$ has a branch cut for $x \leq -s$. To find the monodromy contour, we track what happens to the original integration contour as we take $s$ in a circle around $s=0$, see Fig.~\ref{fig:monodromies_tracking}. As we do so, the branch cut in $x$ space winds around $x=0$, dragging the integration contour along. Finally, after the analytic continuation, the difference between the original contour and the one after analytic continuation in $s$ can be represented as a circle around the branch cut. Since the branch cut is a fractional one in our example, we can write
\begin{equation}
    (1 - \mathscr{M}_s) \I  = \int_{\mathcal{C}} \frac{\d x}{(x+s)^{1-\epsilon}} = - 2 i \sin(\pi \epsilon) \int_0^{-s} \frac{\d x}{(-x-s)^{1-\epsilon}} \,,
\end{equation}
where $\mathcal{C}$ is the monodromy contour from Fig.~\ref{fig:monodromies_tracking}.
The integration contour that vanishes as we analytically continue the external kinematics close to the branch point is called the \emph{vanishing cell}.

Since the monodromy is still captured by an analytic function, we can iterate this process and take sequential monodromies, still tracking the integration contours as we keep traversing $s=0$. Every time the branch point is encircled, we pick up a factor of $-2 i \sin(\pi \epsilon)$, resulting in the monodromies
\begin{equation}
    (1 - \mathscr{M}_s)^N \I = (-2 i \sin(\pi \epsilon))^{N} \int_0^{-s} \frac{\d x}{(-x-s)^{1-\epsilon}} \,.
\end{equation}

In summary, we have demonstrated that as we subtract the analytic continuation of $\I$ around  its branch point at $s=0$ from $\I$, the resulting monodromy can be represented using the same integrand as before, but integrated over a modified integration contour. Next, we will identify all possible contours in the homology class for a Feynman integral, and relate the modified contours from monodromies to cuts of Feynman integrals.

\subsection{Cuts in parameter space}
We can generalize the Feynman integral of~\eqref{eq:FP-gen} by considering various integration contours denoted by $\Gamma$. 
Instead of integrating over the positive orthant, where $\alpha_i \geq 0$ for all $i$, we can instead consider contours bounded by $\alpha_i=0$ for a subset of $i$'s, along with $\F=0$ and/or $\cU=0$. The claim of \cite{Britto:2023rig} is that such integrals compute Cutkosky cuts~\cite{Cutkosky:1960sp} of $\cI$ (up to a factor of $\frac{\sin(\pi \lambda)}{\pi}$, where $\lambda=\nu-LD/2$ as before, which are closely connected to discontinuities and monodromies, see \cite{Britto:2024mna} for a recent overview of these concepts.\footnote{When comparing monodromies to cut integrals via the cutting rules, the overall sign depends on whether we encircle the branch point clockwise or counterclockwise. The sign ambiguity can easily be fixed by numerical evaluation at one point, so we do not worry about it in our definition of cut integrals. See~\cite{pham_thesis,pham1968singularities,Hannesdottir:2022xki} for a detailed discussion of the sign.} Then, we define parametric cut integrals $\cC_\Gamma$ as\footnote{We also assume  $\D=4-2\epsilon$ spacetime dimensions, allowing us to write $\sin(\pi \lambda) = \pm \sin(L\pi \epsilon)$.}
\begin{equation}
\label{eq:cut-integrals}
\cC_\Gamma \cI = 
\frac{\sin(L \pi \eps)}{\pi}
\frac{e^{L\gamma_E\epsilon} \mu^{2\lambda} \Gamma\left(\nu-\frac{L\D}{2}\right)}{\prod_{i=1}^E \Gamma(\nu_i)}
\int_{\Gamma}  \frac{\rd^\E \alpha}{\text{GL}(1)}
\left(\prod_{i=1}^E \alpha_i^{\nu_i-1} \right)
\,\frac{\cU^{\nu-(L+1)\D/2}}
{(-\F)^{\nu-L\D/2}}\,.
\end{equation}
The choice of monodromy determines the specific contour $\Gamma$, in particular, which subset of hyperplanes $\alpha_i=0$ are included in its boundary. 
Note, however, that specifying the boundaries is not sufficient to identify the contour uniquely, since there may be multiple chambers with the same set of boundary elements. In our examples, we will use the contours obtained by taking monodromies around external kinematic invariants.  
Because the integrand is multivalued in dimensional regularization, the integration contours should most properly be constructed as elements of twisted homology \cite{AomotoKita} to include information about the Riemann sheet. 

\paragraph{Example.}

To illustrate the key ideas, we introduce a running example, an asymmetric one-loop triangle with three scales, which we denote by $\I^{\trithree}_{\tri}$:
\begin{equation}
\I^{\trithree}_{\tri} =
    \begin{gathered}
    \begin{tikzpicture}[line width=1,scale=0.85]
    	\coordinate (v1) at (0,0);
        \coordinate (v2) at (0.866025,0.5);
        \coordinate (v3) at (0.866025,-0.5);
        \draw[] (v1) -- (v2) -- (v3) -- (v1);
        \draw[line width=2] (v3) -- (v1) node[below,yshift=-5,xshift=3]{$m_1$};
        \node[scale=0.7] at (0.5,-0.1) {$\alpha_1$};
        \node[scale=0.7] at (1.1,0.3) {$\alpha_2$};
        \node[scale=0.7] at (0.3,0.4) {$\alpha_3$};
        \draw[line width=2] (v2) -- (v3) node[right,yshift=8]{$m_2$};
        \draw[] (v1) -- ++(-190:0.7);
        \draw[] (v1) -- ++(-170:0.7);
        \draw[] (v3) -- ++(-45:0.7);
        \draw[] (v2) -- ++(45:0.7);
        \node[] at ($(v1)-(1.0,0.2)$) {$p$};
    \end{tikzpicture}
    \end{gathered}
\end{equation}
For this triangle, we have the Symanzik polynomials
\begin{equation}
    \cU = \alpha_1 + \alpha_2 + \alpha_3, 
    \qquad \F= -\alpha_1 \alpha_3 p^2 + \cU \left(\alpha_1 m_{1}^2 + \alpha_2 m_{2}^2 \right) \,,
\end{equation}
and the Feynman integral in $\D=4-2\epsilon$ dimensions evaluates to \cite{Abreu:2015zaa} 
\begin{align}
\I^{\trithree}_{\tri}
& = \frac{e^{\gamma_E\epsilon} \mu^{2+2\epsilon} \Gamma(1{+}\epsilon)}{\epsilon(1-\epsilon)\left(m_{1}^2{-}m_{2}^2\right)}\Bigg[\left(m_{2}^2\right)^{-\epsilon}\hypgeo{1}{1}{2-\epsilon}{\frac{p^2}{m_{1}^2{-}m_{2}^2}}
\nonumber
\\
& \hspace{5cm}
-\left(m_{1}^2\right)^{-\epsilon}\,F_1\left(1;1,\epsilon;2-\epsilon;\frac{p^2}{m_{1}^2{-}m_{2}^2};\frac{p^2}{m_{1}^2}\right) \Bigg]
\nonumber
\\
& = \frac{\mu^2}{p^2}\left[\text{Li}_2\left(\frac{m_{1}^2}{m_{2}^2}\right)-\text{Li}_2\left(\frac{m_1^2{-}p^2}{m_{2}^2}\right)
-\log \left(1-\frac{m_1^2{-}p^2}{m_{2}^2}\right) \log\left(1-\frac{p^2}{m_{1}^2}\right)
\right.\\
&\left. \hspace{3cm}
+\log\left(\frac{m_{2}^2}{m_{1}^2}\right) \log \left(1-\frac{p^2}{m_{1}^2{-}m_{2}^2}\right)\right]+\mathcal{O}\left(\epsilon\right)\, ,
\nonumber
\end{align}
where $\text{Li}_2$ is the dilogarithm function, ${}_2F_1$ is the Gauss hypergeometric function, and $F_1$ is the first Appell hypergeometric function.
As we can see from the expansion in $\epsilon$, this triangle is finite in four dimensions. Since we will eventually consider divergent integrals, we consider also the weight 3 terms in the series expansion in $\epsilon$.\footnote{In this paper, we assign $\epsilon$ a weight of 0, not $-1$ which is sometimes conventional in discussing functions of uniform transcendental weight.}

From these expressions, we can work out the symbols. At weight 2 (i.e.\ in $\D=4$), it is
\begin{equation}
\label{eq:t3-sym-wt2}
\mathcal{S}_2(\I^{\trithree}_{\tri})
=
m_{1}^2\otimes\left(\frac{m_{1}^2{-}m_{2}^2}{m_{2}^2}\right)
+m_{2}^2\otimes\left(\frac{m_{1}^2{-}m_{2}^2{-}p^2}{m_{1}^2{-}m_{2}^2}\right)
-(p^2{-}m_1^2)\otimes\left(\frac{
m_{1}^2{-}m_{2}^2{-}p^2
}{m_{2}^2}\right)
\, ,
\end{equation}
while the $\mathcal{O}(\epsilon)$ term at weight 3 has the symbol 
\begin{align}
\mathcal{S}_3(\I^{\trithree}_{\tri}) = &
m_{1}^2 \otimes \frac{p^2}{p^2{-}m_1^2} \otimes
  \frac{m_{1}^2{-}m_{2}^2{-}p^2}{m_{2}^2}
   +m_{1}^2\otimes
   m_{1}^2\otimes \frac{m_{2}^2}{m_{1}^2{-}m_{2}^2}
   \nonumber
   \\ &
   \hspace{-1.9cm}
   +m_{1}^2\otimes \frac{m_{1}^2{-}m_{2}^2}{m_{2}^2} \otimes
   \frac{p^2}{m_{2}^2(m_{1}^2{-}m_{2}^2{-}p^2)}
   \label{eq:t3-sym-wt3}
    \\ &
    \hspace{-1.9cm}
   + m_{2}^2 \otimes m_{2}^2 \otimes \frac{m_{1}^2{-}m_{2}^2}{m_{1}^2{-}m_{2}^2{-}p^2}
   + m_{2}^2 \otimes \frac{m_{1}^2{-}m_{2}^2{-}p^2}{m_{1}^2{-}m_{2}^2} \otimes \frac{p^2}{m_{2}^2(m_{1}^2{-}m_{2}^2{-}p^2)}
   \nonumber
   \\ &
   \hspace{-1.9cm}
  -\left(p^2{-}m_1^2\right)\otimes \frac{m_{1}^2{-}m_{2}^2{-}p^2}{m_{2}^2} \otimes  \frac{p^2}{m_{2}^2(m_{1}^2{-}m_{2}^2{-}p^2)}
   +\left(p^2{-}m_1^2\right)\otimes \frac{\left(p^2{-}m_1^2\right)^2}{p^2} \otimes \frac{m_{1}^2{-}m_{2}^2{-}p^2}{m_{2}^2}\,.
   \nonumber
\end{align}

\begin{figure}[t]
\centering
\captionsetup[subfigure]{labelformat=empty}
\begin{subfigure}{0.32\textwidth}
\includegraphics[width=0.9\linewidth]{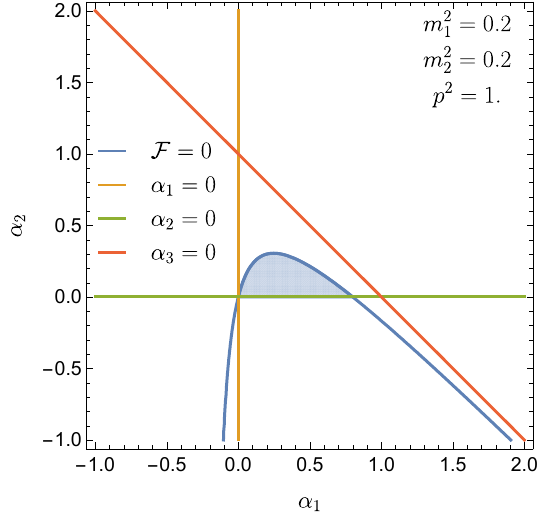} 
\caption{(a) $\Gamma_{p^2{-}m_1^2}$ in  $R^{p^2{-}m_1^2}$}
\label{fig:subim1}
\end{subfigure}
\begin{subfigure}{0.32\textwidth}
\includegraphics[width=0.9\linewidth]{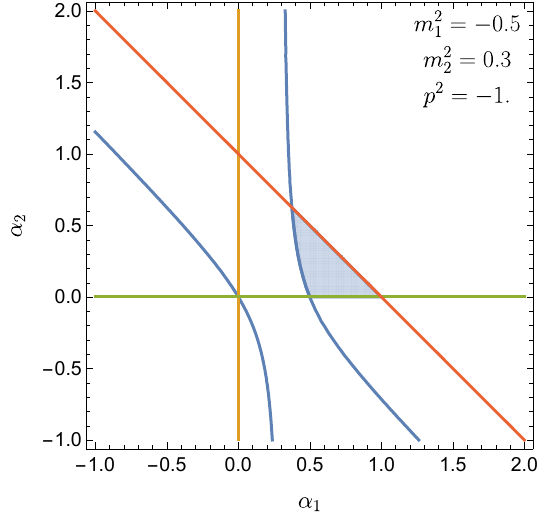}
\caption{(b) $\Gamma_{m_1^2}$ in $R^{m_1^2}$}
\label{fig:subim2}
\end{subfigure}
\begin{subfigure}{0.32\textwidth}
\includegraphics[width=0.9\linewidth]{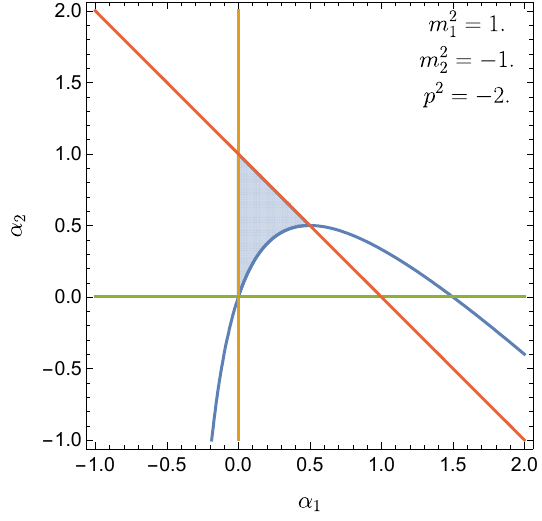}
\caption{(c) $\Gamma_{m_2^2}$ in $R^{m_2^2}$}
\label{fig:subim3}
\end{subfigure}
\caption{The vanishing cells $\Gamma$ are shaded, for the respective discontinuities in $p^2$, $m_1^2$, and $m_2^2$, each shown in their preferred kinematic region, as seen by the values listed in the plots. }
\label{fig:image2}
\end{figure}

To illustrate the cut integration contours of $\I^{\trithree}_{\tri}$, we set
$\alpha_3=1-\alpha_1-\alpha_2$ so that we can plot them in the plane of $\alpha_1$ and $\alpha_2$.
In the region of phase space where $p^2<m_1^2,  m_{1}^2>0, ~m_{2}^2>0$, which we denote by $R_0$, the integral has no discontinuities, and the variety $\F=0$ does not intersect the interior of the standard simplex.
In short, because $\F$ is always positive within the integration domain, the integral has no branch cuts in the $R_0$ region.

Starting in $R_0$, we can find the discontinuity in the $p^2$ variable by analytically continuing using an infinitesimal imaginary part until it crosses the physical threshold at $p^2=m_{1}^2$.
We denote by $R^{p^2{-}m_1^2}$ the region of phase space where $p^2>m_{1}^2>0$ and $m_{2}^2>0$. As shown in Figure \ref{fig:subim1}, $\F=0$ now intersects the Feynman-integral integration region. The vanishing cell $\Gamma_{p^2{-}m_1^2}$ is the contour that shrinks to zero as we approach the singularity, and is bounded by $\F=0$ and $\alpha_2=0$. To compute the discontinuity, we integrate over the contour $\Gamma_{p^2{-}m_1^2}$ with the appropriate phase factor.

Similarly, the discontinuity in the mass variable $m_1^2$ is obtained by starting in $R_0$ and varying $m_1^2$ until it takes negative values. We denote by $R^{m_1^2}$ the region of phase space where $p^2<m_{1}^2<0, ~m_{2}^2>0$. The contour $\Gamma_{m_1^2}$, shown in Figure \ref{fig:subim2}, is bounded by $\F=0$, $\alpha_2=0$, and $\alpha_3=0$. Finally, the discontinuity in  $m_2^2$ is obtained by varying $m_2^2$ out of the region $R_0$ until it takes negative values. We denote by $R^{m_2^2}$ the region of phase space where $p^2<0,~m_{1}^2>0, ~m_{2}^2<0$. The contour $\Gamma_{m_2^2}$, shown in Figure \ref{fig:subim3}, is bounded by $\F=0$, $\alpha_1=0$, and $\alpha_3=0$.

The integrals over the contours reproduce the explicit expressions for cuts obtained in \cite{Abreu:2015zaa}:
\begin{align}
\cC_{\Gamma_{p^2{-}m_1^2}} \I^{\trithree}_{\tri} &= 
\frac{e^{\gamma_E\epsilon} \mu^{2+2\epsilon} \sin(\pi \eps)}{\pi}
\Gamma\left(3-\frac{\D}{2}\right)
\int_0^{\frac{p^2-m_{1}^2}{p^2}} \d{\alpha_1} \int_0^{\F=0} \d{\alpha_2}\,
{(-\F)^{\frac{\D}{2}-3}} \\
\nonumber
&= 
\frac{e^{\gamma_E\epsilon} \mu^{2+2\epsilon} \sin(\pi \eps)}{\pi}
\frac{\Gamma\left(3-\frac{\D}{2}\right) \Gamma\left(\frac{\D}{2}-1\right) \Gamma\left(\frac{\D}{2}-2 \right)}{\Gamma(\D-2)}
\frac{  (p^2{-}m_1^2)^{\D-3} }{m_2^2 \,(p^2)^{\frac{\D}{2}-1}}
\\ & \hspace{5cm} \times
{}_2F_1\left( 1, \frac{\D}{2}-1,\D-2, \frac{m_1^2-p^2}{m_2^2} \right) \,,
\nonumber
\end{align}
\begin{align}
\cC_{\Gamma_{m_1^2}} \I^{\trithree}_{\tri} &= \frac{e^{\gamma_E\epsilon} \mu^{2+2\epsilon} \sin(\pi \eps)}{\pi}
\Gamma\left(3-\frac{\D}{2}\right)
\int_0^{\frac{m_1^2}{m_1^2-m_2^2}} \d{\alpha_2} \int_{\F=0}^{1-\alpha_2} \d{\alpha_1}\,
{(-\F)^{\frac{\D}{2}-3}} \\
&= 
\frac{e^{\gamma_E\epsilon} \mu^{2+2\epsilon} \sin(\pi \eps)}{\pi}
\frac{2 \Gamma\left(2-\frac{\D}{2}\right) (-m_1^2)^{\frac{\D-2}{2}}}{(\D-2) p^2 (m_2^2-m_1^2)}
F_1\left(1;1,2-\frac{\D}{2};\frac{\D}{2};\frac{m_1^2}{m_1^2-m_2^2}, \frac{m_1^2}{p^2}\right) \,,
\nonumber
\end{align}
\begin{align}
\cC_{\Gamma_{m_2^2}} \I^{\trithree}_{\tri} &= 
\frac{e^{\gamma_E\epsilon} \mu^{2+2\epsilon} \sin(\pi \eps)}{\pi}
\Gamma\left(3-\frac{\D}{2}\right)
\int_0^{\frac{m_2^2}{m_2^2-m_1^2}} \d{\alpha_1} \int_{\F=0}^{1-\alpha_1} \d{\alpha_2}\,
{(-\F)^{\frac{\D}{2}-3}} \\
&= \frac{e^{\gamma_E\epsilon} \mu^{2+2\epsilon} \sin(\pi \eps)}{\pi}
\frac{2 \Gamma\left(2-\frac{\D}{2}\right) (-m_2^2)^{\frac{\D-4}{2}}}{(\D-2)(m_1^2-m_2^2)}
{}_2F_1\left(1,1;\frac{\D}{2};\frac{p^2}{m_1^2-m_2^2} \right) \,.
\nonumber
\end{align}
These integrals can be evaluated cleanly by a change of variables, replacing the first integration variable by $-\F$ with the appropriate Jacobian factor. The results then yield the familiar integral formulas for the Gauss hypergeometric function ${}_2F_1$ and the Appell function $F_1$. In general, it can be complicated to take $\F=0$ as a boundary of the integration contour, because it is an endpoint singularity for the integrand. The techniques of \cite{Jones:2025jzc} address this problem. Other proposals for nonstandard parametric contours are presented in \cite{Chen:2020wsh,Chen:2025kdy}.

We remark that it is also possible to obtain the value of the maximal cut by integrating over a contour $\Gamma_{\rm max}$ bounded only by $\F=0$ and none of the coordinate hyperplanes. However, in the analysis of this paper, we are focusing on constructing the symbol, and the cuts listed above are the only ones corresponding to first-entry symbol letters. The maximal cut of a triangle corresponds to a symbol of weight two. For example, it can be interpreted as a sequence of mass and channel discontinuities.  Our framework allows for sequential discontinuities of this sort as well, and we will comment on it briefly in the following subsection.

\subsection{Sequential discontinuities from tracking integration contours}
\label{sec:seqdisc_from_contours}

We will now follow the integration contours $\Gamma$  as the external kinematics are varied, to predict the singular loci of the discontinuities of the Feynman integral. As we analytically continue the kinematic parameters, thus changing the shape of the contour $\Gamma$ through its dependence on $\F$, there are a few different occurrences that can cause the value of the integral $\cC_{\Gamma} \cI$ to change discontinuously.
\begin{itemize}
    \item[(i)] $\F=0$ becomes tangent to a bounding hyperplane ($\alpha_i=0$) at a point that intersects $\Gamma$. 
    \item[(ii)] The variety $\F=0$ becomes singular at a point that intersects $\Gamma$.
    \item[(iii)]  The boundary of $\Gamma$ moves to infinity and becomes singular there.
\end{itemize}
These are the occurrences we will seek in order to identify the loci and values of sequential discontinuities.

For simplicity, we work at the level of symbol entries.\footnote{More generally, we could frame the discussion in terms of monodromies.} It is convenient to assume knowledge of the symbol alphabet $\cA$ for a given Feynman integral, $\cA=\{a_k\}$. 
We define subsets $\cA_{\rm (i)}, \cA_{\rm (ii)}, \cA_{\rm (iii)}$ of $\cA$ corresponding to the three conditions above. Note that a letter $a_k$ may belong to more than one of the different categories. For polylogarithmic integrals, $\cA_{\rm (i)}$ includes the squared masses and the thresholds in all possible momentum channels.

Let us illustrate which Landau singularities are present for which cuts in our example of the triangle $\I^{\trithree}_{\tri}$.
Its symbol alphabet\footnote{In our analysis, we prefer to work with polynomial letters where possible. It is interesting to consider whether it is preferable to work with the rational combinations of these letters that emerge more naturally from considerations of Landau analysis \cite{Dlapa:2023cvx,He:2024fij}.} is known (for example, from the considerations of diagrammatic coaction \cite{Abreu:2017enx,Abreu:2017mtm} or SOFIA~\cite{Correia:2025yao}) to be $\cA(\I^{\trithree}_{\tri})=\{m_1^2, m_2^2, p^2{-}m_1^2, p^2, m_1^2-m_2^2-p^2, m_1^2-m_2^2\}$ in full dimensional regularization. In $\D=4$, the letter $p^2$ is absent, as seen above in Eq.~\eqref{eq:t3-sym-wt2}. The behavior of each integration contour as it approaches a potential singularity is summarized in Table \ref{tab:seqdiscT3}. We will now explain our observations by focusing on a few entries of the table and showing what happens as the contour is analytically continued through the singularity.
While the contours originally emerged in the positive orthant, we are now exploring analytic continuations of the contours themselves into different kinematic regions, which can result in moving outside the positive orthant.

\begin{figure}[t]
\begin{subfigure}{0.32\textwidth}
\includegraphics[width=1\linewidth]{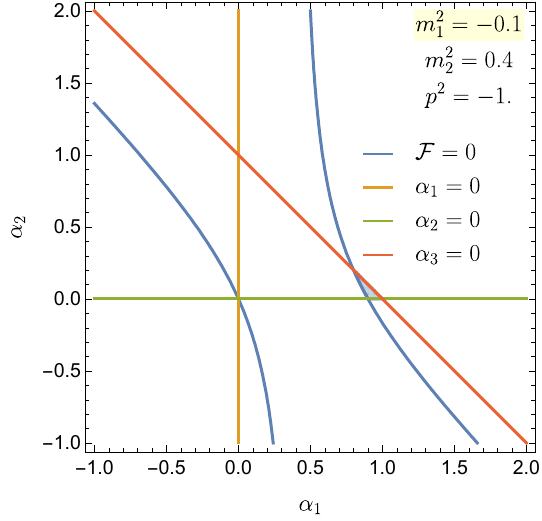} 
\end{subfigure}
\begin{subfigure}{0.32\textwidth}
\includegraphics[width=1\linewidth]{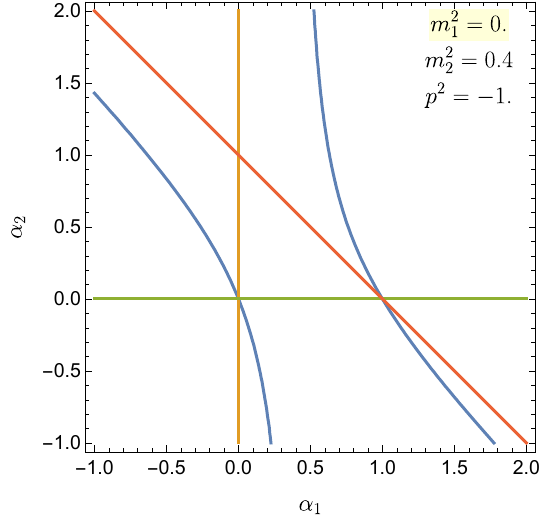} 
\end{subfigure}
\begin{subfigure}{0.32\textwidth}
\includegraphics[width=1\linewidth]{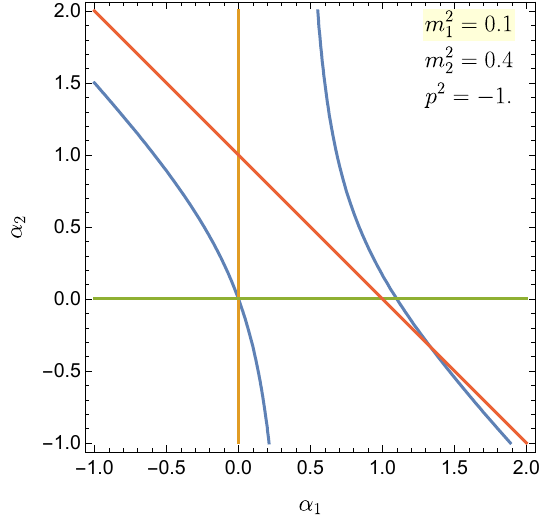} 
\end{subfigure}
\caption{The contour $\Gamma_{m_1^2}$ near the singularity of $m_1^2 \otimes m_1^2$, as the sign of $m_1^2$ is taken from negative to positive. The region shrinks to a point at $m_1^2=0$ and reappears outside the integration region of the Feynman integral. }
\label{fig:G11}
\end{figure}

\begin{figure}[t]
\begin{subfigure}{0.32\textwidth}
\includegraphics[width=1\linewidth]{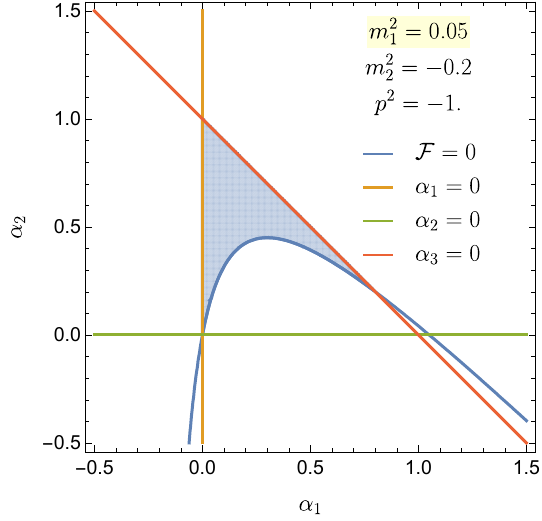} 
\end{subfigure}
\begin{subfigure}{0.32\textwidth}
\includegraphics[width=1\linewidth]{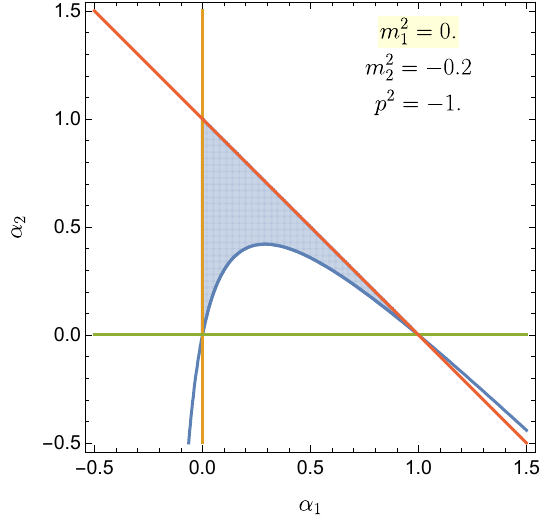}
\end{subfigure}
\begin{subfigure}{0.32\textwidth}
\includegraphics[width=1\linewidth]{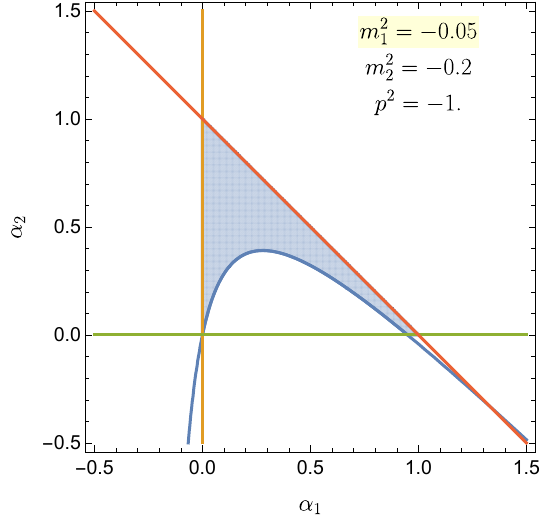}
\end{subfigure}
\caption{The contour $\Gamma_{m_2^2}$ near the singularity of $m_2^2 \otimes m_1^2$. The region experiences no singularity, and the sequence is absent from the symbol.}
\label{fig:G21}
\end{figure}

Case (i) above can be seen in the following examples.
Figure \ref{fig:G11} shows the vanishing of the cycle $\Gamma_{m_1^2}$ at the singular locus $m_1^2=0$, resulting in the symbol sequence $m_1^2 \otimes m_1^2$. Figure \ref{fig:G21} shows that $\Gamma_{m_2^2}$ is not singular at $m_1^2=0$, explaining the absence of the symbol sequence $m_2^2 \otimes m_1^2$. Our interpretation is that the singularity at $m_1^2=0$ is not present when using the integration region corresponding to the discontinuity in $m_2^2$.

In these examples, certain sequential discontinuities vanish because the integration contours, $\Gamma_{m_1^2}$ and $\Gamma_{m_2^2}$, are no longer attached to specific boundaries $\alpha_i=0$ of the integration space. The Feynman integral is defined over a region where $\alpha_i \geq 0$, and its integrand generally has singularities located on the varieties defined by $\F=0$ and/or $\U=0$. The value of the integral depends on the homology class of the integration contour relative to these singular surfaces: $\F=0$, $\U=0$ and $\alpha_i=0$ for all $i$. When we compute the discontinuity with respect to a variable, say $m_2^2$ as presented in Figure \ref{fig:G21}, the original integration contour is replaced by a new contour, $\Gamma_{m_2^2}$. A key feature of this new contour is that it is no longer bounded by the surface $\alpha_2=0$. Consequently, any Landau singularity whose existence requires the condition $\alpha_2=0$ is no longer captured. In other words, the cycle is no longer ``pinched'' by that singularity. While these constraints on singularities overlap with known genealogical relations~\cite{Hannesdottir:2024cnn}, explicitly tracking the integration contours provides a more refined set of constraints as we will now see.

\begin{figure}[t]
\begin{subfigure}{0.32\textwidth}
\includegraphics[width=1\linewidth]{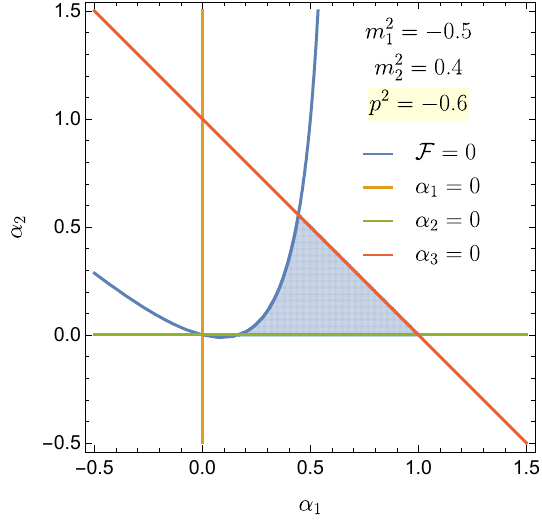} 
\end{subfigure}
\begin{subfigure}{0.32\textwidth}
\includegraphics[width=1\linewidth]{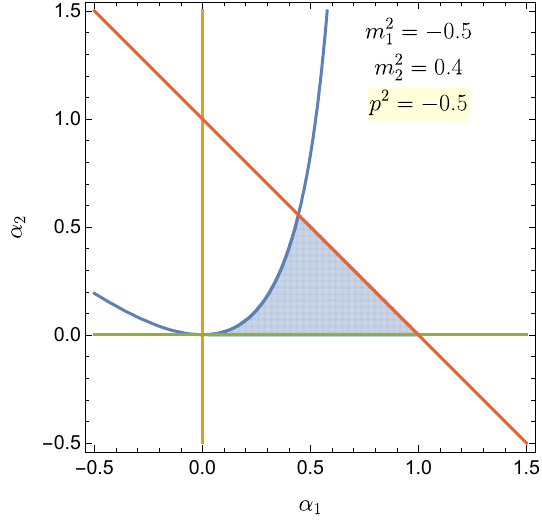} 
\end{subfigure}
\begin{subfigure}{0.32\textwidth}
\includegraphics[width=1\linewidth]{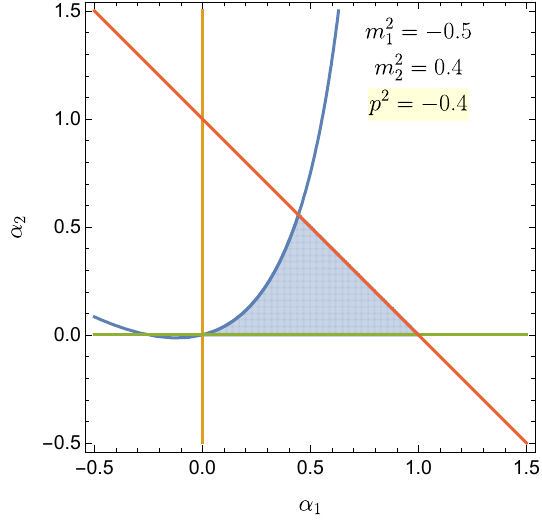} 
\end{subfigure}
\caption{The contour $\Gamma_{m_1^2}$ near the singularity of $m_1^2 \otimes (p^2{-}m_1^2)  $. This sequence is absent in the symbol at weight 2, but present from weight 3.}
\label{fig:G1p}
\end{figure}

 Figure \ref{fig:G1p} shows a more subtle singularity, where the boundary of $\Gamma_{m_1^2}$ remains fixed at $\alpha_1=\alpha_2=0$ for $p^2 \geq m_1^2$.
 In other words, the boundary of the region $\Gamma_{m_1^2}$ must include the point at $\alpha_1=\alpha_2=0$ after crossing the singular locus, but does not contain it before. The corresponding letter sequence $m_1^2 \otimes (p^2{-}m_1^2)$ is present, but only starting from weight 3.  Its absence in the first two entries is an example of a constraint that goes beyond genealogical constraints. We will further discuss how singularities can be absent at certain weights in Section~\ref{subsection:dimensional} below.

\begin{table}
    \centering
    \begin{tabular}{c||c|c|c}
    \backslashbox{$L_2$}{$L_1$}     & $m_1^2$ & \(m_2^2\) & \(p^2{-}m_1^2\) \\ \hline\hline
      \multirow{2}*{\(m_1^2\)}   & \includegraphics[width=0.171\linewidth]{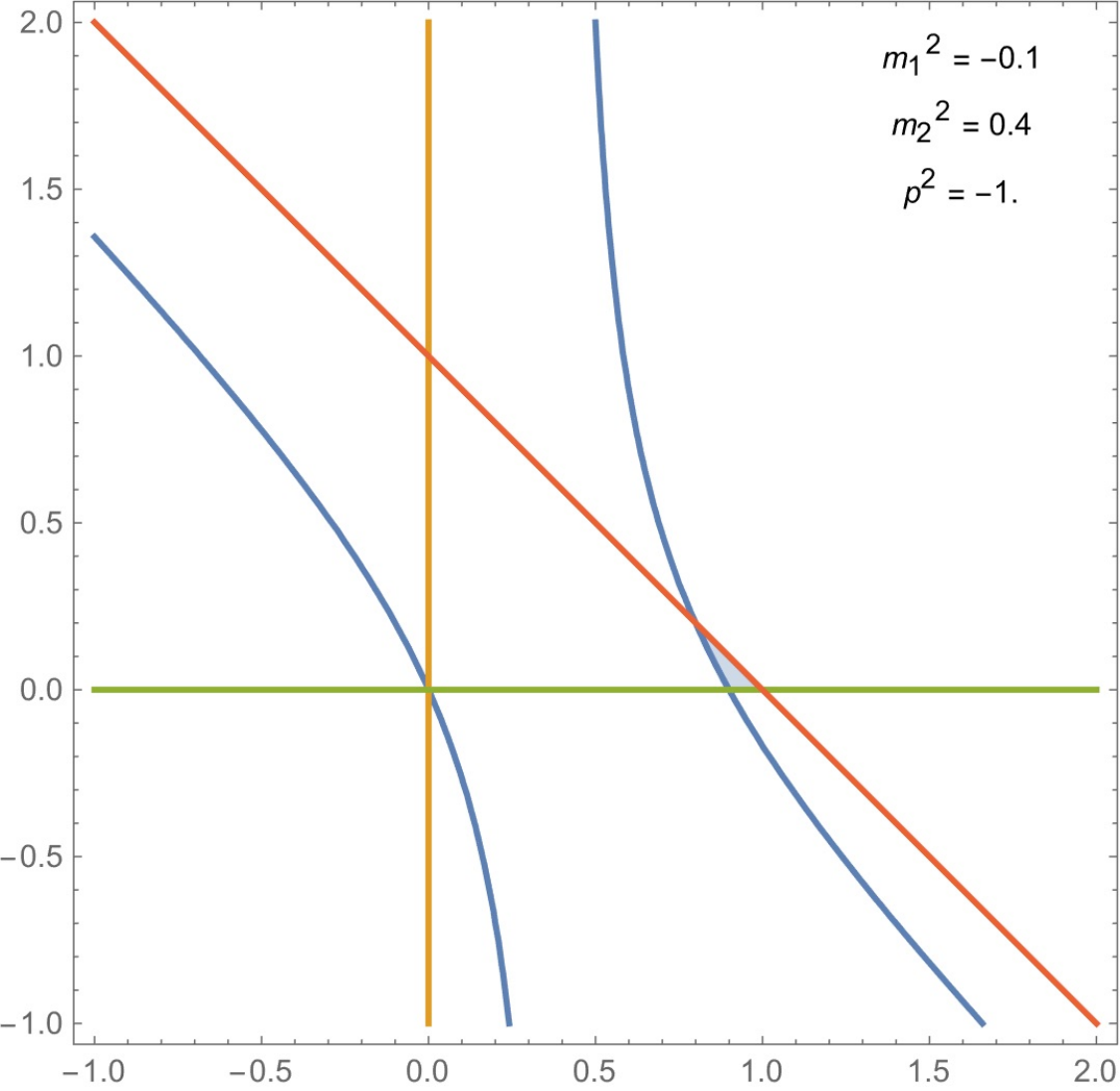}  & \includegraphics[width=0.171\linewidth]{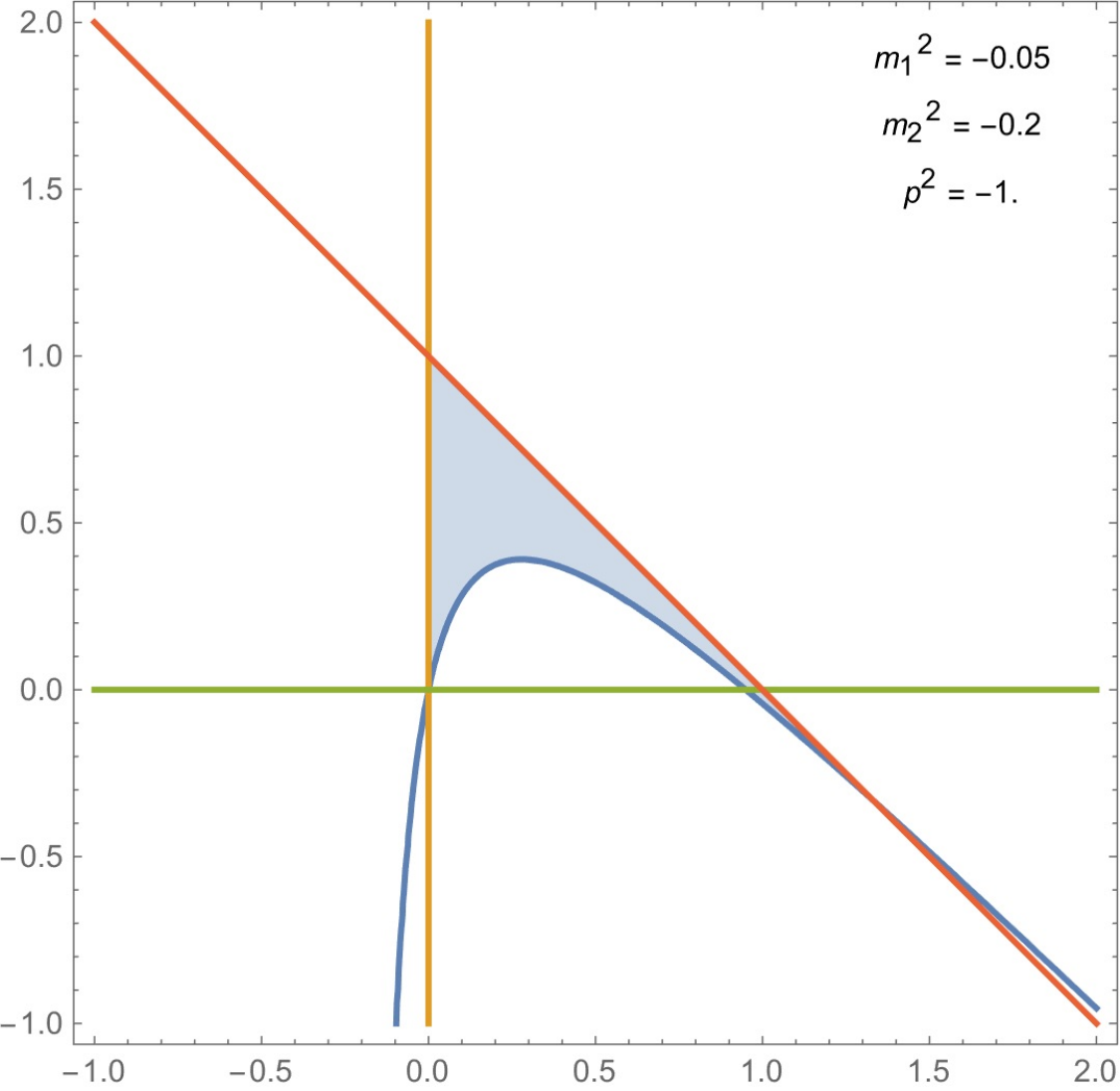}  & \includegraphics[width=0.171\linewidth]{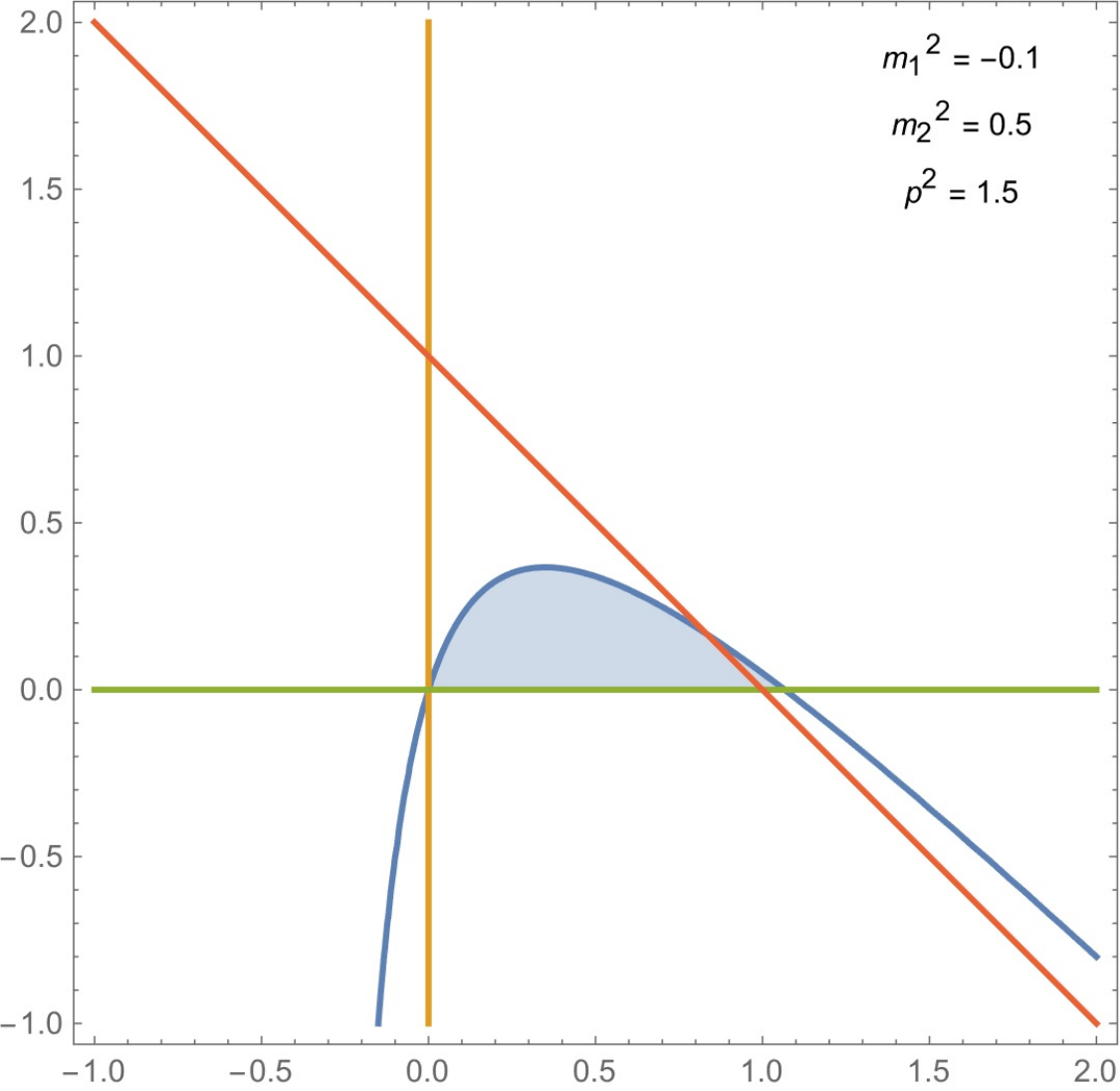} \\ 
       & repeated letter  & not singular  & not singular \\ \hline
       \multirow{2}*{\(m_2^2\)}  & \includegraphics[width=0.171\linewidth]{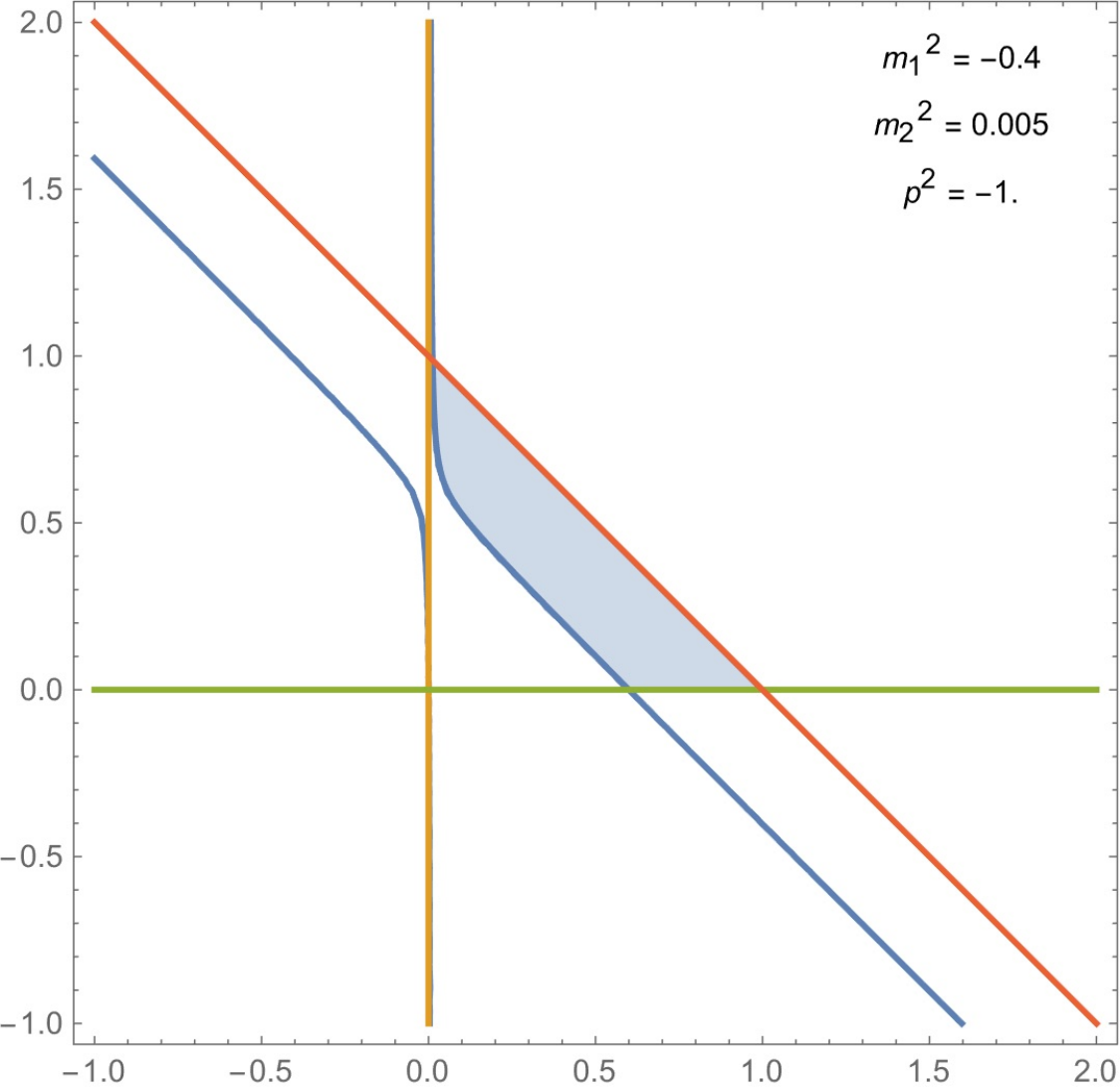}  & \includegraphics[width=0.171\linewidth]{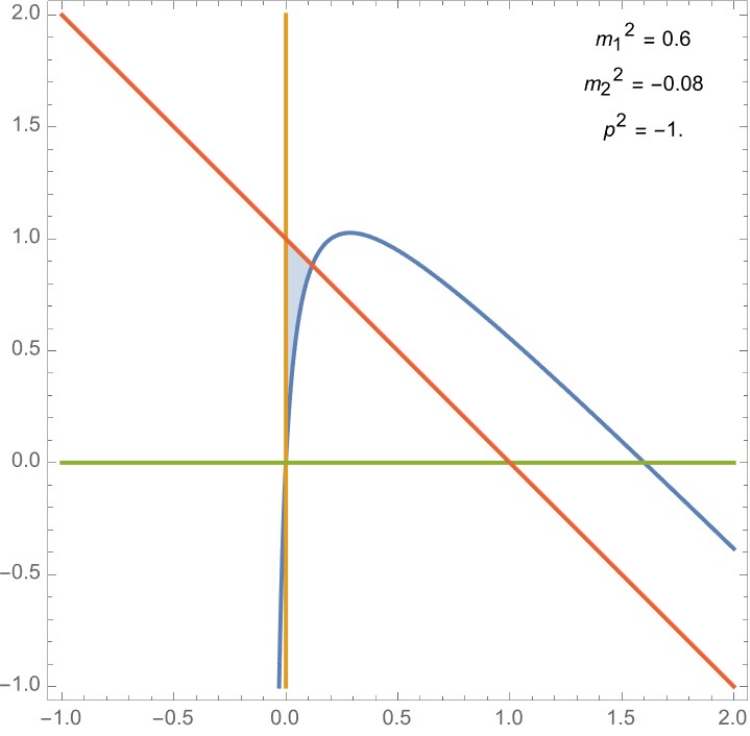}  & \includegraphics[width=0.171\linewidth]{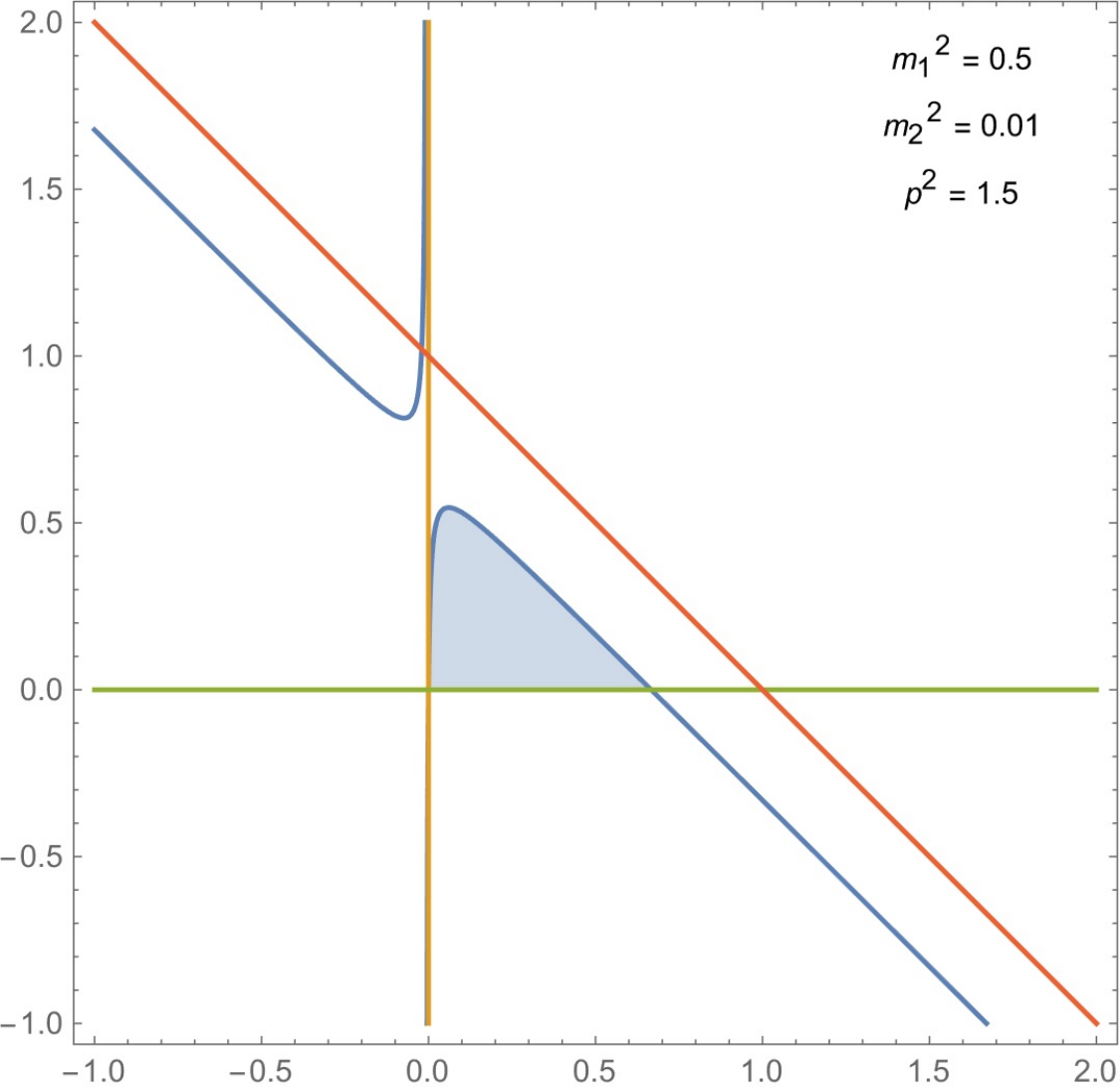}  \\ 
       & singular  & repeated letter  & singular  \\ \hline
    \multirow{2}*{\(p^2{-}m_1^2\)}  & \includegraphics[width=0.171\linewidth]{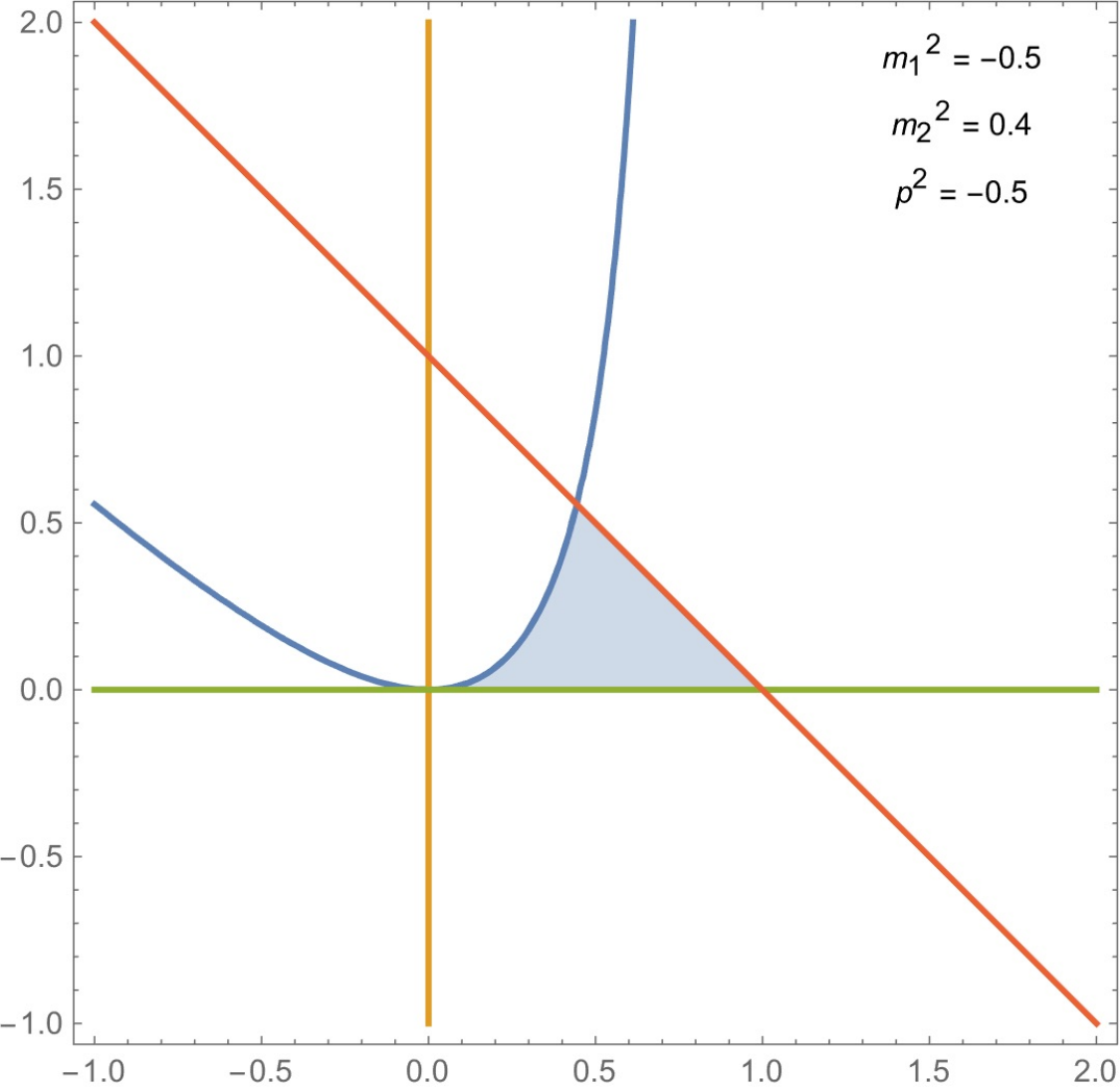}  & \includegraphics[width=0.171\linewidth]{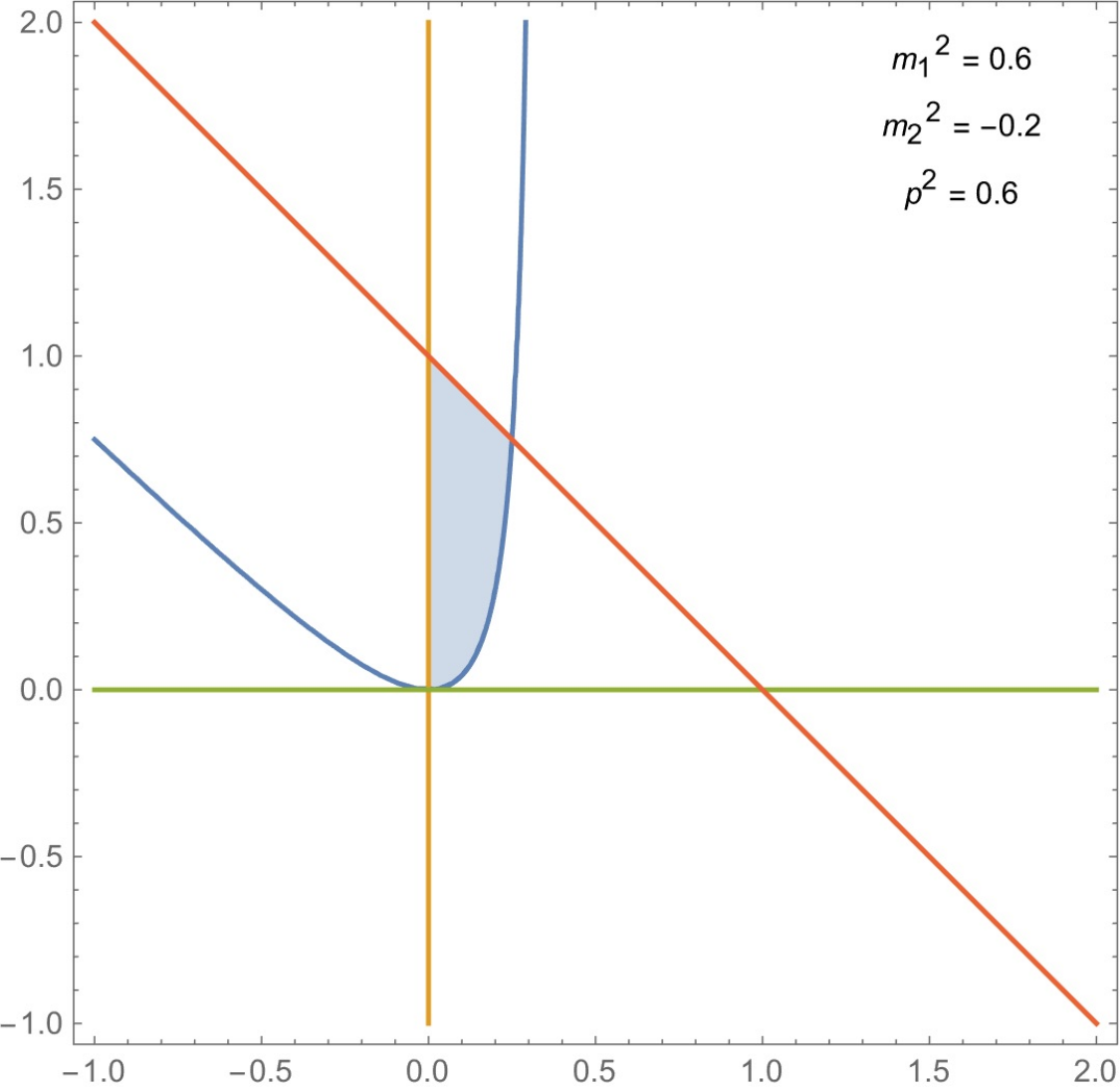}  & \includegraphics[width=0.171\linewidth]{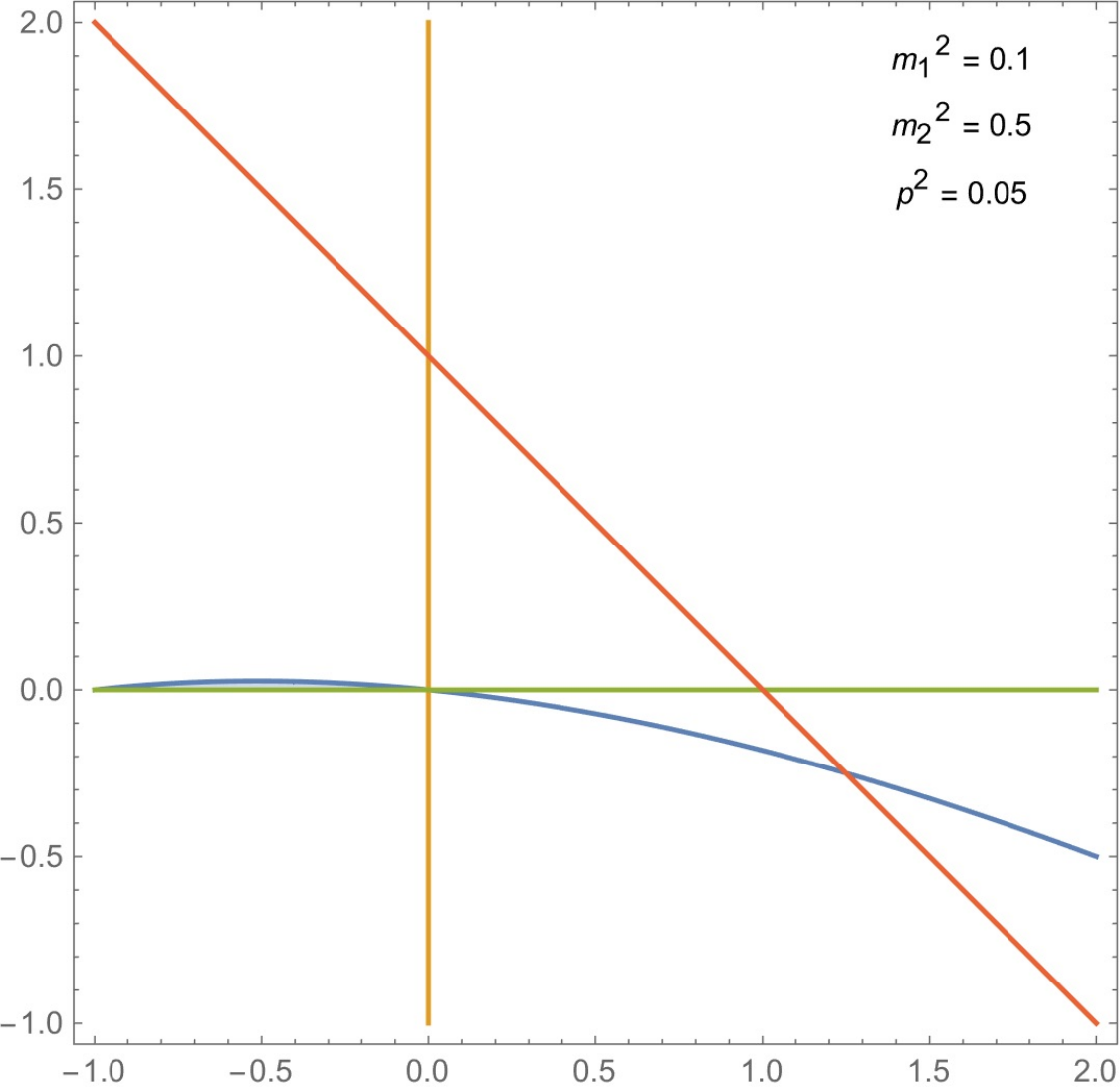} \\ 
    & absent at weight 2  & not singular  & repeated letter \\ \hline
    \multirow{2}*{\(m_1^2-m_2^2-p^2\)}    
    & \includegraphics[width=0.171\linewidth]{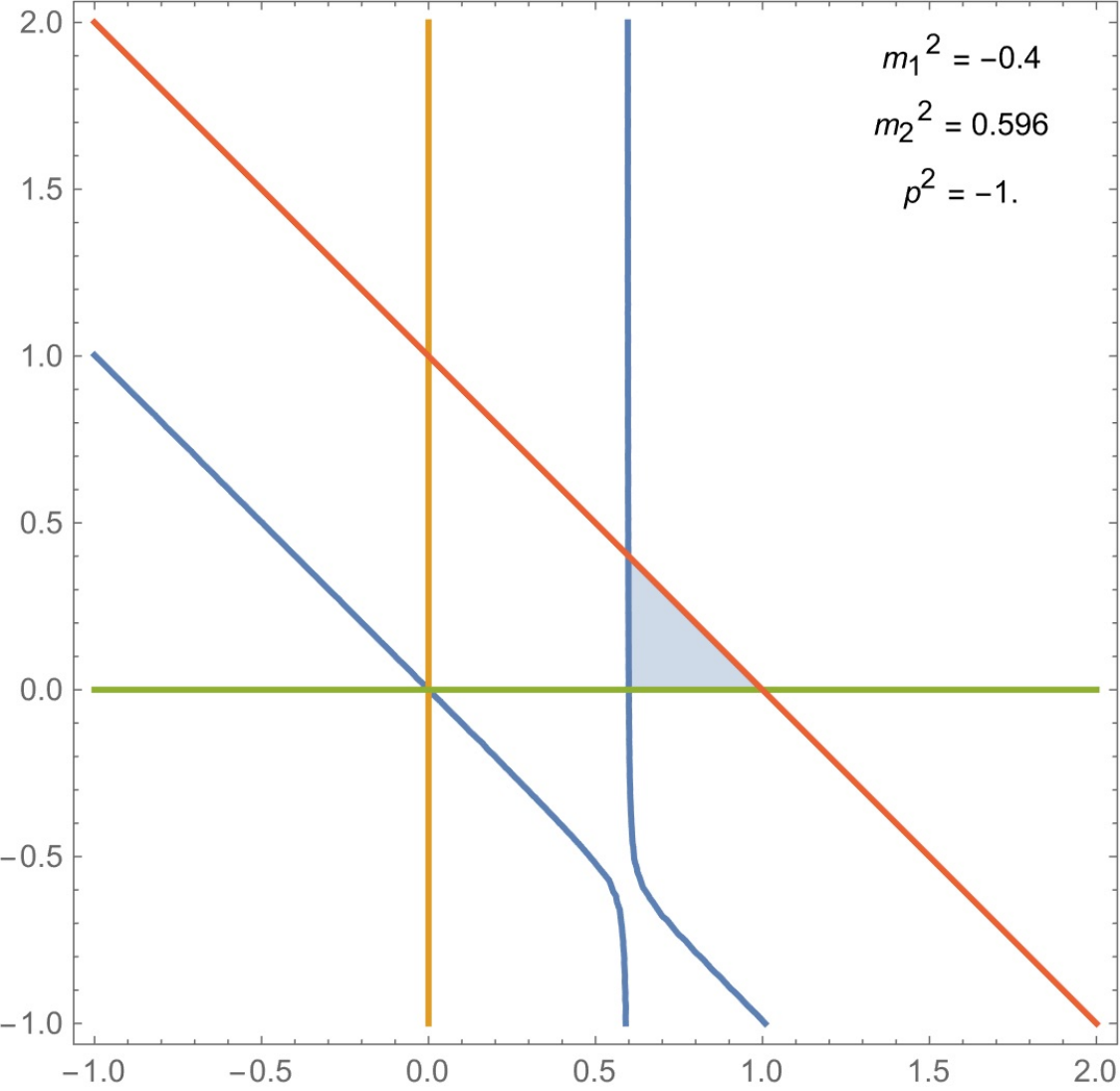} & \includegraphics[width=0.171\linewidth]{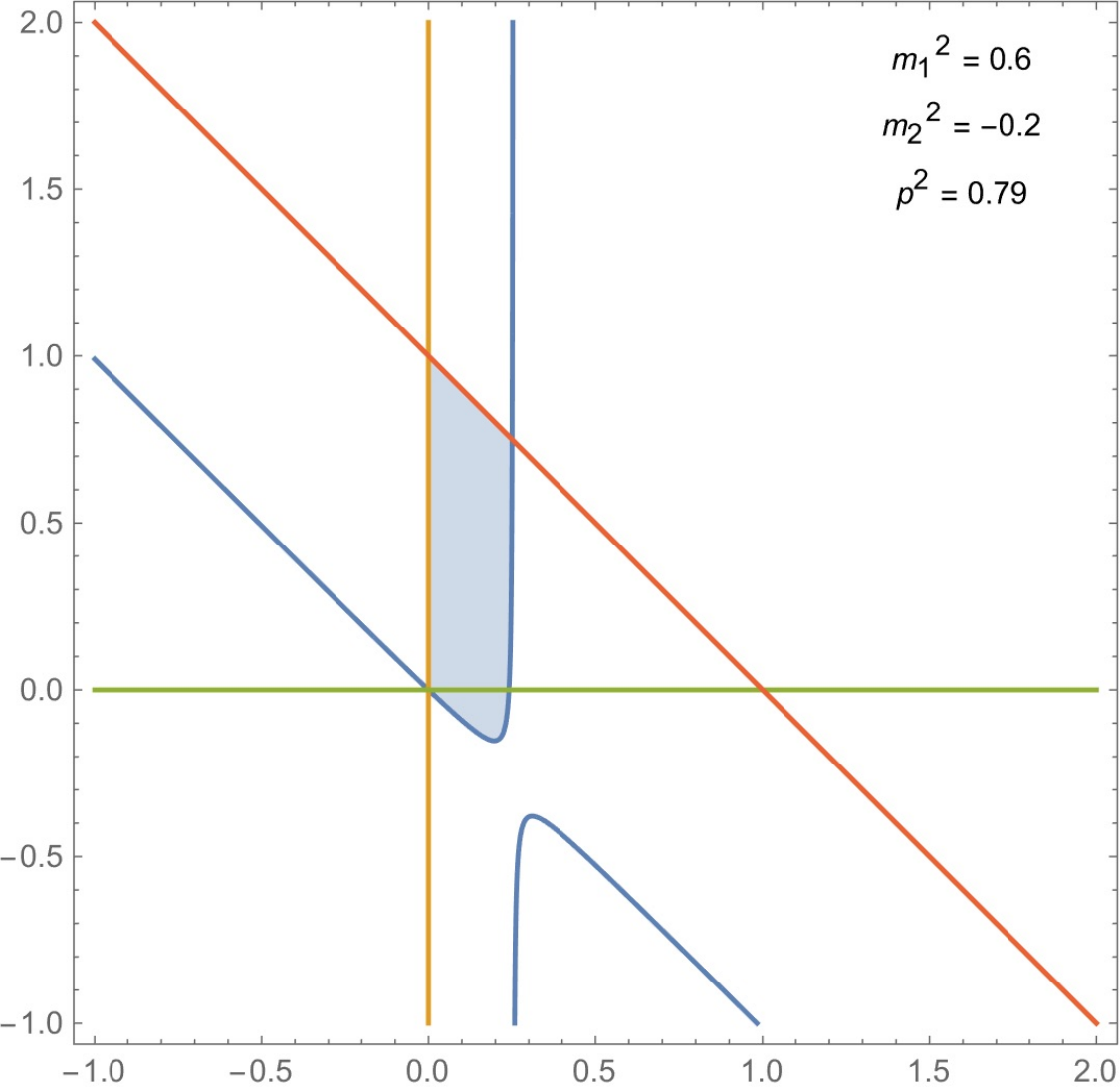}  & \includegraphics[width=0.171\linewidth]{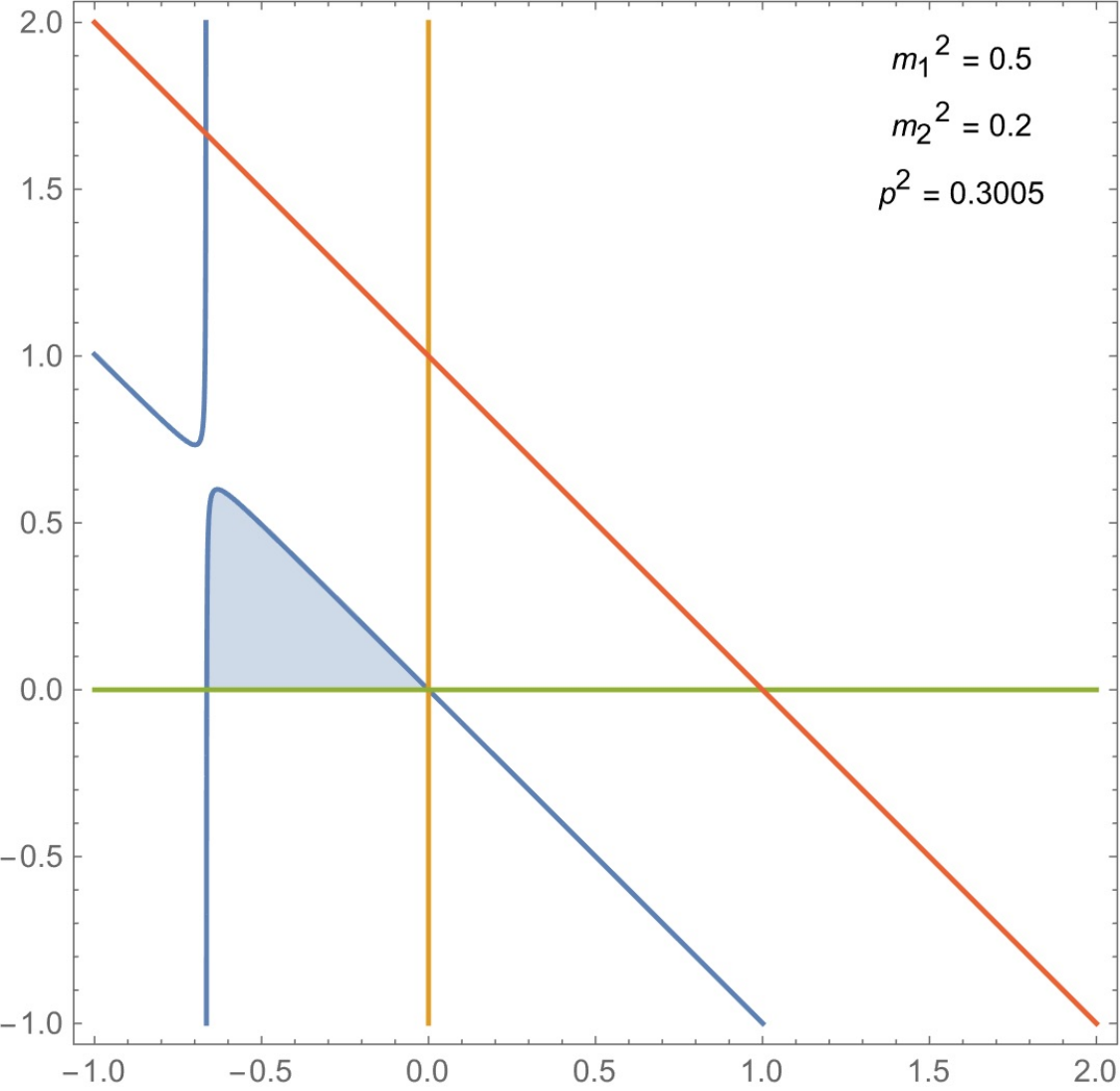} \\ 
    & not singular & singular  & singular \\ \hline
    \multirow{2}*{\(p^2\)}    & \includegraphics[width=0.171\linewidth]{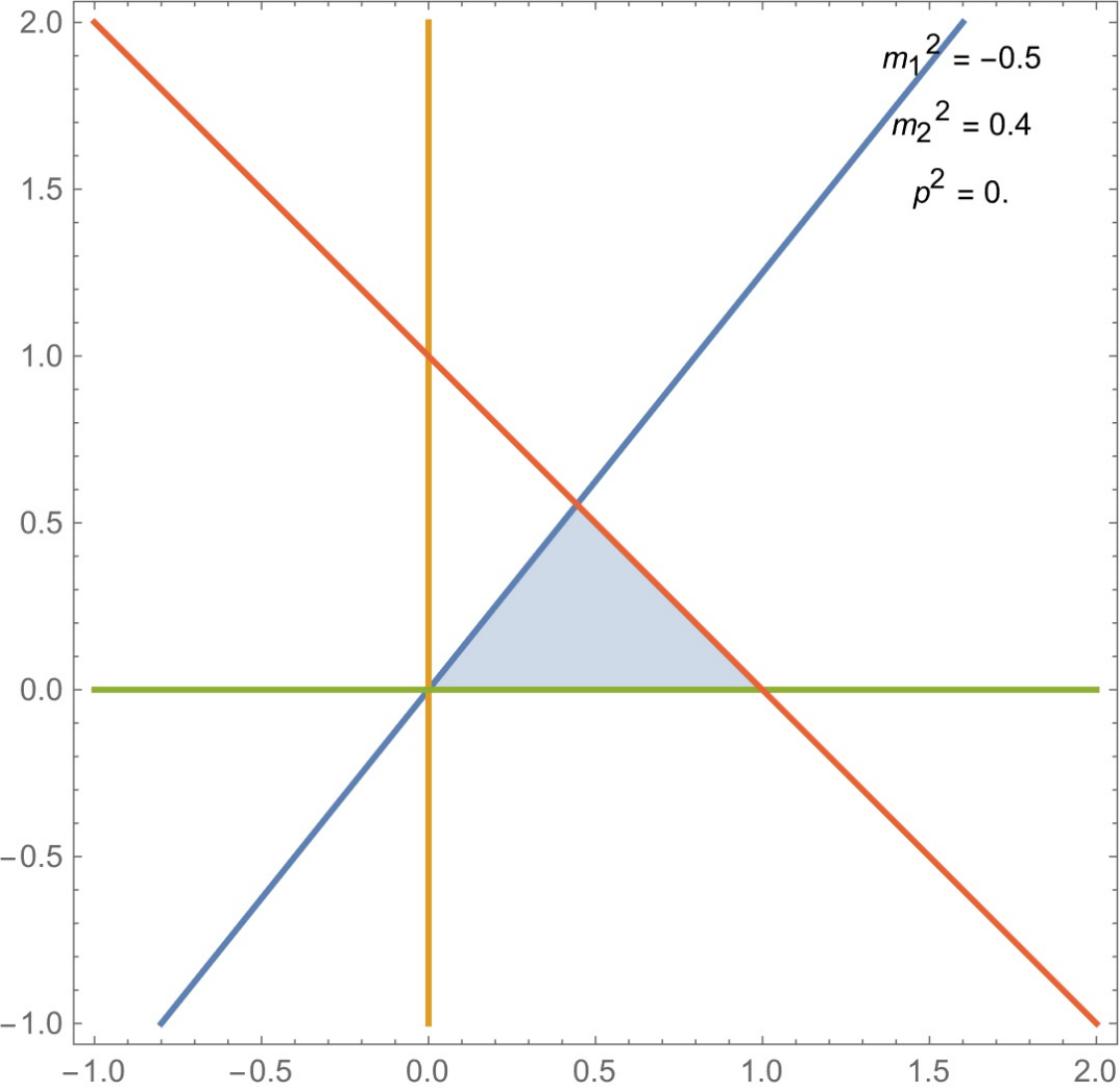} & \includegraphics[width=0.171\linewidth]{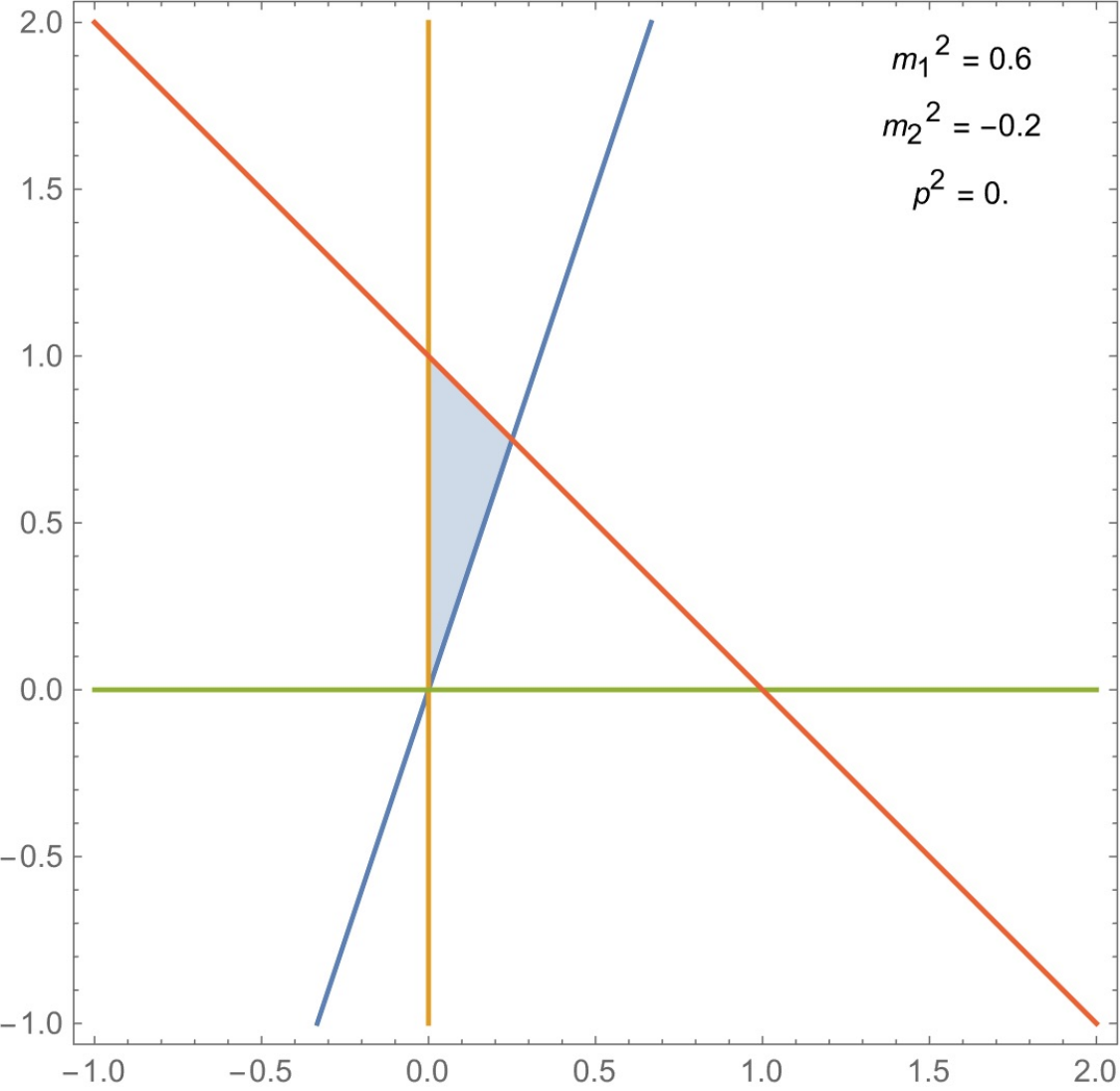}  & \includegraphics[width=0.171\linewidth]{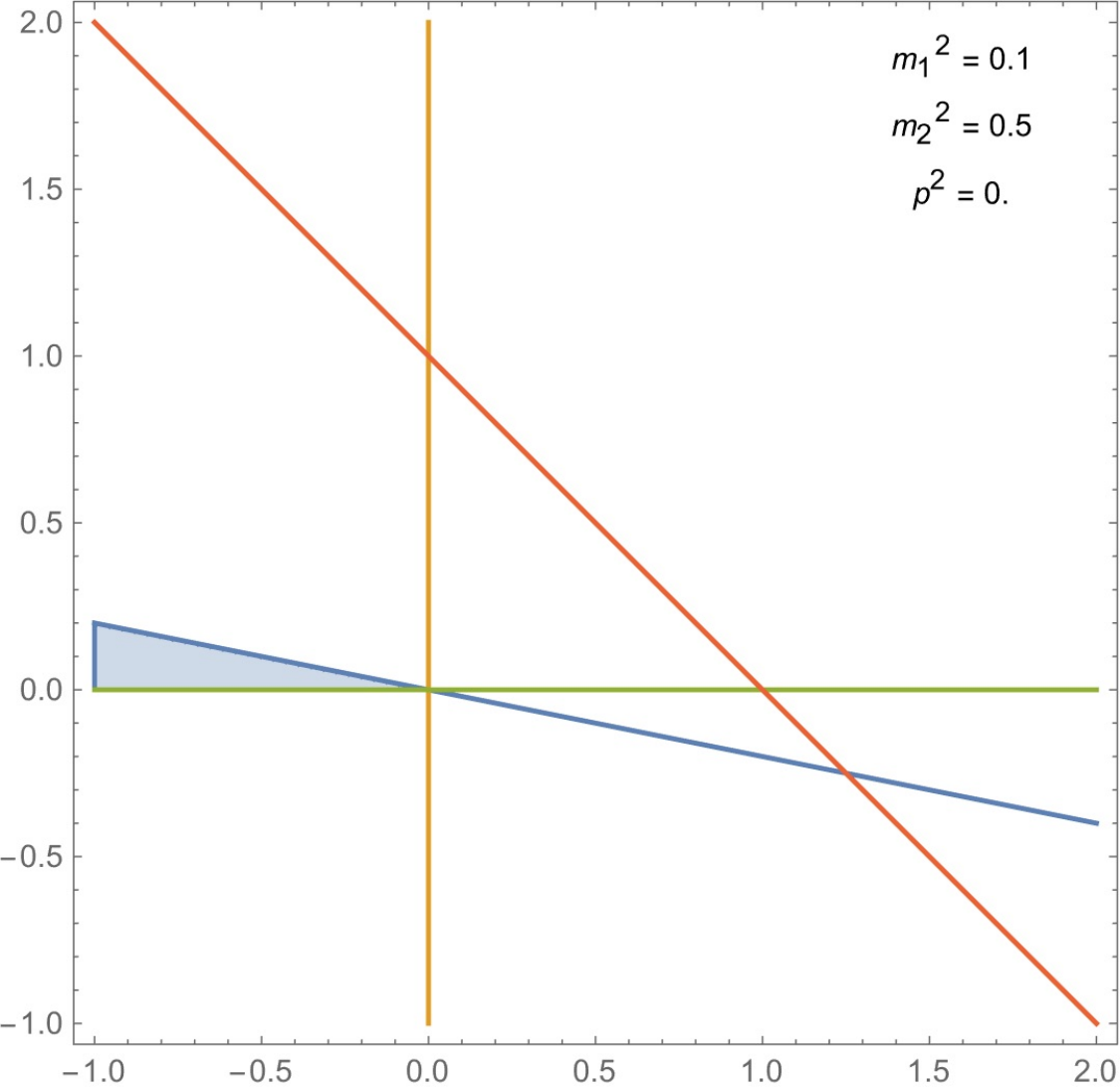} \\  
     & absent at weight 2 & not singular  & absent at weight 2 \\  \hline
   \multirow{2}*{\(m_1^2-m_2^2\)}      & \includegraphics[width=0.171\linewidth]{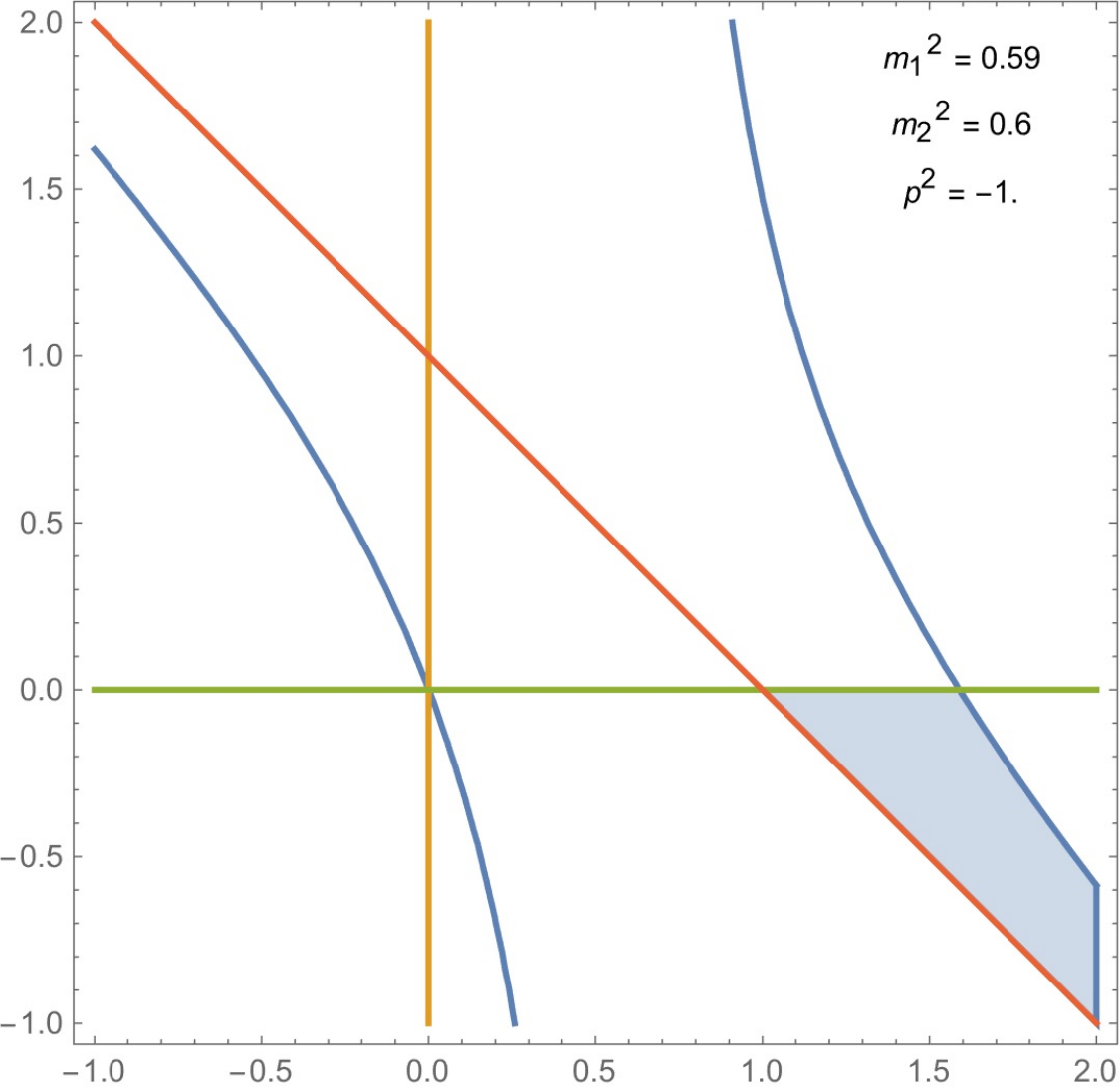} & \includegraphics[width=0.171\linewidth]{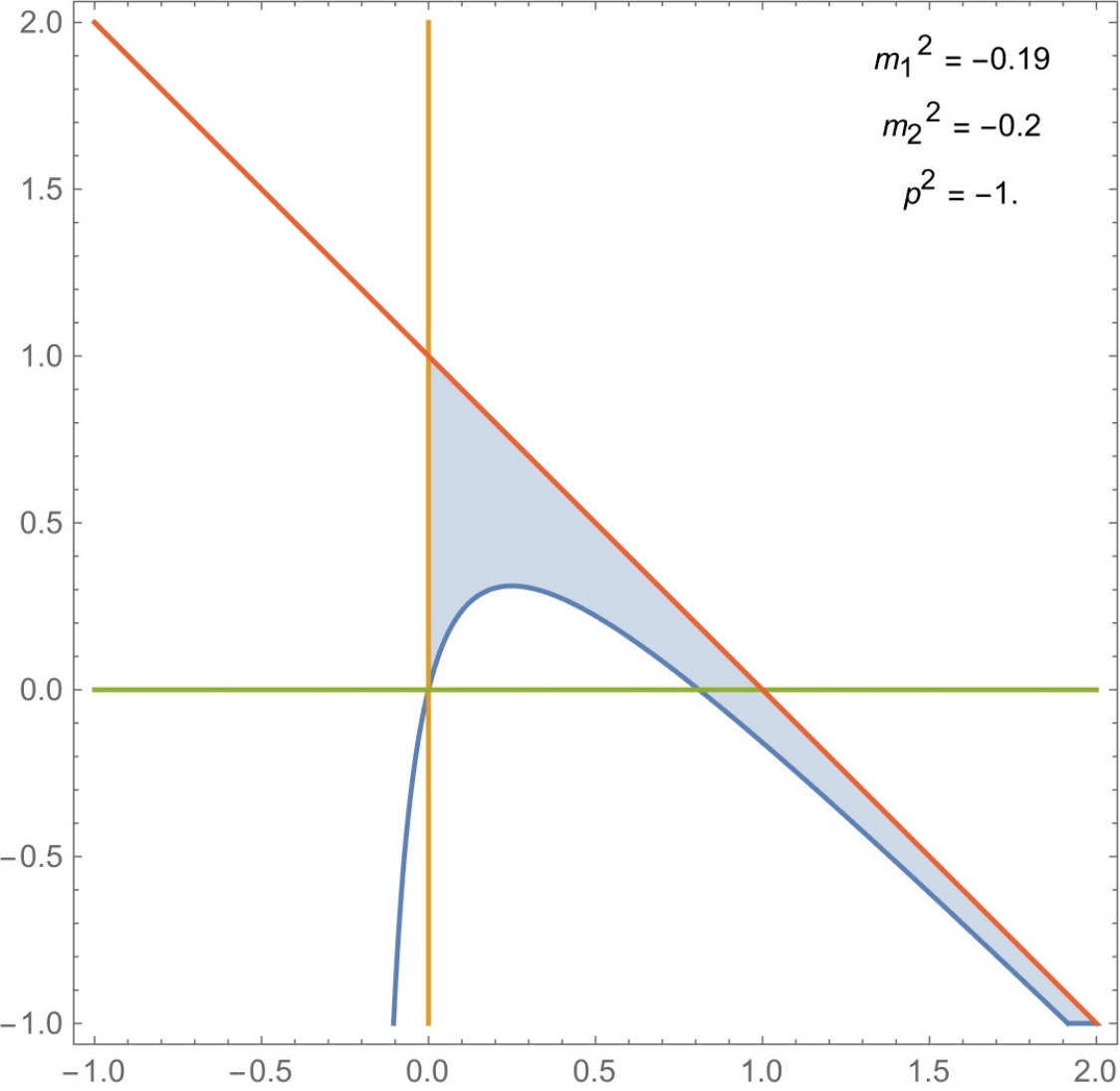}  & \includegraphics[width=0.171\linewidth]{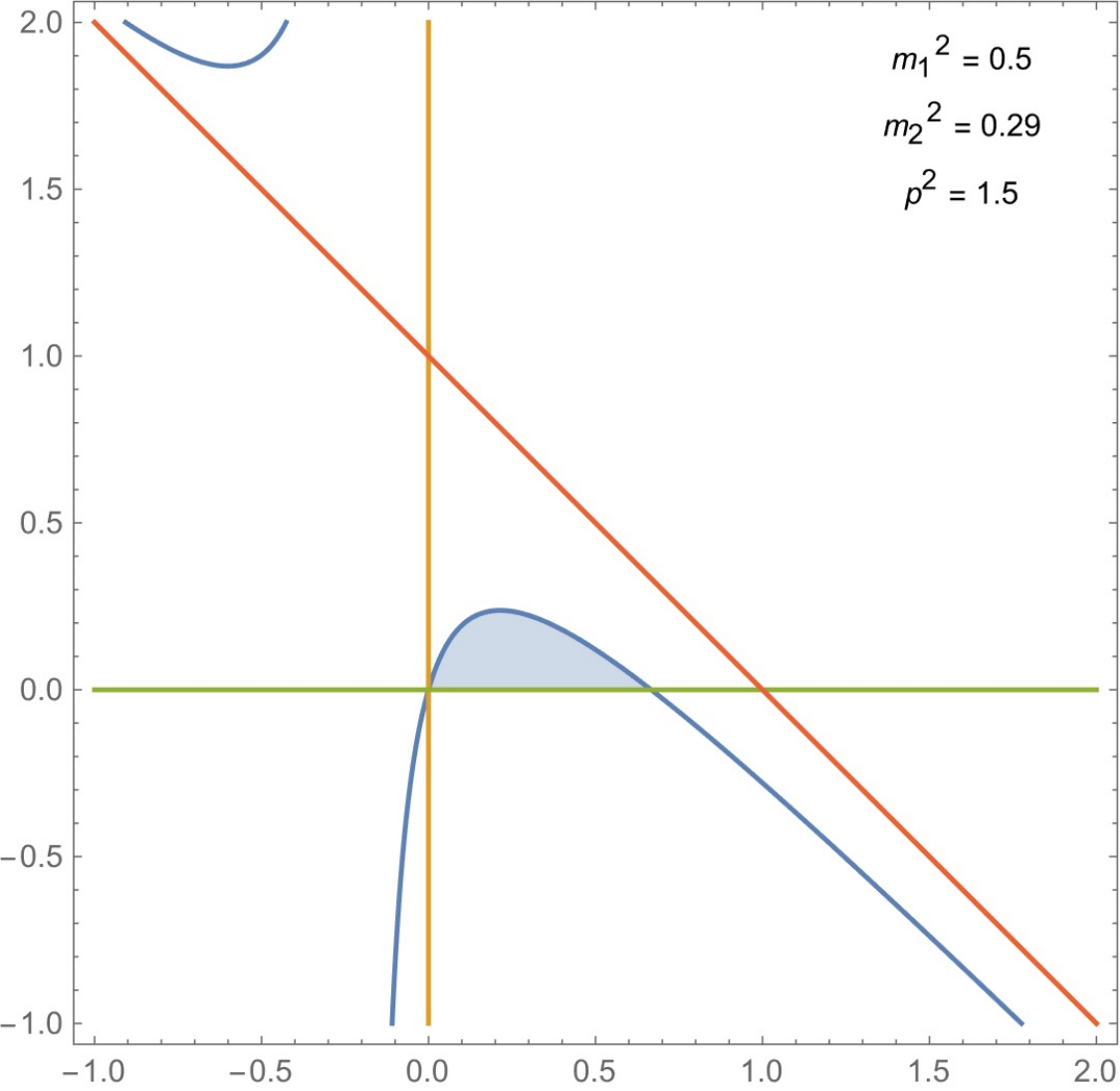} \\
    & singular at $\infty$ & singular at $\infty$ & not singular \\
    \end{tabular}
    \caption{Integration regions for discontinuities of $\I^{\trithree}_{\tri}$, at kinematics near a second discontinuity.}
\label{tab:seqdiscT3}
\end{table}

Case (ii) occurs when the discriminant variety of $\F=0$ vanishes. As mentioned before, this condition may have overlap with case (i).  In the one-loop case, the singular locus for case (ii) is where the discriminant of the quadratic form $\F$ vanishes. In the example of $\I^{\trithree}_{\tri}$, the discriminant of $\F$ is
\begin{equation}
    \frac{1}{4} p^2 m_{2}^2 (m_{1}^2{-}m_{2}^2{-}p^2),
\end{equation}
and so we have $\cA_{\rm (ii)}(\I^{\trithree}_{\tri})=\{m_2^2, p^2, m_1^2-m_2^2-p^2\}$.

\begin{figure}[h]
\centering
\begin{subfigure}{0.32\textwidth}
\includegraphics[width=1\linewidth]{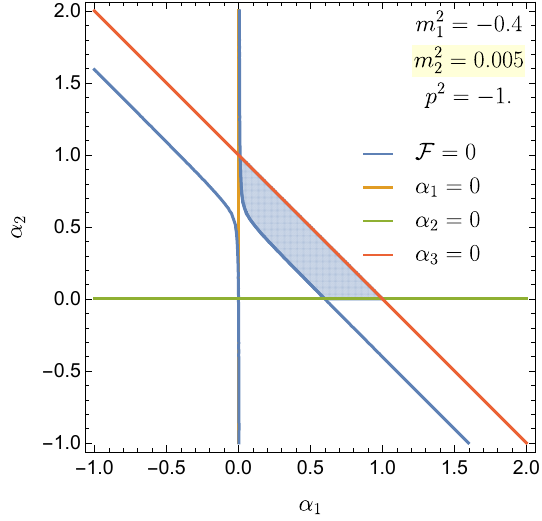} 
\end{subfigure}
\begin{subfigure}{0.32\textwidth}
\includegraphics[width=1\linewidth]{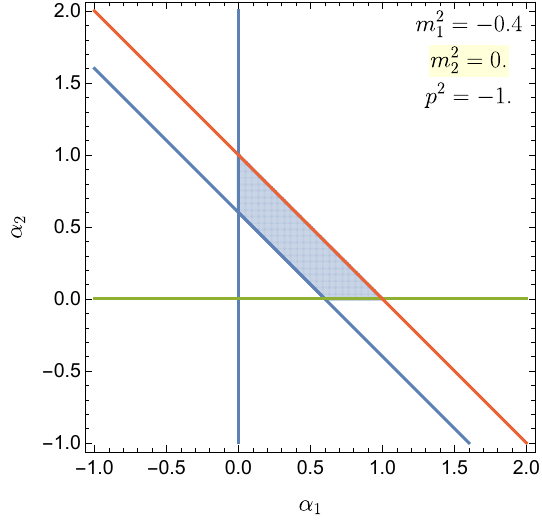} 
\end{subfigure}
\begin{subfigure}{0.32\textwidth}
\includegraphics[width=1\linewidth]{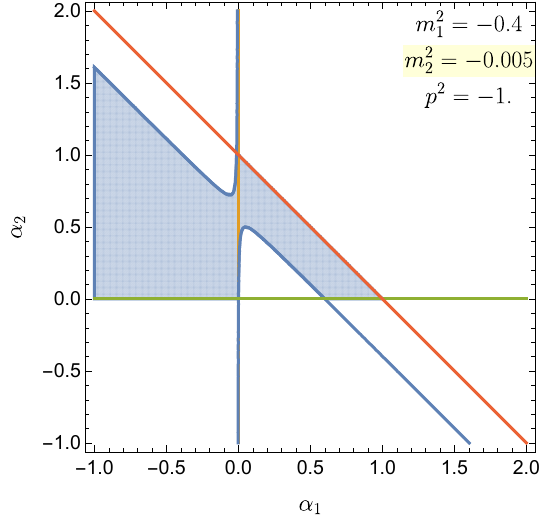} 
\end{subfigure}
\caption{The contour $\Gamma_{m_1^2}$ near the singularity of $m_1^2 \otimes m_2^2$. The region bounded by $\F=0$, $\alpha_2=0$ and $\alpha_3=0$ changes shape due to the self-intersection of $\F=0$ where the discriminant of $\F$ vanishes.  The shaded region in the rightmost plot is one carved out by the same boundaries, but it does not represent an analytic continuation of $\I_{\tri}^{\trithree}$. Instead, the analytically continued contour will be deformed out of the real plane.}
\label{fig:G12}
\end{figure}

\begin{figure}[h]
\centering
\begin{subfigure}{0.32\textwidth}
\includegraphics[width=1\linewidth]{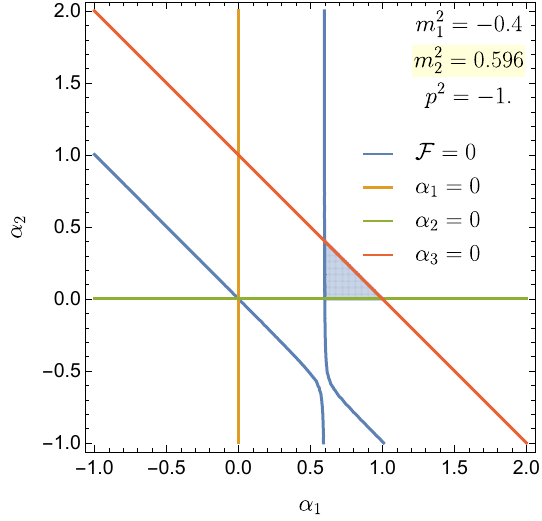} 
\end{subfigure}
\begin{subfigure}{0.32\textwidth}
\includegraphics[width=1\linewidth]{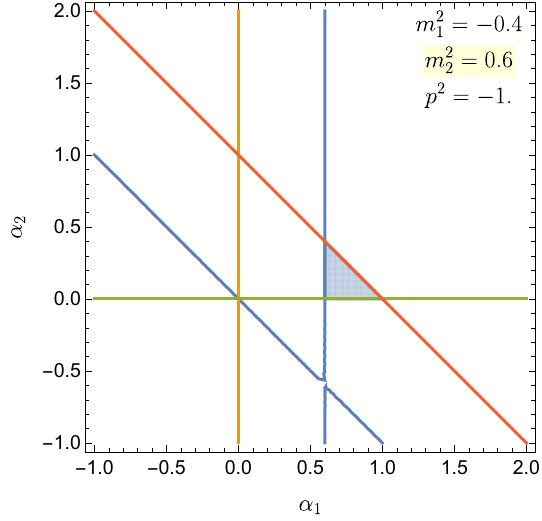} 
\end{subfigure}
\begin{subfigure}{0.32\textwidth}
\includegraphics[width=1\linewidth]{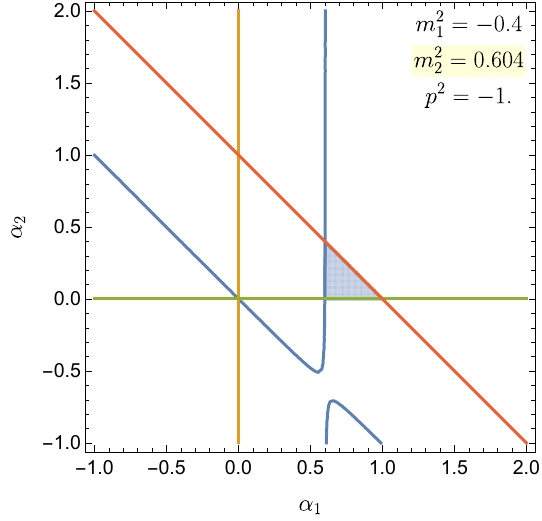} 
\end{subfigure}
\caption{The contour $\Gamma_{m_1^2}$ near the singularity of $m_1^2 \otimes (m_1^2-m_2^2-p^2)$. The region experiences no singularity, and the sequence is absent from the symbol.}
\label{fig:G1F}
\end{figure}

We present two examples in detail. Figure \ref{fig:G12} shows how the region $\Gamma_{m_1^2}$ changes shape at $m_2^2=0$ as it merges with a previously separate region,  resulting in the letter sequence $m_1^2 \otimes m_2^2$. On the other hand, Figure \ref{fig:G1F} shows how the region $\Gamma_{m_1^2}$ is unaffected at the singularity $m_1^2-m_2^2-p^2=0$, because the self-intersection of $\F=0$ occurs away from $\Gamma_{m_1^2}$. In fact, all of the illustrated values of kinematic parameters are within the region of phase space where the discontinuity of $\I^{\trithree}_{\tri}$ in $m_1^2$ is ordinarily computed, and which is directly accessible from the region $R_0$.

\begin{figure}[h]
\centering
\begin{subfigure}{0.32\textwidth}
\includegraphics[width=1\linewidth]{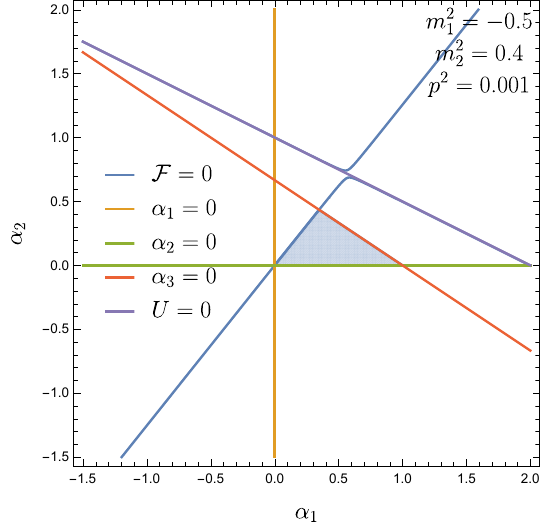} 
\end{subfigure}
\begin{subfigure}{0.32\textwidth}
\includegraphics[width=1\linewidth]{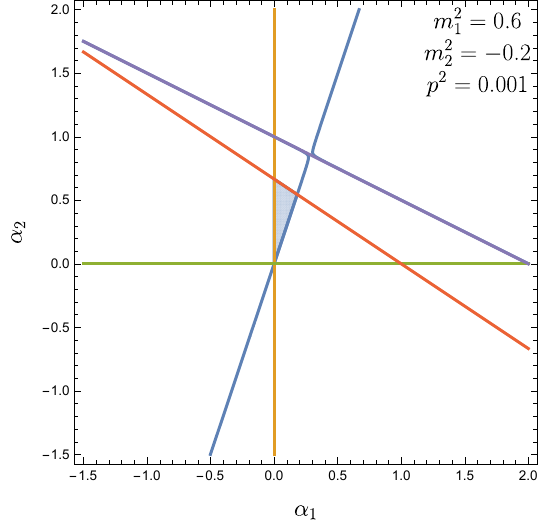} 
\end{subfigure}
\begin{subfigure}{0.32\textwidth}
\includegraphics[width=1\linewidth]{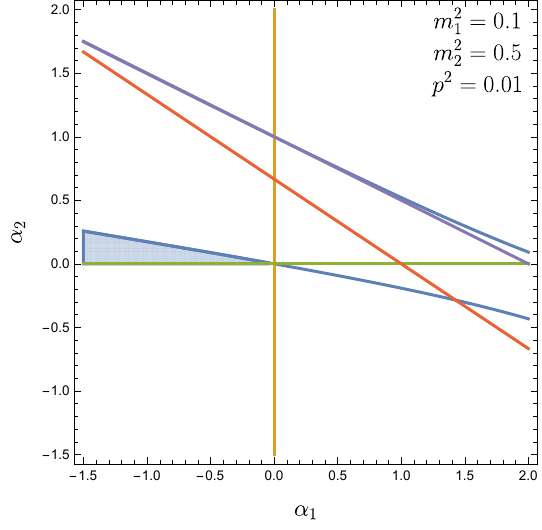} 
\end{subfigure}
\caption{The fifth row of Table \ref{tab:seqdiscT3} represented in the plane $\alpha_3=2-2\alpha_1-3\alpha_2$.}
\label{fig:G0-special}
\end{figure}

We note also that the letter $p^2$ is absent in the second entry at weight 2, although this is not readily visible from Table \ref{tab:seqdiscT3}, because one of the factors of $\F$ at $p^2=0$ is $\cU$ itself. The figures in Table \ref{tab:seqdiscT3} have all been drawn with the usual convention of $\alpha_1+\alpha_2+\alpha_3=1$, so we do not see the location of $0=\cU=\alpha_1+\alpha_2+\alpha_3$. If we choose instead a different slice of projective space, such as $\alpha_3=2-2\alpha_1-3\alpha_2$, compensated by the appropriate Jacobian factor in the integrand, we find the similar plots in Figure \ref{fig:G0-special}, where it becomes clear that $\F=0$ factorizes in a way that does not affect our regions of interest.

Case (iii) occurs where $\Gamma$ reaches infinity. In the example of $\I^{\trithree}_{\tri}$, the intersection of  $\F$ with $\alpha_3=0$ moves to infinity at $m_1^2-m_2^2=0$, giving rise to the last letter of the alphabet, $\cA_{\rm (iii)}(\I^{\trithree}_{\tri})=\{m_1^2-m_2^2\}$.

\begin{figure}
    \centering
    \begin{tabular}{ccc}
  \includegraphics[width=0.3\linewidth]{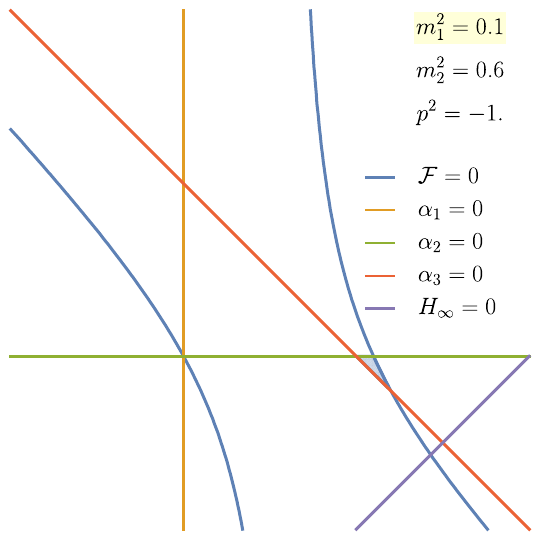}  & \includegraphics[width=0.3\linewidth]{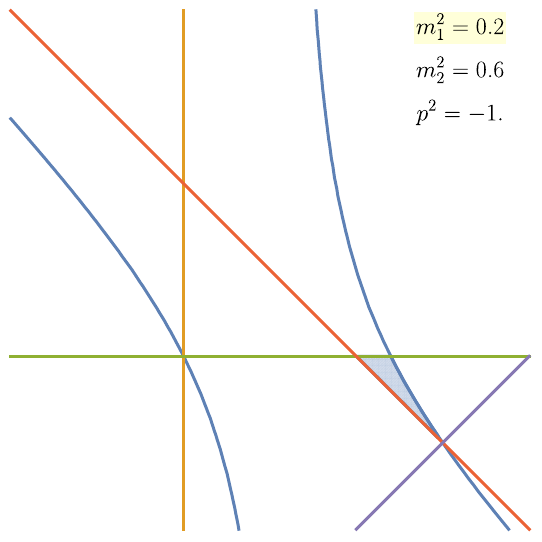}  & \includegraphics[width=0.3\linewidth]{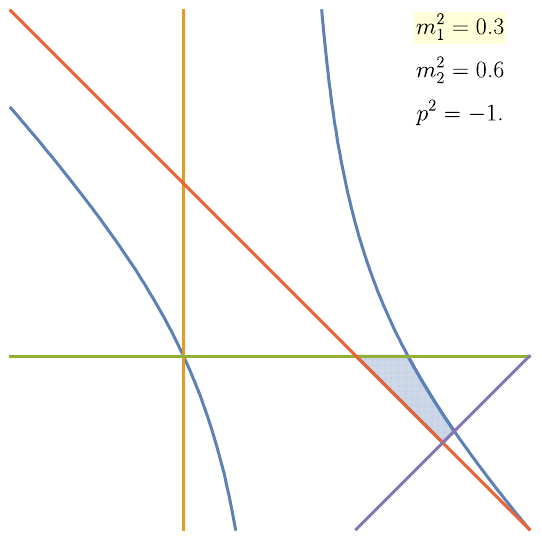} \\ 
 & \includegraphics[width=0.3\linewidth]{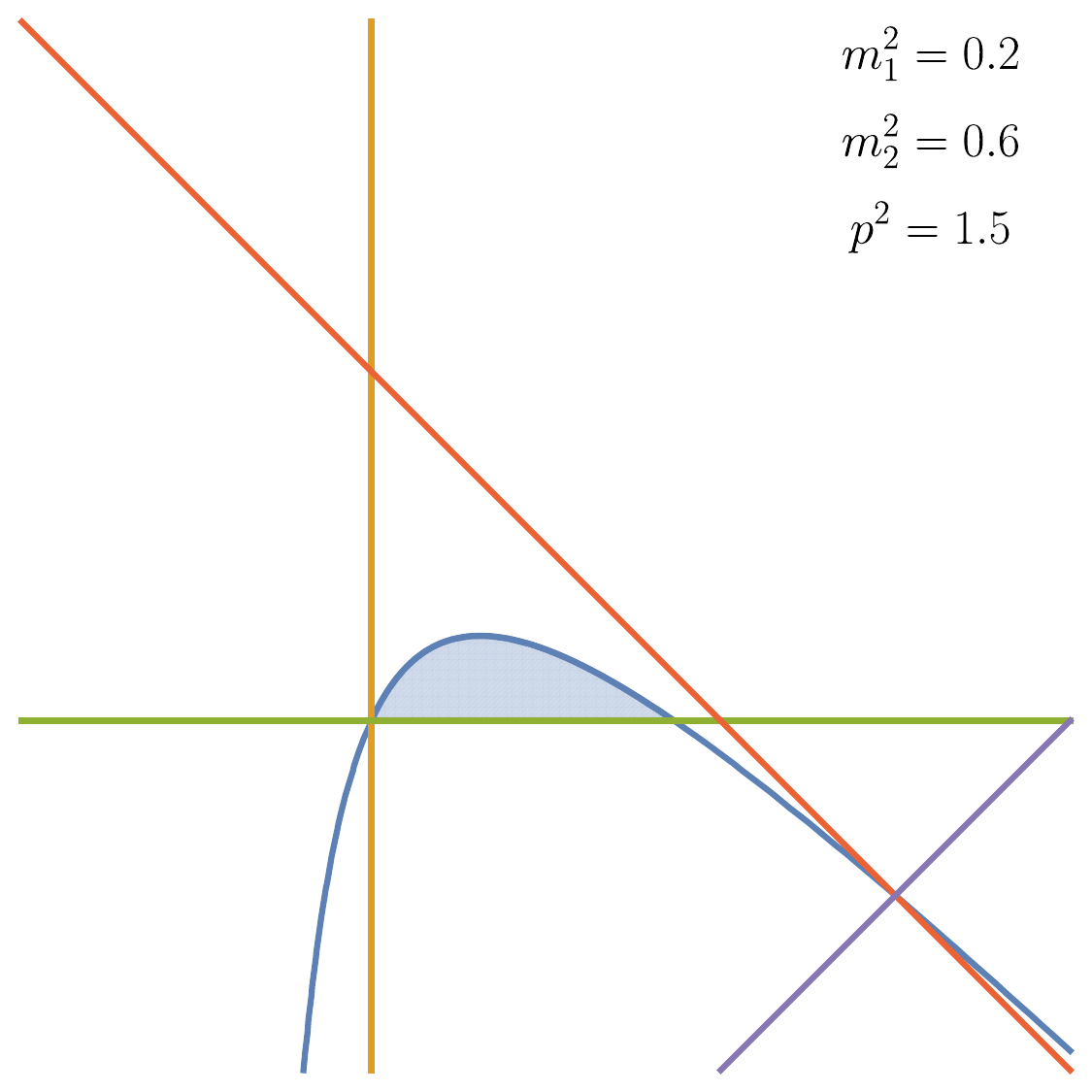}
     \end{tabular}
    \caption{Integration regions for discontinuities of $\I^{\trithree}_{\tri}$, with a compactification making the hyperplane at infinity visible. The top row is for $m_1^2 \otimes (m_1^2-m_2^2)$, corresponding to the bottom-left figure of Table \ref{tab:seqdiscT3}. The hyperplane at infinity becomes a new boundary of the contour.
    The bottom row is for $(p^2{-}m_1^2) \otimes (m_1^2-m_2^2)$, corresponding to the bottom-right figure of Table \ref{tab:seqdiscT3}. The contour is unaffected by the intersection at infinity. 
    }
\label{fig:seqdiscT3-H}
\end{figure}

Singularities of type (iii) 
can be studied with a compactification introducing a visible hyperplane at infinity. 
For the example of the triangle $\I^{\trithree}_{\tri}$, see Figure \ref{fig:seqdiscT3-H}. The hyperplane at infinity, $H_\infty$, appears in the lower right-hand corner. The three rows correspond respectively to the three regions in the fifth row of Table \ref{tab:seqdiscT3}. We can see now that the singularity $m_1^2-m_2^2=0$ is where the curve $\F=0$ crosses the intersection of $\alpha_3=0$ and the hyperplane at infinity. In the regions corresponding to the discontinuities at $m_1^2=0$ and $m_2^2=0$, the singularity at $m_1^2-m_2^2=0$ changes the shape of the integration region and involves $H_\infty$ as a new boundary. By contrast, the region corresponding to the discontinuity at $p^2{-}m_1^2=0$ is not affected by $H_\infty$. We conclude that the words $m_1^2 \otimes (m_1^2-m_2^2)$ and $m_2^2 \otimes (m_1^2-m_2^2)$ do appear in the symbol of $\I^{\trithree}_{\tri}$, and that  $(p^2{-}m_1^2) \otimes (m_1^2-m_2^2)$ does not. This conclusion is consistent with the symbols computed explicitly in equations (\ref{eq:t3-sym-wt2}) and (\ref{eq:t3-sym-wt3}).

In summary, this example of the triangle $\I^{\trithree}_{\tri}$ illustrates the predictions we wish to make in general, about the possible pairs of letters appearing in the first two entries of the symbol. When the region corresponding to the first letter experiences a change of boundary components at the singular locus of a different second letter, the pair is present in the symbol. When the region corresponding to the first letter does not experience a change of boundary components at the singular locus of a different second letter, the pair is absent in the symbol at weight 2. A word of length 2 with a repeated letter is generically present, but absent in this example because the integral is finite in $D=4$. 

We would like to remark that the analysis of sequential discontinuities from tracking contours can be extended to higher-weight truncations. In this example, the maximal cut truncates the first two entries of the symbol, not just one, for example as the sequence of the threshold singularity at $p^2=m_1^2$ and the mass singularity at $m^2=0$.
The corresponding integration contour $\Gamma_{\rm max}$ is bounded only by $\F=0$. Therefore, the singularities of $\Gamma_{\rm max}$ are of type (ii), located where $\F$ factorizes. Indeed we find that the third letter following $\left(p^2{-}m_1^2\right)\otimes m_{2}^2$ in equation (\ref{eq:S-3scale-w3}) is $\frac{p^2}{m_2^2 (m_1^2-m_2^2-p^2)}$, which contains the same letters as the discriminant of $\F$, from the set $\mathcal{A}_{(ii)}$.\footnote{More generally, singularities of the maximal cut include singularities occurring where $\F+\U$ factorizes. In this example, no new letters are involved.}

\subsection{Dimensional dependence of integration contours}
\label{subsection:dimensional}

In the discussion above and in \cite{Britto:2023rig}, we have focused on Feynman integrals in dimensional regularization, where the exponent of $\F$ in the parametric integral is non-integer. Then, the integrand is multi-valued, and the contours $\Gamma$ are elements of a twisted homology. That is, where $\Gamma$ has a boundary at a branch point of the integrand, it should be regulated with a construction similar to a Pochhammer contour \cite{AomotoKita}.

When the dimension $\D$ is instead an even integer, the polynomial $\F$ instead appears to an integer power in the integral. Thus, there is no longer a branch cut starting at $\F=0$, but instead only a \emph{pole} at $\F=0$. We can think about the qualitative difference between non-integer and even integer dimensions by looking at Eq.~\eqref{eq:discF}, where we saw that 
\begin{equation}
     (\F-i \varepsilon)^{-\lambda} - (\F+i \varepsilon)^{-\lambda} \propto
     \begin{cases}
        \delta^{\lambda-1} (-\F) &  \lambda \in \mathbb{Z}_+ \,,
        \\[5pt]
        \theta(-\F) (-\F)^{-\lambda}  & \lambda \not\in \mathbb{Z}_+\,, 
     \end{cases}
     \label{eq:cFdiff}
\end{equation}
and recalling that $\lambda = E - L \D/2$.

\begin{figure}
\centering

    \begin{subfigure}[t]{0.5\textwidth}
    \centering
    
    \begin{tikzpicture}[scale=1]
    \draw[->,thick] (-2,0) -- (2,0);
    \draw[->,thick] (0,-1) -- (0,2);
    \draw[thick] (1.7,1.7) -- (2,1.7);
    \draw[thick] (1.7,1.7) -- (1.7,2);
    \draw[thick] (1.65,1.65) node[above right] {$x$};
    \fill[Maroon] (0,0) circle[radius=0.07] node[below left, yshift=1.2em] {\footnotesize$0$};
    \fill[Maroon] (1,0) circle[radius=0.07] node[below right, yshift=1.2em] {\footnotesize$1$};
    \draw[->, line width=0.9, Maroon] (0.8,0.6) -- (0.7,0.4);
    \node[Maroon, scale=0.7, align=center] at (1,0.8) {original contour}; 
    \fill[RoyalBlue] (0.6,0) circle[radius=0.07] node[above, yshift=-20] {\footnotesize$x=-s$};
    \node[] at (-1,1) {$\lambda \in \mathbb{Z}$};
    \coordinate (start2) at (0,0); 
    \coordinate (end2) at (1,0); 
    \coordinate (p1) at (0.25,0.2);      
    \coordinate (p2) at (0.5,0.3);     
    \coordinate (p3) at (0.75,0.2);     
    \draw[line width=1.2, Maroon] 
        (start2) to[out=45, in=-135] 
        (p1) to[out=45, in=180] 
        (p2) to[out=0, in=135] 
        (p3) to[out=-45, in=135] 
        (end2);
    \draw[Maroon, line width=1.2, -{>[scale=0.6]}] ($(0.2,0.15)$) -- ($(0.2,0.15)+(0.05,0.025)$);
    \end{tikzpicture}
    \end{subfigure}%
    \begin{subfigure}[t]{0.5\textwidth}
    \centering
    
    \begin{tikzpicture}[scale=1]
    \draw[decorate, decoration={zigzag, segment length=6, amplitude=2}, RoyalBlue!80] (-2,0) -- (0.6,0);
    \draw[->,thick] (-2,0) -- (2,0);
    \draw[->,thick] (0,-1) -- (0,2);
    \draw[thick] (1.7,1.7) -- (2,1.7);
    \draw[thick] (1.7,1.7) -- (1.7,2);
    \draw[thick] (1.65,1.65) node[above right] {$x$};
    \fill[Maroon] (0,0) circle[radius=0.07] node[below left, yshift=1.2em] {\footnotesize$0$};
    \fill[Maroon] (1,0) circle[radius=0.07] node[below right, yshift=1.2em] {\footnotesize$1$};
    \draw[->, line width=0.9, Maroon] (0.8,0.6) -- (0.7,0.4);
    \node[Maroon, scale=0.7, align=center] at (1,0.8) {original contour}; 
    \fill[RoyalBlue] (0.6,0) circle[radius=0.07] node[above, yshift=-20] {\footnotesize$x=-s$};
    \node[] at (-1,1) {$\lambda \not\in \mathbb{Z}$};
    \coordinate (start2) at (0,0); 
    \coordinate (end2) at (1,0); 
    \coordinate (p1) at (0.25,0.2);      
    \coordinate (p2) at (0.5,0.3);     
    \coordinate (p3) at (0.75,0.2);     
    \draw[line width=1.2, Maroon] 
        (start2) to[out=45, in=-135] 
        (p1) to[out=45, in=180] 
        (p2) to[out=0, in=135] 
        (p3) to[out=-45, in=135] 
        (end2);
    \draw[Maroon, line width=1.2, -{>[scale=0.6]}] ($(0.2,0.15)$) -- ($(0.2,0.15)+(0.05,0.025)$);
    \end{tikzpicture}
    \end{subfigure}

    \begin{subfigure}[t]{0.5\textwidth}
    \centering
    
    \begin{tikzpicture}[scale=1]
    \draw[->,thick] (-2,0) -- (2,0);
    \draw[->,thick] (0,-1) -- (0,2);
    \draw[thick] (1.7,1.7) -- (2,1.7);
    \draw[thick] (1.7,1.7) -- (1.7,2);
    \draw[thick] (1.65,1.65) node[above right] {$x$};
    \fill[Maroon] (0,0) circle[radius=0.07] node[below left, yshift=1.2em,xshift=-0.3em] {\footnotesize$0$};
    \fill[Maroon] (1,0) circle[radius=0.07] node[below right, yshift=1.2em] {\footnotesize$1$};
    \draw[->, line width=0.9, Maroon, yshift=10] (0.8,0.6) -- (0.7,0.4);
    \node[Maroon, scale=0.7, align=center, yshift=20] at (1,0.8) {after analytic\\continuation}; 
    \fill[RoyalBlue] (0.6,0) circle[radius=0.07] node[above, yshift=-25] {\footnotesize$x=-s$};
    \node[] at (-1,1) {$\lambda \in \mathbb{Z}$};
    \coordinate (start2) at (0,0); 
    \coordinate (end2) at (1,0); 
    \coordinate (p3) at (0.8,0);
    \coordinate (p5) at (0,-0.3);
    \coordinate (p55) at (-0.2,0);
    \coordinate (p6) at (0,0.5);
    \coordinate (p7) at (0.5,0.7);
    \draw[line width=1.2, Maroon] 
        (start2) to[out=25, in=100] 
        (p3) to[out=-90, in=-30] 
       (p5) to[out=150, in=-90] 
       (p55) to[out=90, in=-135] 
        (p6) to[out=45, in=180] 
        (p7) to[out=0, in=90] 
        (end2);
    \draw[Maroon, line width=1.2, -{>[scale=0.6]}] ($(0.35,0.15)$) -- ($(0.35,0.15)+(0.05,0.025)$);    
    \end{tikzpicture}
    \end{subfigure}%
    \begin{subfigure}[t]{0.5\textwidth}
    \centering
    
  \begin{tikzpicture}[scale=1]
    \draw[decorate, decoration={zigzag, segment length=6, amplitude=2}, RoyalBlue!80] (-2,0) -- (-0.4,0);
    \draw[->,thick] (-2,0) -- (2,0);
    \draw[->,thick] (0,-1) -- (0,2);
    \draw[thick] (1.7,1.7) -- (2,1.7);
    \draw[thick] (1.7,1.7) -- (1.7,2);
    \draw[thick] (1.65,1.65) node[above right] {$x$};
    \fill[Maroon] (0,0) circle[radius=0.07] node[below left, yshift=1.2em,xshift=-0.3em] {\footnotesize$0$};
    \fill[Maroon] (1,0) circle[radius=0.07] node[below right, yshift=1.2em] {\footnotesize$1$};
    \draw[->, line width=0.9, Maroon, yshift=10] (0.8,0.6) -- (0.7,0.4);
    \node[Maroon, scale=0.7, align=center, yshift=20] at (1,0.8) {after analytic\\continuation}; 
    \fill[RoyalBlue] (0.6,0) circle[radius=0.07] node[above, yshift=-25] {\footnotesize$x=-s$};
    \coordinate (start2) at (0,0); 
    \coordinate (end2) at (1,0); 
    \coordinate (p3) at (0.8,0);
    \coordinate (p5) at (0,-0.3);
    \coordinate (p55) at (-0.2,0);
    \coordinate (p6) at (0,0.5);
    \coordinate (p7) at (0.5,0.7);
    \draw[line width=1.2, Maroon] 
        (start2) to[out=25, in=100] 
        (p3) to[out=-90, in=-30] 
       (p5) to[out=150, in=-90] 
       (p55) to[out=90, in=-135] 
        (p6) to[out=45, in=180] 
        (p7) to[out=0, in=90] 
        (end2);
    \draw[Maroon, line width=1.2, -{>[scale=0.6]}] ($(0.35,0.15)$) -- ($(0.35,0.15)+(0.05,0.025)$);
    \coordinate (start) at (-0.4,0); 
    \coordinate (end) at (0.6,0); 
    \coordinate (q1) at (0,-0.5);  
    \coordinate (q2) at (0.87,0);   
    \coordinate (q3) at (0.4,0.45);   
    \coordinate (q4) at (-0.1,0);   
    \draw[decorate, decoration={zigzag, segment length=6, amplitude=0.8}, RoyalBlue!80]
    (start) to [out=-45, in=180] 
    (q1) to [out=0, in=-90]
    (q2) to [out=90, in=0]
    (q3) to [out=180, in=90]
    (q4) to [out=-90, in=-125]
    (end);
    \end{tikzpicture}
    \end{subfigure}

    \begin{subfigure}[t]{0.5\textwidth}
    \centering
    
    \begin{tikzpicture}[scale=1]
    \draw[->,thick] (-2,0) -- (2,0);
    \draw[->,thick] (0,-1) -- (0,2);
    \draw[thick] (1.7,1.7) -- (2,1.7);
    \draw[thick] (1.7,1.7) -- (1.7,2);
    \draw[thick] (1.65,1.65) node[above right] {$x$};
    \fill[Maroon] (0,0) circle[radius=0.07] node[below left, yshift=1.2em,xshift=-0.3em] {\footnotesize$0$};
    \fill[Maroon] (1,0) circle[radius=0.07] node[below right, yshift=1.2em] {\footnotesize$1$};
    \draw[->, line width=0.9, Maroon, yshift=10] (0.9,0.6) -- (0.76,0.2);
    \node[Maroon, scale=0.7, align=center, yshift=20] at (1,0.8) {monodromy\\contour}; 
    \fill[RoyalBlue] (0.6,0) circle[radius=0.07] node[above, yshift=-25] {\footnotesize$x=-s$};
    \node[] at (-1,1) {$\lambda \in \mathbb{Z}$};
    \draw[Maroon, line width=1.2] (0.6,0) circle[radius=0.2];
    \draw[Maroon, line width=1.2, -{<[scale=0.6]}] ($(0.52,0.2)$) -- ($(0.57,0.2)+(0.1,0)$);
    \end{tikzpicture}
    \end{subfigure}%
    \begin{subfigure}[t]{0.5\textwidth}
    \centering
    
    \begin{tikzpicture}[scale=1]
    \draw[decorate, decoration={zigzag, segment length=6, amplitude=2}, RoyalBlue!80] (-2,0) -- (0.6,0);
    \draw[->,thick] (-2,0) -- (2,0);
    \draw[->,thick] (0,-1) -- (0,2);
    \draw[thick] (1.7,1.7) -- (2,1.7);
    \draw[thick] (1.7,1.7) -- (1.7,2);
    \draw[thick] (1.65,1.65) node[above right] {$x$};
    \fill[Maroon] (0,0) circle[radius=0.07] node[below left, yshift=1.2em,xshift=-0.3em] {\footnotesize$0$};
    \fill[Maroon] (1,0) circle[radius=0.07] node[below right, yshift=1.2em] {\footnotesize$1$};
    \draw[->, line width=0.9, Maroon, yshift=10] (0.9,0.6) -- (0.76,0.2);
    \node[Maroon, scale=0.7, align=center, yshift=20] at (1,0.8) {monodromy\\contour}; 
    \fill[RoyalBlue] (0.6,0) circle[radius=0.07] node[above, yshift=-25] {\footnotesize$x=-s$};
    \node[] at (-1,1) {$\lambda \not\in \mathbb{Z}$};
    \coordinate (start2) at (0,0); 
    \coordinate (p1) at (0.5,0.2);  
    \coordinate (p2) at (0.85,0);   
    \coordinate (p3) at (0.5,-0.2);
    \draw[line width=1.2, Maroon] 
        (start2) to[out=45, in=180] 
        (p1) to[out=0, in=90] 
        (p2) to[out=-90, in=0] 
        (p3) to[out=180, in=-45]
        (start2);
    \draw[Maroon, line width=1.2, -{>[scale=0.6]}] ($(0.25,0.15)$) -- ($(0.25,0.15)+(0.05,0.025)$);
    \end{tikzpicture}
    \end{subfigure}
    \caption{\textbf{Top row.} The original integration contour for $\I$, for integer \textbf{(left)} and non-integer \textbf{(right)} values of $\lambda$ encircles the singularity of the integrand by a deformation in the upper half-plane. \textbf{Middle row.} After analytically continuing $s$ in the complex plane in a counterclockwise circle around $s=0$, the integration contour has been suitably modified. \textbf{Bottom row.} The contour representing the monodromy of $\I$ around $s=0$ is given as either a contour around the pole, or, when $\lambda$ is not an integer, as a contour around the branch cut. 
    }
    \label{fig:monodromies}
\end{figure}
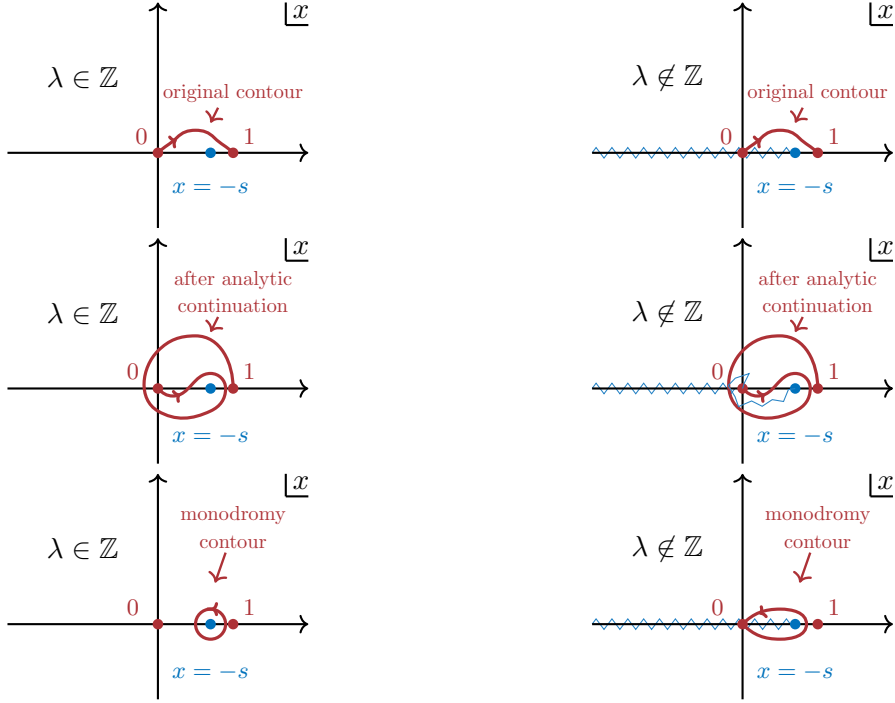

To illustrate what happens in integer dimensions, let us get back to the toy model from Sec.~\ref{sec:monodromy_theory}, but set $\epsilon=0$:
\begin{equation}
    \I = \int_0^1 \frac{\d x}{s + x + i \varepsilon} \,.
\end{equation}
In the complex $x$-plane, the singularity at $x=-s$ is represented by a pole, and when $-1<s<0$, we can compute the integral by deforming the integration contour around this pole via the $i\varepsilon$, see Fig.~\ref{fig:monodromies}. When analytically continuing in a counterclockwise circle around $s=0$, the root at $x=-s$ also traverses in a counterclockwise circle around $x=0$ and the integration contour gets deformed. The monodromy, i.e., the difference between the original integration contour and the one after analytic continuation becomes $2\pi i$ times the residue at $x=-s$, or
\begin{equation}
    (1 - \mathscr{M}_s) \, \I = 2 \pi i \, \text{Res}_{s=-x} \left( \frac{1}{s+x} \right) = 2 \pi i\,.
\end{equation}

We can contrast this result with the one when, say $\epsilon=1/2$, and we have to take into account that the branch cut obstructs the integration contour between $0$ and $x=-s$ canceling out. Instead, they differ by a phase factor, which in the case of a square-root branch cut is simply $-1$. That is, the values just below the branch cut are the negatives of those just above the branch cut. Taking into account the orientation of the contour, the integral over the monodromy contour simply gives \emph{twice} the integral from $0$ to $x=-s$:
\begin{equation}
    (1 - \mathscr{M}_s) \, \I = -2 \int_0^{-s} \frac{\d x}{(s+x+i\varepsilon)^{\epsilon}}= -4 i \sqrt{-s}\,, \qquad \text{when } \epsilon = 1/2 \,.
\end{equation}
Note that if $\epsilon \neq 1/2$, the phase factor is different, as represented by the $-\sin (\pi \lambda)$ in~\eqref{eq:discF}.

We immediately notice a qualitative difference between these two answers. When $\epsilon = 0$, we obtain a discontinuity in $s$ that has \emph{no further discontinuities} in $s$. However, when $\epsilon = 1/2$, we can freely keep taking discontinuities in $s$, and each time the monodromy gets multiplied by a factor of two. This is reflected in the monodromy contours in Figure~\ref{fig:monodromies}: the monodromy contour for $\lambda \in \mathbb{Z}$ can be freely analytically continued around $x=0$ or $x=1$ and it will always simply go back to itself. Thus, sequential monodromies will be zero. On the other hand, the monodromy contour for $\lambda \not\in \mathbb{Z}$ will drag the branch cut around for each circle that the root at $x=-s$ traverses around the endpoints at $x=0$ or $x=1$, as shown previously in Figure~\ref{fig:monodromies_tracking}. Hence, sequential monodromies around $s=0$ will not vanish in this case. This observation is reflected in the full answers, which are given by
\begin{equation}
    \I = \begin{cases}
        \log(1 + \frac{1}{s+i\varepsilon}) & \text{when } \epsilon = 0 \,,
        \\
        2 \sqrt{s+i \varepsilon +1}-2 \sqrt{s+i \varepsilon }
        & \text{when } \epsilon = 1/2 \,.
    \end{cases} 
\end{equation}
In these results, we see that analytically continuing around $s=0$ collapses the log when $\epsilon=0$, but leaves a $-4\sqrt{s}$ when $\epsilon = 1/2$.

Note that integer dimensions that are a multiple of 2 do not necessarily imply that a singularity is logarithmic. As a toy example, equivalent to a bubble Feynman integral in $\D=2$ spacetime dimensions with some reinterpretation of the parameters, consider the integral
\begin{equation}
    \I = \int_0^1 \frac{\d x}{s+x^2+i\varepsilon} \,.
\end{equation}
In this case, there are two roots of the integral, at $x = \pm \sqrt{s} \mp i \varepsilon$. For $0<s<1$, these roots are slightly displaced off the real axis, with the integration contour going straight between them. As $s$ is analytically continued around $s=0$, the two roots switch places. If we now take a sequential monodromy, the roots will again switch places, but the integration contours have the opposite orientations. This shows that the monodromy contour for the \emph{sequential} discontinuity is represented by \emph{twice} the monodromy contour for the first discontinuity, consistent with a square-root singularity. Indeed, the result of the integral is
\begin{equation}
    \I = \frac{i \log \left(1-\frac{i}{\sqrt{s}}\right)}{2 \sqrt{s}}-\frac{i \log \left(1+\frac{i}{\sqrt{s}}\right)}{2 \sqrt{s}}\,,
\end{equation}
which behaves as $\I \sim \frac{\pi}{2 \sqrt{s}}$ around $s=0$. This shows that even though integer powers of the denominators lead to circular contours (or tubular in higher dimensions), the singularity need not be logarithmic. For a further discussion of this specific example, see~\cite[Sec.~3]{hannesdottir:2022bmo}.

Looking ahead, analyzing what types of contours are obtained for discontinuities and sequential discontinuities of Feynman-integrals are crucial in determining their form from first principles. Since non-integer dimensions often involve analytic continuations around branch cuts in the space of integration variables, we generally expect sequential discontinuities to be nonvanishing. At the level of the symbol, a nonvanishing sequential discontinuity can correspond to a square-root type singularity, and/or to a repeated letter of the symbol. Nevertheless, in both integer and non-integer dimensions, the monodromy contours can result in ones analogous to the left bottom panel of Fig.~\ref{fig:monodromies}, where there are no sequential discontinuities and hence no repeated letters.

\subsection{Using constraints in a bootstrap procedure}
\label{sec:dimless-ratios}

In the examples presented in the following sections, we fix the integrated function as much as possible using the constraints on sequential discontinuities from tracking the integration contours in parameter space. When possible, we will start in a Euclidean kinematic region (i.e., where $\cF > 0$ manifestly on the integration contour), and analytically continue to other kinematic regions containing points with $\cF < 0$ by changing one kinematic variable at a time. Then, we have a well-defined integration contour for the corresponding monodromy, which we continue to track as we  analytically continue around another branch point. At this point, we only determine whether the sequential discontinuity is zero or not.

These types of constraints determine what letters can follow each other in the symbol. Using that information, we go through the procedure of the Landau bootstrap~\cite{Hannesdottir:2024hke}, aiming to fully determine the Feynman integral from its singularities, sequential discontinuities, limits, symmetries and other constraints. We work with examples that are known to have constant leading singularities, i.e.~there is an overall algebraic factor multiplying a pure function.

We will see that it proves to be powerful to consider symbol letters that are dimensionless ratios of kinematic variables. Some of the examples will be carried out in dimensional regularization, as an expansion around $\D=4-2\epsilon$ dimensions, in which case we have a new kinematic scale $\mu^2$. Since we know that $\mu^2$ appears as a prefactor to the power of $2 L \epsilon$, we can still get meaningful predictions order by order in $\epsilon$, by relying on lower-order results.

We will try to get as far as possible using the constraints from sequential discontinuities and ones that are straightforward to compute. In this paper, we will not always chase the goal of fixing the full answer, which can for example be done numerically (see~\cite{Barrera:2025uin}).

\section{One-loop examples}
\label{sec:one-loop}

In this section, we will show how to leverage the tracking of integration contours to constrain and sometimes determine one-loop integrals. We will start with a simple example of a two-scale triangle integral in $\D=4$ spacetime dimensions of uniform transcendental weight, and then move on to non-integer dimensions, non-uniform weight, and more scales.

\subsection{Triangle with one leg off shell, mass opposite}
\paragraph{Uniform transcendental weight}
We start with the simple example of a scalar triangle integral with one off-shell external leg with incoming momentum $p$ and a nonzero internal mass $m$ on the opposite edge to the off-shell leg. Let $s=p^2$.
The Feynman integral is
\begin{equation}
    \I_{\tri}^{\triopm} =
    \begin{gathered}
    \begin{tikzpicture}[line width=1,scale=0.75]
    	\coordinate (v1) at (0,0);
        \coordinate (v2) at (0.866025,0.5);
        \coordinate (v3) at (0.866025,-0.5);
        \draw[] (v1) -- (v2) -- (v3) -- (v1);
        \draw[line width=2] (v2) -- (v3) node[right,yshift=12]{$m$};
        \draw[] (v1) -- ++(-190:0.7);
        \draw[] (v1) -- ++(-170:0.7);
        \draw[] (v3) -- ++(-45:0.7);
        \draw[] (v2) -- ++(45:0.7);
        \node[] at ($(v1)-(1.0,0.2)$) {$p$};
    \end{tikzpicture}
    \end{gathered}
    =
    e^{\gamma_E \epsilon} \mu^{2 (3-D/2)} \, \Gamma\Big(3 - \frac{\D}{2}\Big)
    \int_0^\infty \frac{\d^3 \alpha}{\GL(1)}  \frac{\left(\U_{\tri}^{\triopm}\right)^{3-\D}}{\left(\F_{\tri}^{\triopm}-i\varepsilon\right)^{3-\D/2} } \,,
    \label{eq:tri-op-m}
\end{equation}
where
\begin{align}
    \F_{\tri}^{\triopm} & = -s \alpha_1 \alpha_2 + m^2 \alpha_3 (\alpha_1+\alpha_2+\alpha_3) \,, \\
    \U_{\tri}^{\triopm} & = \alpha_1+\alpha_2+\alpha_3 \,.
\end{align}
We henceforth set $\mu^2=1$.
As a triangle with unit propagators, this integral is of uniform transcendental weight  and is UV and IR finite in $\D=4$ spacetime dimensions. Moreover, it depends on only one dimensionless ratio,
\begin{equation}
    x = \frac{s}{m^2}\,,
\end{equation}
which makes the analysis below particularly simple.

Let us start by identifying the Landau singularities \cite{Fevola:2023kaw,Fevola:2023fzn,Helmer:2024wax,Caron-Huot:2024brh,Correia:2025yao} and tracking the integration contours as we take monodromies around each of them. 
In this example, the discriminant of $\F_{\tri}^{\triopm}$ is $-m^2 s (s+m^2)/4$, and we classify the singularities as follows.
\begin{itemize}
    \item $s=0$: A physical threshold singularity where the vanishing cell is bounded by $\F_{\tri}^{\triopm}$ and $\alpha_3=0$.
    \item $m^2=0$: A physical singularity where the vanishing cell is bounded by $\F_{\tri}^{\triopm}$, $\alpha_1=0$ and $\alpha_2=0$. Additionally, $\F_{\tri}^{\triopm}$ intersects itself at this point, so we expect this singularity to survive to the last entries of the symbol.
    \item $s+m^2=0$: An unphysical singularity (i.e.\ a singularity on the second sheet).
\end{itemize}

\begin{figure}[t]
    \centering
    \includegraphics[width=0.32\linewidth]{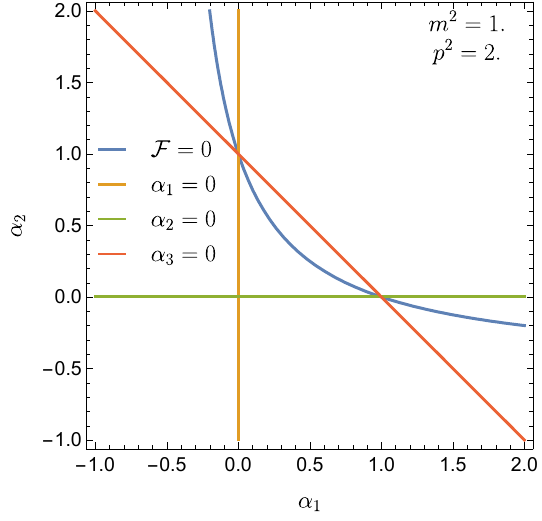}
    \includegraphics[width=0.32\linewidth]{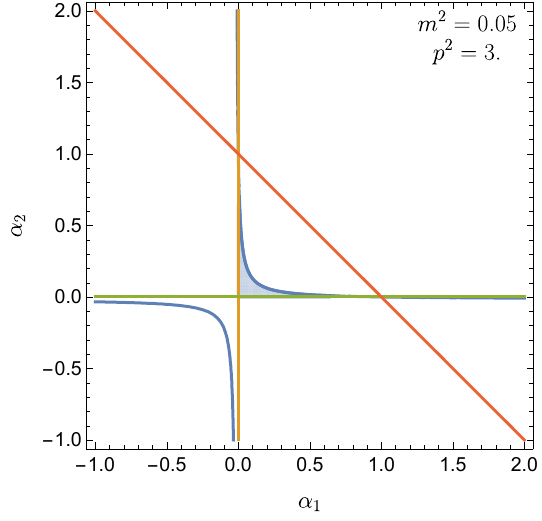}
    \includegraphics[width=0.32\linewidth]{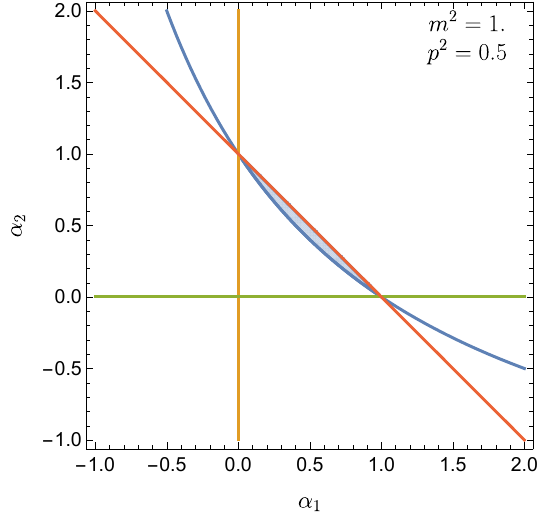}
    \caption{Tracking the integration contours of a two-scale triangle with one off-shell leg and an opposite mass. \textbf{Left.} Generic values of $m^2$ and $s$.\textbf{Middle.} When encircling $m^2 = 0$, the monodromy contour is bounded by $\alpha_1=0$, $\alpha_2=0$ and $\F=0$. \textbf{Right.} When encircling $s = 0$, the monodromy contour is bounded by $\alpha_3=0$ and $\F=0$. After taking a monodromy around either $s=0$ or $m^2=0$, the other one becomes disallowed.}
    \label{fig:two-scale_tri-opp}
\end{figure}

Since there is just one dimensionless ratio $x$, the symbol letters can be given as functions of $x$.
In fact, the Landau singularities only allow for two possible independent dimensionless letters, which we take to be $x$ and $1+x$. Of these two letters, only $x$ can appear in the first entry, because $1+x$ contains a non-physical singularity.
We track the monodromy contours in Fig.~\ref{fig:two-scale_tri-opp} and see that the integration contour obtained by encircling $m^2=0$ is bounded by $\alpha_1=0$, $\alpha_2=0$ and $\F=0$, which also coincides with the case when $\F=0$ degenerates itself. However, the integration contour for the $s=0$ singularity is bounded by only $\alpha_3=0$ and $\F=0$. This means, according to the analysis in Sec.~\ref{sec:seqdisc_from_contours}, that once we have taken a monodromy around $m^2=0$, the $s=0$ singularity is no longer accessible. In other words, the symbol cannot have a term of the form $m^2 \otimes s$. Since $x = \frac{s}{m^2}$, this analysis shows that 
a term of the form $x \otimes x$ cannot appear, since it would lead to a sequential discontinuity in $s$ following $m^2$.   
However, after taking a monodromy around $s=0$, we can still access the $m^2=0$ singularity when it corresponds to $\cF=0$ degenerating itself, so sequences of the form $x \otimes 1+x$ or $1+x \otimes 1+x$ are allowed. 

\begin{table}
\centering
\begin{tabular}{c|c}
\diagbox{$a_2$}{$a_1$}  & $x$\\
\hline
$x$ & $\times$ \\
$1+x$ & $\checkmark$
\end{tabular}
\caption{Summary of the sequential-discontinuity tracking for the one off-shell leg and opposite mass triangle.}
\label{tab:tri_first}
\end{table}

With these constraints, the only possible symbol term at weight 2 for the integral from~\eqref{eq:tri-op-m} is
\begin{equation} \label{eq:sym-tri-op-m}
    \mathcal{S}^2 (\I_{\tri}^{\triopm}) \propto x \otimes (1+x) \,.
\end{equation}
It is easy to identify this symbol as the one of the dilogarithm $-\Li_2(1 + x)$.
To fix the answer for the full integral, we also have to compute the leading singularity, which we find to be
\begin{equation}
    \LS \, \I_{\tri}^{\triopm} = \frac{1}{s} \,.
    \label{eq:maxCut-triopm}
\end{equation}
Then, we can identify the function to be
\begin{equation}
\label{eq:Itriopm}
    \I_{\tri}^{\triopm} = \frac{1}{s} \left[ \Li_2 \left(1 + \frac{s}{m^2} \right) 
    + c_1 \right] \,,
\end{equation}
where $c_1$ is a numerical constant. We check that this branch of the dilogarithm is suitable for physical interpretation: the values on other branches would introduce singularities at $s+m^2=0$, which is not on the physical sheet.
The constant $c_1$ can be determined, for example, by taking the limit of $\I_{\tri}^{\triopm}$ as $s \to - m^2$, which shows that $c_1 = -\frac{\pi^2}{6}$. This analysis fixes the full expression to be 
\begin{equation}
    \I_{\tri}^{\triopm} = \frac{1}{s} \left[ \Li_2 \left(1 + \frac{s}{m^2} \right) - \frac{\pi^2}{6} \right] \,.
\end{equation}

\paragraph{Non-uniform weight}
It seems intriguing that we could fix the weight-two part symbol of  $\I_{\tri}^{\triopm}$ just by looking at the physical singularities and sequential discontinuities of $\I_{\tri}^{\triopm}$. Thus, the integral in $\D=4$ spacetime dimensions with general powers of the propagators,
\begin{equation}
    {\I_{\tri, \, (\nu_1,\nu_2,\nu_3)}^{\triopm}} 
    =
    \int_0^\infty \frac{\d^3 \alpha_e}{\GL(1)}  \frac{\alpha_1^{\nu_1-1} \alpha_2^{\nu_2-1} \alpha_3^{\nu_3-1} }{\left(\F_{\tri}^{\triopm}-i\varepsilon\right) \U_{\tri}^{\triopm} } \,,
\end{equation}
should have the same symbol at weight 2 as given in~\eqref{eq:sym-tri-op-m}. However, a key property changes when the $\nu_i$ are not all 1: the integral is no longer of uniform transcendental weight. Furthermore, the leading singularity changes depending on the propagator powers.

To explore what happens in this case, let us look at a concrete example where we take $(\nu_1, \nu_2, \nu_3) = (2,1,1)$. Then,
\begin{equation}
    {\I_{\tri, \, (2,1,1)}^{\triopm}}
    =
    \int_0^\infty \frac{\d^3 \alpha_e}{\GL(1)}  \frac{\alpha_1}{\left(\F_{\tri}^{\triopm}-i\varepsilon\right) \U_{\tri}^{\triopm} } \,.
\end{equation}
The leading singularity is now given by $\LS \, \I_{\tri}^{\triopm} = -\frac{m^2}{s^2}$, and according to our previous analysis, the highest-weight term of the symbol is the same as in~\eqref{eq:sym-tri-op-m}. With the first-entry condition, our ansatz for the non-uniform-weight
symbol becomes 
\begin{equation}
{\mathcal S}(\I_{\tri}^{\triopm})
\propto 
\left[x \otimes (1+x) + c_1 \,\otimes x + c_2 \right]
\end{equation}
leading to a function of the form
\begin{equation}
    {\I_{\tri, \, (2,1,1)}^{\triopm}} = - \frac{m^2}{s^2} \left[ \Li_2 \left(1 + \frac{s}{m^2} \right)   + c_1 \log \left( - \frac{m^2}{s}  \right)+ c_2\right] \,.
\end{equation}
Again we check that the branch of the dilogarithm is appropriate: it is the one exhibiting regularity at $s = -m^2$. We can also use that the $m^2 \to \infty$ limit should have leading terms that go like $\log(m^2)/m^2$, as can be checked e.g.\ using the method of regions~\cite{Jantzen:2012mw}. Our ansatz, on the other hand, behaves as
\begin{equation}
    \lim_{m^2 \to \infty} {\I_{\tri, \, (2,1,1)}^{\triopm}}= -\frac{m^2 \left[  c_1 \log (-m^2/s)+ c_2 +\pi ^2/6\right]}{s^2}+\frac{\log (-m^2/s)-1}{s} + \mathcal{O} \left( \frac{\log(-m^2/s)}{m^2}  \right)  \,,
\end{equation}
showing that we can fix the values of $c_1$ and $c_2$ as the ones that eliminate the superleading terms in the $m^2 \to \infty$ limit.
To eliminate the first two terms, we set
\begin{align}
    c_1 & =  - \frac{s}{m^2} + \mathcal{O} \left(  \log \Big( -\frac{m^2}{s} \Big) \frac{s^2}{m^4} \right) \,,\\
    c_2 & = - \frac{s}{m^2} - \frac{\pi^2}{6} + \mathcal{O} \left( \log \Big( -\frac{m^2}{s} \Big) \frac{s^2}{m^4} \right) \,. 
\end{align}
We need one more condition to fix the solution uniquely since we have still not excluded terms that are suppressed in the $m^2 \to \infty$ limit. For that, we can use a bound on the growth of ${\I_{\tri, \, (2,1,1)}^{\triopm}}$ as $s \to \infty$, which we can compute to be at most $\log(s)/s$, to forbid all terms of order $ \mathcal{O} \left(  \log \left(\frac{-m^2}{s} \right) \frac{s^2}{m^4} \right)$ in $c_1$ and $c_2$. So, we have uniquely fixed the answer to be
\begin{equation}
    {\I_{\tri, \, (2,1,1)}^{\triopm}}= - \frac{m^2}{s^2} \left[ \Li_2 \left(1 + \frac{s}{m^2} \right)  - \frac{s}{m^2} \log \left( - \frac{m^2}{s} \right)
    - \frac{s}{m^2} - \frac{\pi^2}{6}\right] \,.
\end{equation}

\subsection{Triangle with one leg off-shell, mass adjacent}
\paragraph{Non-integer dimensions}
Our next example is a triangle integral with one off-shell leg with incoming momentum $p$, and one massive internal edge, adjacent to the off-shell leg. The Feynman integral takes the form
\begin{equation}
\I^{\triadjm}_{\tri} =
    \begin{gathered}
    \begin{tikzpicture}[line width=1,scale=0.75]
    	\coordinate (v1) at (0,0);
        \coordinate (v2) at (0.866025,0.5);
        \coordinate (v3) at (0.866025,-0.5);
        \draw[] (v1) -- (v2) -- (v3) -- (v1);
        \draw[line width=2] (v3) -- (v1) node[below,yshift=-3,xshift=2]{$m$};
        \draw[] (v1) -- ++(-190:0.7);
        \draw[] (v1) -- ++(-170:0.7);
        \draw[] (v3) -- ++(-45:0.7);
        \draw[] (v2) -- ++(45:0.7);
        \node[] at ($(v1)-(1.0,0.2)$) {$p$};
    \end{tikzpicture}
    \end{gathered}
    =
    e^{\gamma_E \epsilon} \mu^{2 (3-D/2)} \,
    \Gamma\left(3 - \frac{\D}{2}\right)
    \int_0^\infty \frac{\d^3 \alpha_e}{\GL(1)}  \frac{\left(\U^{\triadjm}_{\tri}\right)^{3-\D}}{\left(\F^{\triadjm}_{\tri}-i\varepsilon\right)^{3-\D/2} } \,,
    \label{eq:tri-adj-m}
\end{equation}
with
\begin{equation}
    \F^{\triadjm}_{\tri} = - s \alpha_1 \alpha_2 + m^2 \alpha_1 (\alpha_1+\alpha_2+\alpha_3) \,, \qquad \U^{\triadjm}_{\tri} = \alpha_1+\alpha_2+\alpha_3 \,.
\end{equation}
Let us gauge fix the $\GL(1)$ covariance by setting $\U^{\triadjm}_{\tri}=1$. Then,
\begin{equation}
    \F^{\triadjm}_{\tri} = - \alpha_1 \left(s \alpha_2-m^2\right) \,.
\end{equation}
In this representation,  $\F^{\triadjm}_{\tri}$ is always factorized into a product of two hyperplanes, regardless of the values of $s$ and $m^2$. Moreover, the hyperplane at $\alpha_1=0$ is on the boundary of the integration contour. Thus, this configuration signals a possible \emph{divergence} of the integral. Indeed, we see that since the $\alpha_1$ integral factors out, ~\eqref{eq:tri-adj-m} can be written as
\begin{equation}
\label{eq:div-triadjm}
    \I^{\triadjm}_{\tri} =
    \int_0^1 \d \alpha_2 \int_0^{1-\alpha_2}  \frac{\d \alpha_1}{\alpha_1^{3-\D/2}} \times \ldots \,,
\end{equation}
where the dots represent the $\alpha_1$-independent factors.
This form shows that the integral $\I^{\triadjm}_{\tri}$ will be logarithmically divergent in $\D=4$ due to the integration region around $\alpha_1=0$. Our goal is to determine the integral order by order in an expansion around $\D=4-2\epsilon$ spacetime dimensions, where $\epsilon\to 0$.

We start by identifying the Landau singularities. They are at 
\begin{equation}
    s = 0\,, \qquad s - m^2 = 0 \,, \qquad m^2 = 0 \,,
\end{equation}
of which the $s=0$ singularity is not on the physical sheet. For the first entries, we thus have:
\begin{itemize}
    \item $s-m^2=0$: A physical threshold singularity where the vanishing cell is bounded by $\alpha_3 = 0$ and $\F = 0$. Note that $\F=0$ includes the coordinate hyperplane $\alpha_1 = 0$ as one of its factors. 
    \item $m^2 = 0$: A physical mass
    singularity for which the vanishing cell is bounded by  $\alpha_2 = 0$, $\alpha_3=0$ and $\F = 0$, again noting that $\F=0$ includes the hyperplane $\alpha_1 = 0$.
\end{itemize}
Just as in the previous example, we only have one dimensionless ratio, $x = s/m^2$, and the singularities written in terms of $x$ are at
\begin{equation}
    x = 0, \qquad x-1 = 0, \qquad x^{-1} = 0 \,.
\end{equation}
The solution at $s = 0$ (or $x = 0$) is of second type, so it is not a physical singularity of the integral. We therefore only have one possible first entry in the symbol, namely $x-1$. 

\begin{figure}[t]
    \centering
    \includegraphics[width=0.32\linewidth]{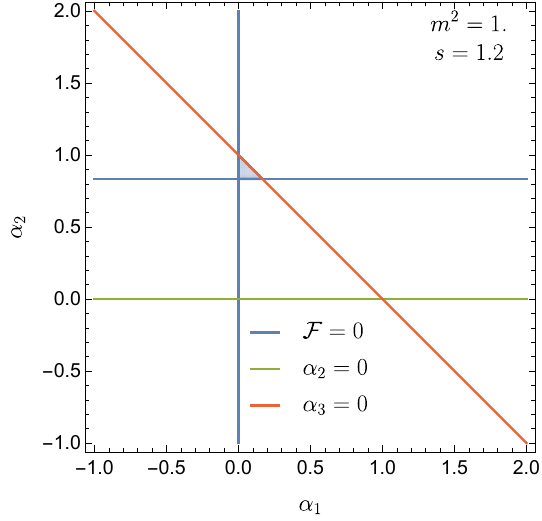}
    \includegraphics[width=0.32\linewidth]{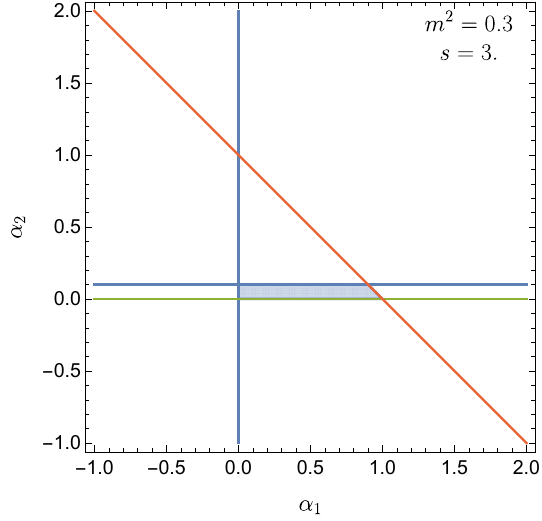}
    \includegraphics[width=0.32\linewidth]{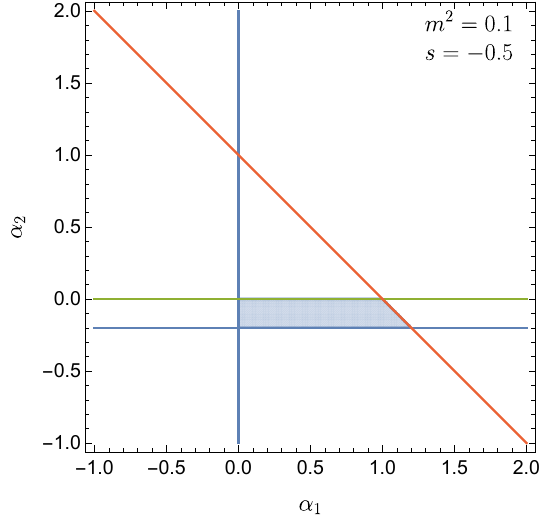}
    \caption{Tracking the integration contours of a two-scale triangle with one off-shell leg and an adjacent mass. \textbf{Left.} When encircling $s = m^2$, the monodromy contour is bounded by $\alpha_3=0$ and $\F=0$. \textbf{Middle.} When encircling $m^2 = 0$, the monodromy contour is bounded by $\alpha_2=0$, $\alpha_3=0$ and $\F=0$. \textbf{Right.} If we start in the region where $s<m^2$, the monodromy contour when encircling $m^2=0$ lands outside the physical integration region.}
    \label{fig:two-scale_tri-adj}
\end{figure}

Next, we work out the sequential discontinuities. This can readily be done from Fig.~\ref{fig:two-scale_tri-adj}. While the singularity at $m^2=0$ retains all the integration boundaries in its discontinuity contour, including $\alpha_2=0$, the singularity at $s-m^2$ is not bounded by $\alpha_2=0$. Thus, the singularity at $m^2=0$ cannot follow the one at $s-m^2$. This information is summarized in Table~\ref{tab:tri_adjmass}.
\begin{table}
\centering
\begin{tabular}{c|c|c}
\diagbox{$a_2$}{$a_1$}  & $s-m^2$ & $m^2$ \\
\hline
$s-m^2$ & $\checkmark$ & $\checkmark$ \\
$m^2$ & $\times$ & $\checkmark$ \\
$s$ & $\checkmark$ & $\checkmark$
\end{tabular}
\caption{Summary of the sequential-discontinuity tracking for the one off-shell leg and adjacent mass triangle.}
\label{tab:tri_adjmass}
\end{table}
Last, we determine the leading singularity from the maximal cut, which will determine the prefactor of the integral, 
\begin{equation}
    \LS \, \I^{\triadjm} \propto \frac{(s-m^2)^{\D-4}}{s^{\D/2-1}} \,.
    \label{eq:maxCut-triadjm}
\end{equation}
We now have enough information to start bootstrapping this integral.

We start with the coefficient of $1/\epsilon$, coming from the logarithmic divergence of~\eqref{eq:div-triadjm}. We find a prefactor of $1/s$ as we plug $\D=4$ into ~\eqref{eq:maxCut-triadjm}. The bootstrap is extremely simple for this term: since we assume that the maximum weight for this integral in $\D=4$ dimensions is equal to 2, and $1/\epsilon$ has a weight of 1, this term can only multiply a single logarithm, and there is only one letter that can appear. So we already have
\begin{equation}
\label{eq:tri-adjm_divergence}
    \I^{\triadjm} \propto \frac{1}{\epsilon} \frac{\log(x-1)}{s} + \mathcal{O} (\epsilon^0) \,.
\end{equation}

Next, we bootstrap the $\mathcal{O} (\epsilon^0)$ term. The bootstrap is now made a bit more complicated due to the fact that we have an extra scale in the problem, $\mu^2$ from~\eqref{eq:tri-adj-m}. So we cannot simply form dimensionless ratios of Landau singularities as our symbol letters (or, in other words, $\mu^2=0$ is an extra Landau singularity). Instead, we use the alphabet,
\begin{equation}
    L_1 = \frac{s-m^2}{\mu^2}\,, \qquad L_2 = \frac{m^2}{\mu^2}\,, \qquad L_3 = \frac{s}{\mu^2} \,.
\end{equation}
and form the symbol ansatz at weight 2,
\begin{equation}
    \mathcal{S}_2 (\I^{\triadjm}) = \sum_{i \leq j} c_{ij} \, L_i \otimes L_j \,.
\end{equation}
Imposing integrability fixes two of the $c_{ij}$'s, bringing the number of undetermined coefficients down to 7. Imposing that $L_3$ cannot appear in the first entry brings us down to 4 coefficients, and imposing the sequential-discontinuity relations gets us down to only 3 undetermined coefficients. This information is summarized in Table~\ref{tab:tri-adjm_bootstrap}.

\begin{table}
\centering
\begin{tabular}{ c | c }
 Constraint & Undetermined coefficients\\ 
 \hline
 Landau singularities & 9 \\  
 Integrability & 7 \\
 Physical branch cuts & 4 \\
 Sequential discontinuities & 3 \\
 Dim-reg consistency & 2 \\
 Leading singularity & 0 \\
\end{tabular}
\caption{Constraints for the adjacent-mass triangle.}
\label{tab:tri-adjm_bootstrap}
\end{table}

Even though we cannot use to our advantage the dimensionless arguments of symbol letters due to the extra scale $\mu^2$, we can still constrain the symbol. Equation (\ref{eq:tri-adjm_divergence}) shows that we do not have a $1/\epsilon^2$ term. Since the prefactor is given by $\mu^{2+2\epsilon}$, we know that there cannot be a term of the form $\mu^2 \otimes \mu^2$ in the symbol. Imposing this constraint gets us down to two undetermined coefficients. In fact, we could also impose that at $\mathcal{O}(\epsilon^0)$, the term multiplying $\log(\mu^2)$ must be proportional to $\log(x-1)/s$, but this condition does not impose additional constraints.

At this point, the $\mathcal{O}(\epsilon^0)$ part of the symbol is given by 
\begin{align}
    \mathcal{S}_2 (\I^{\triadjm}) & = c_1  \left[ \frac{s-m^2}{\mu^2} \otimes \frac{s-m^2}{\mu^2} - \frac{m^2}{\mu^2} \otimes \frac{m^2}{\mu^2} \right] 
    \\
    & \hspace{2cm}
    + c_2 \left[ \frac{s-m^2}{m^2} \otimes \frac{s}{\mu^2} + \frac{m^2}{\mu^2} \otimes \frac{s-m^2}{m^2} \right] \,.
\end{align}
To fix the remaining coefficients $c_1$ and $c_2$, we can resort to the leading singularity from~\eqref{eq:maxCut-triadjm}. Recall that this leading singularity is related to the maximal cut and is thus proportional to the discontinuity around the $s-m^2$ threshold, and the discontinuity in the symbol is obtained by clipping off the relevant first entry:
\begin{equation}
    \disc_{s-m^2} \mathcal{S}_2 (\I^{\triadjm}) = c_1 \frac{s-m^2}{\mu^2} + c_2 \frac{s}{\mu^2} \,.
\end{equation}
In order to get an answer proportional to $\log[(s-m^2)^2/s]$, we take $c_1 = -2 c_2$. The rational prefactor is simply the leading singularity $1/s$, which is found from the leading term of~\eqref{eq:maxCut-triadjm} expanded around $D=4$.

This result is consistent with the full expression \cite{Abreu:2015zaa}, 
\begin{align}
\I^{\triadjm}= & e^{\gamma_E\epsilon} \mu^{2+2 \epsilon} \Gamma(\epsilon-1)(m^2)^{-1-\epsilon}\hypgeo{1}{1+\epsilon}{2-\epsilon}{\frac{s}{m^2}}\\
=&\frac{\mu^2}{s}\left[\frac{1}{\epsilon}\log\left(1-\frac{s}{m^2}\right)-\text{Li}_2\left(\frac{s}{m^2}\right)-\log ^2\left(1-\frac{s}{m^2}\right)\right.
\\
& \hspace{4cm} \left.-\log \left(\frac{m^2}{\mu^2} \right) \log
   \left(1-\frac{s}{m^2}\right)\right]+\mathcal{O}\left(\epsilon\right)\,,
\end{align}
with the symbol 
\begin{align}
\mathcal{S} (\I^{\triadjm})
\propto
&
\frac{1}{\epsilon}\frac{s-m^2}{m^2}
+m^2\otimes\frac{m^2\left(s-m^2\right)}{s}
+\left(s-m^2\right)\otimes
\frac{s}{\left(s-m^2\right)^2}
+\mathcal{O}\left(\epsilon\right)\,.
\label{eq:symbol-t-p1-m12}
\end{align}

\subsection{Massless off-shell triangle}
\label{subsection:massless-off-shell}
\paragraph{Letters with square roots}
Next, we look at the off-shell (``three-mass'') triangle with massless internal edges. The Feynman integral is
\begin{equation}
    \mathcal{I}_\tri^{\threem} =
    \begin{gathered}
    \begin{tikzpicture}[line width=1,scale=0.75]
    \coordinate (v1) at (0,0);
        \coordinate (v2) at (0.866025,0.5);
        \coordinate (v3) at (0.866025,-0.5);
        \draw[] (v1) -- (v2) -- (v3) -- (v1);
        \draw[] (v3) -- (v1);
        \draw[] (v1) -- ++(-190:0.7);
        \draw[] (v1) -- ++(-170:0.7);
        \draw[] (v3) -- ++(-35:0.7);
        \draw[] (v3) -- ++(-55:0.7);
        \draw[] (v2) -- ++(35:0.7);
        \draw[] (v2) -- ++(55:0.7);
        \node[] at ($(v1)-(1.1,0.2)$) {$p_1$};
        \node[] at ($(v2)+(0.7,1)$) {$p_2$};
        \node[] at ($(v3)+(0.7,-1)$) {$p_3$};
    \end{tikzpicture}
    \end{gathered}
    =
    e^{\gamma_E \epsilon} \mu^{2 (3-D/2)}
    \Gamma\left(3-\frac{\D}{2}\right) \int_0^\infty \frac{\d^3 \alpha_e}{\GL(1)} \frac{{\U_\tri^{\threem}}^{3-\D} }{(\F_\tri^{\threem}-i\varepsilon)^{3-\D/2}} \,,
\end{equation}
with
\begin{equation}
    \F_\tri^{\threem} = - p_1^2 \alpha_1 \alpha_2 - p_2^2 \alpha_2 \alpha_3 - p_3^2 \alpha_3 \alpha_1 \,, \qquad \U_\tri^{\threem} = \alpha_1+\alpha_2+\alpha_3\,.
\end{equation}
The Landau singularities are at
\begin{equation}
    p_1^2=0\,, \qquad p_2^2=0\,, \qquad p_3^2=0\,, 
\end{equation}
where $\F_{\tri}^{\threem}$ factorizes, 
as well as at
\begin{equation}
    \lambda \equiv p_1^4 + p_2^4+ p_3^4 - 2 p_1^2 p_2^2 -2 p_1^2 p_3^2 - 2 p_2^2 p_3^2 = 0\,,
\end{equation}
which is an unphysical singularity. When forming the symbol letters, we allow for singularities at $\lambda=0$, as well as at $p_i^2=0$ for $i=1,2,3$. From these constraints, we form the following dimensionless letters.
\begin{equation}
    L_{ij} = \frac{p_i^2}{p_j^2}\,, \qquad
    L_{ijk} = \frac{p_i^2+p_j^2-p_k^2 + \sqrt{\lambda}}{p_i^2+p_j^2-p_k^2 - \sqrt{\lambda}}\,, \qquad L_
    {\lambda} = \frac{\sqrt{\lambda}}{p_1^2}
    \label{eq:Lijk}
\end{equation}
Here $i,j,k$ are distinct labels from the set $\{1,2,3\}$. Furthermore, the leading singularity of $\I_{\text{tri}}$ in $\D=4$ dimensions can be read from the maximal cut, which is proportional to $\frac{1}{\sqrt{\lambda}}$.

We can rationalize the alphabet by defining new variables $z$ and $\zb$ according to
\begin{equation}
    z \zb = \frac{p_2^2}{p_1^2}\,, \qquad (1-z)(1-\zb) = \frac{p_3^2}{p_1^2} \,.
\end{equation}
Then, $\lambda = z-\zb$, and the letters from~\eqref{eq:Lijk} take the form
\begin{equation}
    L_{21} = L_{12}^{-1} = z \zb\,, \qquad
    L_{31} = L_{13}^{-1} = (1-z)(1-\zb) \,, \qquad
    L_{23} = L_{32}^{-1} = \frac{z \zb}{(1-z)(1-\zb)}\,,
\end{equation}
and
\begin{equation}
    L_{123} = \frac{z}{\zb}\,, \qquad L_{231} = \frac{\zb (1-z)}{z (1-\zb)} \,, \qquad  L_{312} = \frac{1-\zb}{1-z} \,, \qquad L_{\lambda} = z-\zb \,.
\end{equation}
It is manifest from these equations that only five of the letters are multiplicatively independent. Let us choose a basis of letters of
\begin{equation}
    L_1 = z\,, \quad L_2 = \zb\,, \quad L_3 = 1-z\,, \quad L_4 = 1-\zb\,, \quad L_5 = z-\zb \,.
\end{equation}

\begin{figure}[t]
    \centering
    \includegraphics[width=0.45\linewidth]{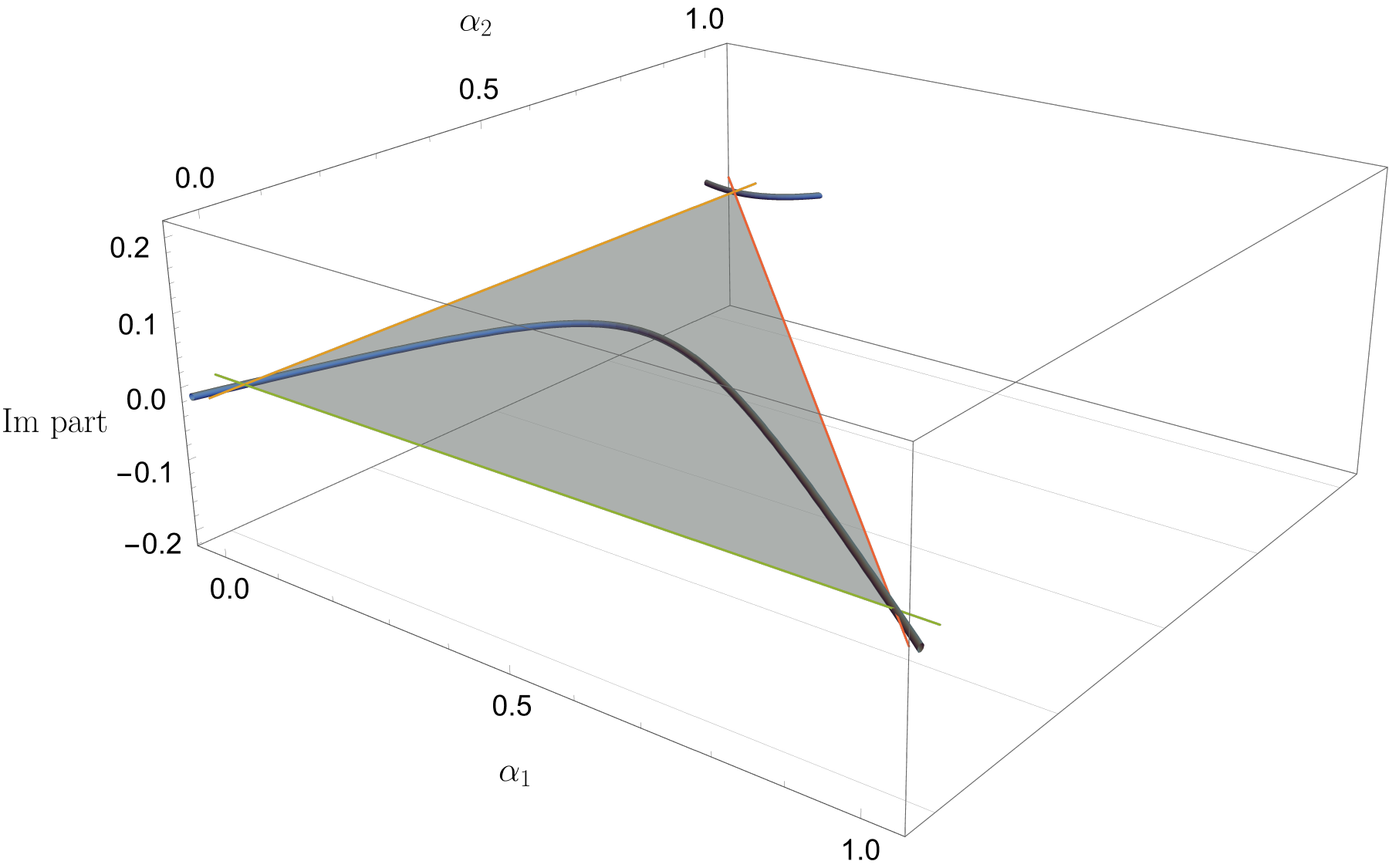}
    \includegraphics[width=0.45\linewidth]{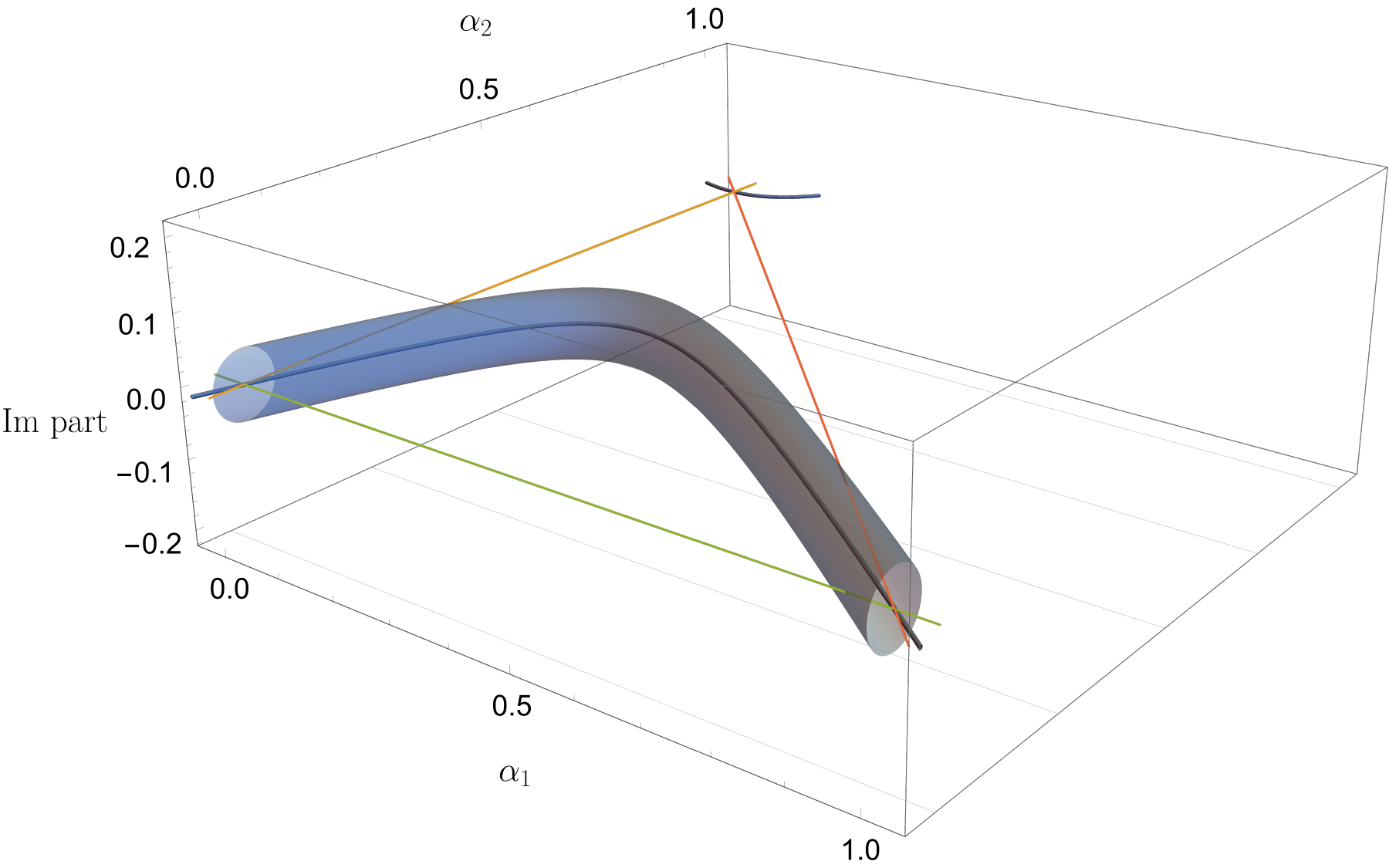}
    \includegraphics[width=0.45\linewidth]{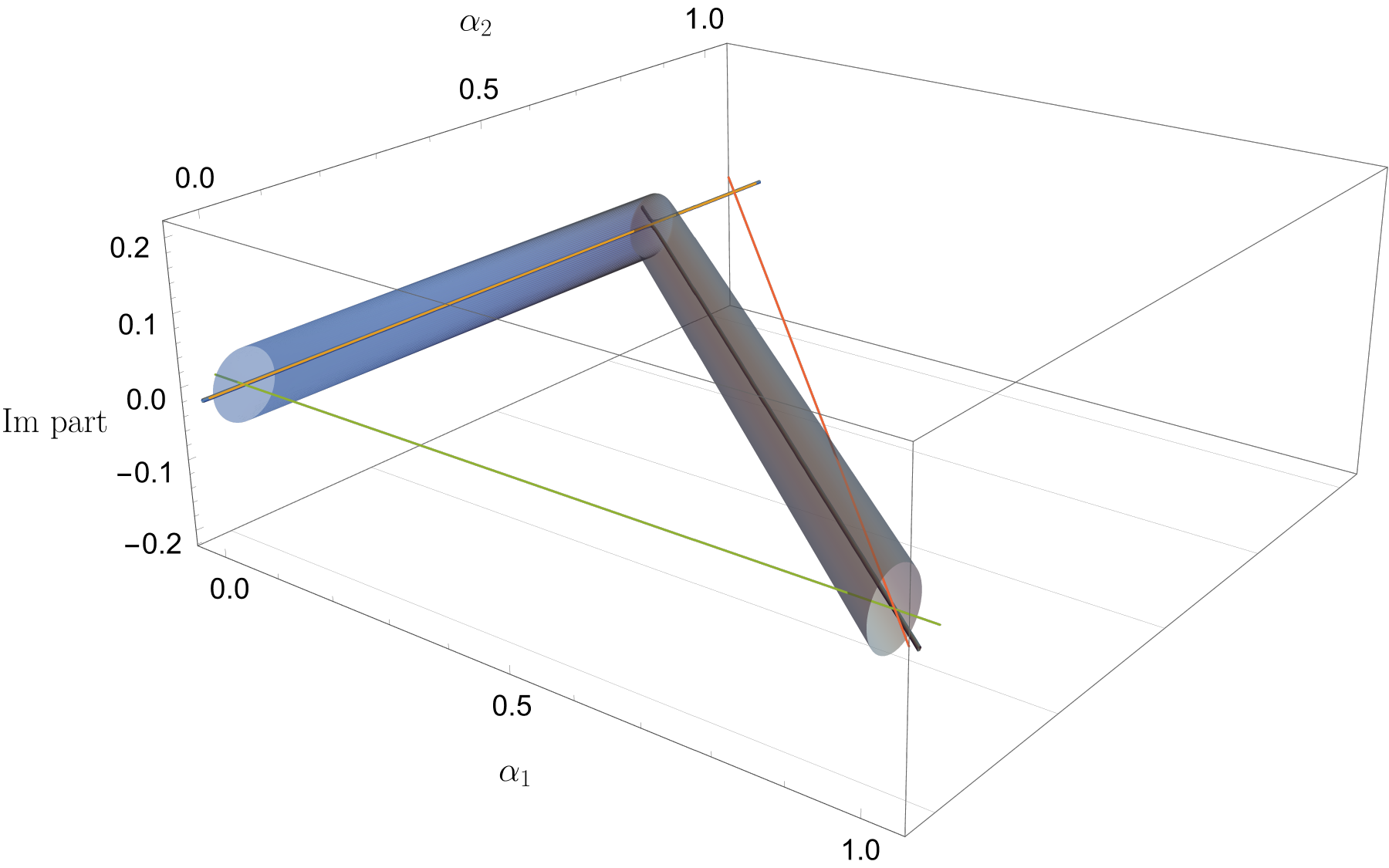}
    \caption{Tracking the integration contours of the off-shell triangle. \textbf{Top left.} The original integration contour (shaded region) is bounded by $\alpha_1=0$, $\alpha_2=0$ and $1-\alpha_1-\alpha_2=0$. For the values $z>1$ and $\zb<0$, the singular curve at $\F_\tri^{\threem}=0$ is slightly displaced to have a slight negative imaginary part in $z$, so that it does not intersect the integration contour. \textbf{Top right.} When encircling $z = 0$ in the positive direction, the monodromy contour is obtained as a tubular contour bounded by $\alpha_1=0$ and $\alpha_1=1$, and encircling the $\F_{\tri}^{\threem}=0$ curve. \textbf{Bottom.} A kinematic configuration showing the self-intersection of $\F_\tri^{\threem}$ when $z=1$, leading to a singularity. }
    \label{fig:off-shell_tri}
\end{figure}

It turns out that there are no nontrivial constraints on the vanishing of sequential discontinuities in this example if we simply ask which boundaries are included in the discontinuity integral. Instead of checking whether the new integration regions have the same boundaries or not, we will here be more careful in our analysis and track the contours directly in integer dimensions, as discussed in Sec.~\ref{subsection:dimensional}. Thus, the discontinuity contour will not be proportional to an area bounded by the coordinate hyperplanes and $\F_{\tri}^{\threem}=0$, but rather a tubular contour around the singular surface $\F_{\tri}^{\threem}=0$. Say we start in the region where $z>1$ and $\zb<0$. This region corresponds to where $p_1^2 >0$, while $p_2^2<0$ and $p_3^2<0$, and the physical value is obtained by taking $z \to z - i\varepsilon$ and $\zb \to \zb + i\varepsilon$ (see~\cite{Chavez:2012kn,Bourjaily:2020wvq} for more detailed discussions of different kinematic regions and analytic continuations). As we analytically continue $z$ in a circle in the complex $z$-plane counterclockwise around $z=0$ while holding $\zb$ fixed, we need to drag the integration contour, originally given by $0<\alpha_i<1$ for $i \in \{1,2,3\}$ as the $\F_{\tri}^{\threem}=0$ singular surface moves. So after the analytic continuation, we are left with the original integration contour plus the tubular integration contour from Figure~\ref{fig:off-shell_tri}. The discontinuity contour (i.e.\ the result of the analytic continuation with the original integration contour subtracted) is therefore simply the tubular integration contour itself.

From this sequential-discontinuity tracking, we furthermore learn that if we keep continuing $z=0$ in a circle around zero, the tubular integration contour simply winds around in a circle. We learn that a sequential discontinuity in $z=0$ is absent, and thus the symbol cannot have a term of the form $z \otimes z$.  Note that this statement only holds true for the original Feynman integral, since $z=0$ leads to a self-intersection of $\F_{\tri}^{\threem}=0$ outside the standard simplex. An analogous argument shows that $\zb$ cannot follow $z$ for the Feynman integral in $\D=4$ spacetime dimensions. Similarly, the letter $1-z$ cannot follow $1-z$, and similarly for bars on either letter. We have summarized these constraints, along with those that follow from the symmetry in $z$, $\zb$, in Table~\ref{tab:offshelltri}. We emphasize that this analysis is specific to $\D=4$ spacetime dimensions, where the integral is UV and IR finite. In non-integer dimensions, the analysis is different and the discontinuities excluded here may become allowed.

\begin{table}
\centering
\begin{tabular}{c|c|c|c|c}
\diagbox{$a_2$}{$a_1$}  & $z$ & $\zb$ & $1-z$ & $1-\zb$  \\
\hline
$z$ & $\times$ & $\times$ & $\checkmark$ & $\checkmark$  \\
$\zb$ & $\times$ & $\times$ & $\checkmark$ & $\checkmark$  \\
$1-z$ & $\checkmark$ & $\checkmark$ & $\times$ & $\times$ \\
$1-\zb$ & $\checkmark$ & $\checkmark$ & $\times$ & $\times$ \\
$z-\zb$ & $\checkmark$ & $\checkmark$ & $\checkmark$ & $\checkmark$ \\
\end{tabular}
\caption{Summary of sequential-discontinuity tracking for the off-shell triangle in $\D=4$ spacetime dimensions.}
\label{tab:offshelltri}
\end{table}

\begin{table}
\centering
\begin{tabular}{ c | c }
 Constraint & Undetermined coefficients\\ 
 \hline
 Landau singularities & 25 \\  
 Integrability & 19 \\
 Physical branch cuts &  14 \\
 Sequential discontinuities & 2 \\
 Symmetry & 1 \\
 Leading singularity & 0 
\end{tabular}
\caption{Constraints for the off-shell triangle.}
\label{tab:tri-offshell_bootstrap}
\end{table}

We can gain further insight from the tracking of the integration contours by being careful about the kinematic regions. Suppose that we want to compute the discontinuity in the Mandelstam variable $p_2^2$, in the kinematic region $R^{(2)}$, defined by $p_2^2>0$, $p_1^2<0$ and $p_3^2<0$. This discontinuity is equal to the unitarity cut in the $p_2^2$ channel. In our $z$, $\zb$ variables, however, we have a choice of how to implement this analytic continuation in $p_2^2$. In the region $R^{(2)}$ where we have $\zb<0$, $z>0$, we take $z \to z + i\varepsilon$ and $\zb \to \zb - i\varepsilon$. To achieve a counterclockwise rotation in $p_2^2$, we can either take $z \to z_0 e^{-i \theta}$ or $\zb \to \zb_0 e^{-i \theta}$; both lead to a rotation corresponding to $p_2^2 \to (p_2^2)_0 e^{i \theta}$ (we assume that $(p_2^2)_0$ is infinitesimally small so that $p_1^2$ and $p_3^2$ are closed to being fixed during the rotation). Thus, the unitarity cut in the $p_2^2$ channel can be obtained either by taking a discontinuity in $z$ or $\zb$, with the same orientation for both variables. This suggests that the discontinuity in $z$ and $\zb$ must be the same. In other words, if our symbol is given by $\mathcal{S} (\I_{\tri}^{\threem}) = c_1 \, z \otimes f(z,\zb) + c_2 \, \zb \otimes g(z,\zb) + \ldots$, where $f$ and $g$ are functions while $c_1$ and $c_2$ are numbers, then we must have $f=g$ and $c_1 = c_2$.

We now use this information to bootstrap the symbol. That is, we write,
\begin{equation}
    \mathcal{S}(\I_{\tri}^{\threem}) \propto \sum_{i,j=1}^5 c_{ij} L_i \otimes L_j \,,
\end{equation}
where we start with 25 undetermined coefficients $c_{ij}$. Next, we impose integrability, i.e.\,
\begin{equation}
    \partial_z \partial_{\zb} \mathcal{S}(\I_{\tri}^{\threem}) = \partial_{\zb} \partial_z \mathcal{S}(\I_{\tri}^{\threem}) \,,
\end{equation}
which results in 19 undetermined coefficients. The first-entry condition restricts the first entries to not include $L_5$, and further imposing the sequential discontinuity constraints listed in Table~\ref{tab:offshelltri} reduces the number of undetermined coefficients to 2, and the last two can be fixed by symmetry in $z$ and $\zb$ and imposing the correct leading singularity.\footnote{The symmetry group due to the permutation symmetry in the $p_i^2$ was discussed and identified as $S_3$ in~\cite{Chavez:2012kn}.} 
This information is summarized in Table~\ref{tab:tri-offshell_bootstrap}. The sequential-discontinuity constraint is particularly powerful in this case, since we carefully analyzed the nontrivial constraints in $\D=4$. The symbol of $\I_{\tri}^{\threem}$ becomes 
\begin{equation}
    \mathcal{S}(\I_{\tri}^{\threem}) \propto (1-z)(1-\zb) \otimes \frac{\zb}{z} + +z \zb \otimes \frac{1-z}{1-\zb}\,,
\end{equation}
and fixing the prefactor using the leading singularity determines the
corresponding function to be 
\begin{equation}
    \I_\tri^{\threem} = -\frac{2}{p_1^2(z-\zb)} \left[ \Li_2(z)-\Li_2(\zb)+\frac{1}{2}\log(z \zb) \log \left( \frac{1-z}{1-\zb} \right)  \right] \,.
\end{equation}
In different kinematic regions, the polylogarithms must be evaluated on the correct branches, which can be implemented by the appropriate infinitesimal imaginary parts of $z$ and $\zb$. See~\cite{Chavez:2012kn} for a more detailed discussion of analytic continuation.

\subsection{A triangle with three scales}
\label{sec:3scaletri}

Next, we return to the triangle with two internal masses $m_1$ and $m_2$, and two external massless legs, which we introduced as our original example in Sec.~\ref{sec:seqdisc_from_contours}. The Feynman integral is
\begin{equation}
\I^{\trithree}_{\tri} =
    \begin{gathered}
    \begin{tikzpicture}[line width=1,scale=0.75]
    	\coordinate (v1) at (0,0);
        \coordinate (v2) at (0.866025,0.5);
        \coordinate (v3) at (0.866025,-0.5);
        \draw[] (v1) -- (v2) -- (v3) -- (v1);
        \draw[line width=2] (v3) -- (v1) node[below,yshift=-2]{$m_1$};
        \draw[line width=2] (v2) -- (v3) node[right,yshift=8]{$m_2$};
        \draw[] (v1) -- ++(-190:0.7);
        \draw[] (v1) -- ++(-170:0.7);
        \draw[] (v3) -- ++(-45:0.7);
        \draw[] (v2) -- ++(45:0.7);
        \node[] at ($(v1)-(1.1,0.2)$) {$p$};
    \end{tikzpicture}
    \end{gathered}
    = e^{\gamma_E \epsilon} \mu^{2 (3-D/2)} \Gamma\left(3-\frac{\D}{2}\right) \int_0^\infty \frac{\d^3 \alpha_e}{\GL(1)} \frac{{\U_\tri^{\trithree}}^{3-\D} }{(\F_\tri^{\trithree}-i\varepsilon)^{3-\D/2}} \,,
\end{equation}
This integral is both UV and IR finite in $\D=4$ spacetime dimensions. Its Symanzik polynomials are given by
\begin{equation}
    \F^{\trithree}_{\tri} = -p^2 \alpha_1 \alpha_3 + (m_1^2 \alpha_1+m_2^2 \alpha_2) (\alpha_1+ \alpha_2 + \alpha_3)\,, \qquad \U^{\trithree}_{\tri} = \alpha_1+\alpha_2+\alpha_3\,,
\end{equation}
and the Landau singularities are at
\begin{align}
    m_1^2 = 0 \,, \qquad m_2^2 = 0\,, \qquad p^2-m_1^2 =0 \,,
    \label{eq:Landau3scale-apos} \\
    m_1^2-m_2^2-p^2=0 \,, \qquad p^2=0 \,, \qquad
    m_1^2-m_2^2=0
    \,. \nonumber
\end{align}
Out of these singularities, the physical ones, that can potentially appear in the first symbol entry, are $\{p^2-m_1^2,m_1^2,m_2^2\}$. The sequential-discontinuity tracking was already carried out in Sec.~\ref{sec:seqdisc_from_contours}, and we summarize the results in Table~\ref{tab:3scaletri},  along with adding the constraints obtained by tracking contours directly in $\D=4$. We are now ready to start bootstrapping this integral.

\begin{table}
\centering
\begin{tabular}{c|c|c|c}
\diagbox{$a_2$}{$a_1$}  & $m_1^2$ & $m_2^2$ & $p^2 - m_1^2 $  \\
\hline
$m_1^2$ & $\checkmark$ ($\times$ in $\D=4$) & $\times$ & $\times$  \\
$m_2^2$ & $\checkmark$ & $\checkmark$ ($\times$ in $\D=4$) & $\checkmark$ \\
$p^2 - m_1^2 $ & $\checkmark$ ($\times$ in $\D=4$) & $\times$  & $\checkmark$ ($\times$ in $\D=4$) \\
$m_1^2-m_2^2-p^2$ & $\times$ & $\checkmark$ & $\checkmark$ \\
$p^2$ & $\checkmark$ ($\times$ in $\D=4$) & $\times$ & $\checkmark$ ($\times$ in $\D=4$) \\
$m_1^2-m_2^2$ & $\checkmark$ & $\checkmark$ & $\times$
\end{tabular}
\caption{Summary of sequential-discontinuity tracking for the three-scale triangle.}
\label{tab:3scaletri}
\end{table}

We will implement the bootstrap in an expansion around $\D=4-2\epsilon$ spacetime dimensions, up to the $\mathcal{O}(\epsilon)$ term. In $\D=4$, we can work with dimensionless quantities in the symbol; to that end, we write $x = m_1^2/p^2$ and $y=m_2^2/p^2$. The possible dimensionless letters that we can form using its singular loci are
\begin{align}
    L_1 & = x\,, \quad  L_2 = y\,, \quad L_3 = x-y\,, \quad  L_4 = x-1\,, \quad L_5 = x-y-1 \,.
    \label{eq:3scaletri-letters}
\end{align}
We factor out the leading divergence and expand order-by-order in $\epsilon$ by writing
\begin{equation} \label{eq:I-tri-splitting-1}
    \frac{p^2}{\mu^2} \left( \frac{\mu^2}{p^2} \right)^{-\epsilon} \I^{\trithree}_{\tri}
    =
    \cJ_{\text{w2}}
    +
    \epsilon \,
    \cJ_{\text{w3}}
    +
    \mathcal{O}(\epsilon^2) \,,
\end{equation}
where $\cJ_{\text{w2}}$ and $\cJ_{\text{w3}}$ denote the weight-2 and weight-3 parts of the rescaled integral. Our task will be to bootstrap the symbols of $\cJ_{\text{w2}}$ and $\cJ_{\text{w3}}$. Note that the right-hand side of~\eqref{eq:I-tri-splitting-1} is dimensionless and independent of $\mu^2$.

\paragraph{Weight 2}
At weight 2, the ansatz for the symbol of $\I^{\trithree}_{\tri}$ becomes
\begin{equation} \label{eq:S-3scale}
    \mathcal{S} \big( \cJ_{\text{w2}} \big)
    \propto
    \sum_{i,j=1}^5 c_{i j} L_i \otimes L_j \,.
\end{equation}
Imposing integrability, i.e.,
\begin{equation}
    \partial_x \partial_y \mathcal{S} \big( \cJ_{\text{w2}} \big)
    = 
    \partial_y \partial_x \mathcal{S} \big( \cJ_{\text{w2}} \big)
\end{equation}
we end up with 19 linearly independent symbol terms.

Next, we impose that only $\alpha$-positive Landau singularities can lead to branch cuts on the first sheet, that is, only $m_1^2$, $m_2^2$ and $p^2-m_1^2$ are singularities of the physical scattering amplitude. The other singular locations can be singularities of the discontinuities of $\I^{\trithree}_{\tri} $, but not of $\I^{\trithree}_{\tri} $ itself. In terms of the letters from~\eqref{eq:3scaletri-letters}, only $L_1$, $L_2$ and $L_4$ can appear in the first entry and furthermore, the $p^2$ singularity has to cancel between the different letters. Doing so, we end up with 6 independent symbol terms. 
\begin{table}
    \centering
    \begin{tabular}{l|c|c|c}
    Constraints imposed & $\substack{\text{Number of}\\ \text{coefficients for } \cJ_{\text{w2}}}$ & $\substack{\text{Number of}\\ \text{coefficients for } \cJ_{\text{w3}}}$ & $\substack{\text{Number of}\\ \text{coefficients for } \mu^{2\epsilon}p^{-2\epsilon}\cJ_{\text{w3}}}$ \\ 
    \hline
    Sequences of five letters \ \ & \ \ 25 \ \ & \ \ 125 \ \ & \ \ - \ \ \\
    Integrable weight-three symbols \ \ & \ \ 19 \ \ & \ \ 65 \ \ & \ \ - \ \   \\
    Physical branch cuts & 6 & 27 & 3 \\
    Sequential-discontinuity constraints & 1 & 7 & 2
    \end{tabular}
    \caption{The number of free parameters $c_{ij}$ in the weight-two and three part of the symbol $\mathcal{S} \big( \I^{\trithree}_{\tri} \big)$ for the three-scale triangle, from imposing different constraints.}
    \label{tab:bootstrap_dbox}
\end{table}

This expression is our starting point for imposing the sequential discontinuity-constraints. Since this was our running example in Sec.~\ref{sec:seqdisc_from_contours}, we have already determined the relevant constraints. We impose the following, corresponding to the summary from Table~\ref{tab:3scaletri}:
\begin{itemize}
    \item $L_i \otimes L_i$ is forbidden for any $i$,
    \item $L_2 \otimes L_1$ is forbidden,
    \item $L_2 \otimes L_4$ is forbidden.
\end{itemize}
It turns out that these constraints suffice to uniquely fix the coefficients $c_{ij}$ from~\eqref{eq:S-3scale}, up to an overall constant that can be determined from the leading singularity, see Table~\ref{tab:bootstrap_dbox}. We find that
\begin{align}
    \mathcal{S} \big( \cJ_{\text{w2}} \big)
    & \propto
    \frac{L_1}{L_4} \otimes L_2 + \frac{L_2}{L_1} \otimes L_3 + \frac{L_4}{L_2} \otimes L_5 \\ 
    & = 
    \frac{m_1^2}{m_1^2-p^2} \otimes \frac{m_2^2}{p^2} + \frac{m_2^2}{m_1^2} \otimes \frac{m_1^2-m_2^2}{p^2} + \frac{m_1^2-p^2}{m_2^2} \otimes \frac{m_1^2-m_2^2-p^2}{p^2}
    \\ 
    & = 
    m_1^2 \otimes \frac{m_2^2}{m_1^2-m_2^2} + m_2^2 \otimes \frac{m_1^2-m_2^2}{m_1^2-m_2^2-p^2}
    + (m_1^2-p^2) \otimes \frac{m_1^2-m_2^2-p^2}{m_2^2}
    \,.
\end{align}

The symbol can easily be integrated to give the answer
\begin{align}
    \I^{\trithree}_{\tri} 
    & = \frac{\mu^2}{p^2} \Bigg[
    \Li_2 \left( \frac{m_2^2-m_1^2+p^2}{m_2^2}  \right)
    -
    \Li_2 \left( \frac{m_2^2-m_1^2}{m_2^2} \right)
    \Bigg] \,,
\end{align}
for $p^2<0$, where the branches were fixed by imposing physical singularities and limits.

\paragraph{Weight 3}

We would now like to reconstruct the symbol of $\cJ_{\text{w3}}$ from~\eqref{eq:I-tri-splitting-1}. We write the weight-3 ansatz as
\begin{equation} \label{eq:S-3scale-w3}
    \mathcal{S} \big(\cJ_{\text{w3}} \big)
    \propto
    \sum_{i,j,k=1}^5 c_{i j k} L_i \otimes L_j \otimes L_k \,,
\end{equation}
where we have used the same letters as before. We now have the extra layer of complication that $\cJ_{\text{w3}}$ is not the Feynman integral itself, but rather has a factor involving $\log(p^2)$ stripped off. We must therefore be careful to not impose restrictions on the sequential-discontinuities or first-entry conditions in $p^2$ until after we have multiplied out the full $\mathcal{O}(\epsilon)$ term in~\eqref{eq:I-tri-splitting-1}, including the $\log \left( \frac{\mu^2}{p^2}\right)$ part.

We start with imposing integrability in adjacent pairs of letters, which results in 65 undetermined coefficients. Next, we impose that neither $L_3$ nor $L_5$ can appear in the first entry. This gets us down to 27 undetermined coefficients. Finally, we impose the following constraints on sequential discontinuities: 
\begin{itemize}
    \item $L_i \otimes L_i \otimes L_i$ is forbidden for any $i$, 
    \item $L_2 \otimes \ldots \otimes L_1$ is forbidden anywhere in the symbol, 
    \item $L_2 \otimes L_4$ is forbidden anywhere in the symbol,
    \item $L_4 \otimes L_1$ is forbidden anywhere in the symbol.
    \item $L_2 \otimes L_1$ is forbidden anywhere in the symbol. 
\end{itemize}
We take the first constraint, forbidding the same letter to appear in every symbol entry, as a conjecture when the $\D=4$ integration-contour tracking shows that repeated letters are not allowed. We leave further investigation of this point to future work. The other constraints are genealogical constraints from the discontinuity contours no longer having a boundary component. We could, in fact, impose stricter constraints by eliminating sequences of the form $L_i \otimes \cdots \otimes L_j$, but here they lead to redundant constraints.
After imposing these constraints, we are left with 7 undetermined coefficients for $\mathcal{S} \big(\cJ_{\text{w3}} \big)$. 

We still have some conditions to impose from the consistency in dimensional regularization -- we know that the prefactor $\mu^{2\epsilon}$ can only produce a cross term with the weight-2 part of the symbol, which we already bootstrapped. The result of this second step of the bootstrap for the $\mathcal{O}(\epsilon)$ part of the integral is shown in the rightmost column of Table~\ref{tab:bootstrap_dbox}. More precisely, we can write down the full $\mathcal{O} (\epsilon)$ part of the symbol at the cost of introducing one extra coefficient:
\begin{equation}
    \widetilde{\mathcal{S} }\big(\cJ_{\text{w3}} \big)  \equiv  c_0 \, \frac{\mu^2}{p^2} \shuffle \mathcal{S} \big(\cJ_{\text{w2}} \big)   + \mathcal{S} \big(\cJ_{\text{w3}} \big) \,,
\end{equation}
where we found above that $\mathcal{S} \big(\cJ_{\text{w3}} \big)$ can be written as a linear combination of 7 terms, $c_i$, with $i \in \{1,2, \ldots,7\}$. We are now free to impose a first-entry condition on $\widetilde{\mathcal{S} }\big(\cJ_{\text{w3}} \big) $, which forbids $p^2$ in the first entry of the symbol. This leaves us with three possible linear combinations of symbol entries:
\begin{align}
    T_1 & = -L_1\otimes L_2\otimes \mu ^2-L_1\otimes \mu ^2\otimes L_2-\mu ^2\otimes L_1\otimes L_2-L_1\otimes L_3\otimes L_2
    \\ & +L_1\otimes L_1\otimes L_2+L_1\otimes L_2\otimes L_2
    +L_1\otimes L_3\otimes \mu ^2+L_1\otimes \mu ^2\otimes L_3
    \nonumber
    \\ & +\mu ^2\otimes L_1\otimes L_3-L_1\otimes L_1\otimes L_3-L_2\otimes L_3\otimes \mu ^2-L_2\otimes \mu ^2\otimes L_3-\mu ^2\otimes L_2\otimes L_3
    \nonumber
    \\ & + L_2\otimes L_2\otimes L_3+L_2\otimes L_3\otimes L_2+L_4\otimes L_2\otimes \mu ^2+L_4\otimes \mu ^2\otimes L_2
    \nonumber
    \\ & +\mu ^2\otimes L_4\otimes L_2+L_4\otimes L_5\otimes L_2-L_4\otimes L_2\otimes L_2-L_4\otimes L_4\otimes L_2
    \nonumber
    \\ & + L_2\otimes L_5\otimes \mu ^2+L_2\otimes \mu ^2\otimes L_5+\mu ^2\otimes L_2\otimes L_5-L_2\otimes L_2\otimes L_5-L_2\otimes L_5\otimes L_2
    \nonumber
    \\ & - L_4\otimes L_5\otimes \mu ^2-L_4\otimes \mu ^2\otimes L_5-\mu ^2\otimes L_4\otimes L_5+L_4\otimes L_4\otimes L_5
    \nonumber
    \\
    T_2
    & =
    -L_1\otimes L_2\otimes L_3-L_1\otimes L_3\otimes L_2+L_1\otimes L_2\otimes L_2 +L_1\otimes L_3\otimes L_3
    \nonumber
    \\ & +L_2\otimes L_2\otimes L_3+L_2\otimes L_3\otimes L_2 -L_2\otimes L_3\otimes L_3 +L_4\otimes L_2\otimes L_5
    \\ & +L_4\otimes L_5\otimes L_2 -L_4\otimes L_2\otimes L_2-L_2\otimes L_2\otimes L_5-L_2\otimes L_5\otimes L_2
    \nonumber
    \\ & +L_2\otimes L_5\otimes L_5-L_4\otimes L_5\otimes L_5
    \nonumber
    \\
    T_3
    & =
    -L_1\otimes L_4\otimes L_2-L_1\otimes L_2\otimes L_5+L_1\otimes L_2\otimes p^2+L_1\otimes p^2\otimes L_2
    \nonumber
    \\ & +L_1\otimes L_3\otimes L_5-L_1\otimes L_3\otimes p^2+L_1\otimes L_4\otimes L_5-L_1\otimes p^2\otimes L_5
    \nonumber
    \\ & -L_2\otimes L_3\otimes L_5+L_2\otimes L_3\otimes p^2+L_4\otimes L_2\otimes L_5-L_4\otimes L_2\otimes p^2
    \\ & -L_4\otimes p^2\otimes L_2+L_4\otimes L_4\otimes L_2-L_2\otimes L_5\otimes p^2+L_2\otimes L_5\otimes L_5
    \nonumber
    \\ & +L_4\otimes L_5\otimes p^2+L_4\otimes p^2\otimes L_5-L_4\otimes L_4\otimes L_5-L_4\otimes L_5\otimes L_5
    \nonumber
\end{align}
The last task is then to figure out which linear combination of $T_1$, $T_2$ and $T_3$ corresponds to the Feynman integral. This can, for example, be done numerically or by computing some cuts. The correct answer turns out to not include $T_2$ at all, but can rather be written as
\begin{equation}
    \widetilde{\mathcal{S} }\big(\cJ_{\text{w3}} \big)  
    =
    T_1 - T_3 \,.
\end{equation}
Note that $T_2$ contains some unique sequences of triple discontinuities, e.g., $L_1 \otimes L_2 \otimes L_3$. If we were able to show that such longer sequences do not appear, even though $L_1 \otimes L_2$ and $L_2 \otimes L_3$ are separately allowed, we would have been able to eliminate $T_2$.


Both the results at weight two and weight three are consistent with the full $\D$-dimensional answer for the three-scale triangle \cite{Abreu:2015zaa}, 
\begin{align}\bsp
\I^{\trithree}_{\tri}
=&\frac{e^{\gamma_E\epsilon \mu^{2+2\epsilon}}\Gamma(1+\epsilon)}{\epsilon(1-\epsilon)\left(m_{1}^2-m_{2}^2\right)}\left[\left(m_{2}^2\right)^{-\epsilon}\hypgeo{1}{1}{2-\epsilon}{\frac{p^2}{m_{1}^2-m_{2}^2}}\right.\\
& \hspace{4cm} \left.-\left(m_{1}^2\right)^{-\epsilon}\,F_1\left(1;1,\epsilon;2-\epsilon;\frac{p^2}{m_{1}^2-m_{2}^2};\frac{p^2}{m_{1}^2}\right)\right] \, .
\esp\end{align}

\section{Two-loop examples}
\label{sec:two-loop}

In this section, we initiate a study of sequential-discontinuity relations for two-loop integrals in dimensional regularization. We will analyze two versions of the ice cream cone diagram with massless internal legs and two off-shell external legs. First, we look at the symmetric configuration of the off-shell legs, and then the asymmetric one.

\subsection{Symmetric ice cream cone}
\label{sec:ice-symmetric}

We start with analyzing the sequential discontinuities of the symmetric 2-scale ice cream cone, 
\begin{equation}
\I_{\icecream} =
    \begin{gathered}
    \begin{tikzpicture}[line width=1,scale=0.75]
    	\coordinate (v1) at (0,0);
        \coordinate (v2) at (0.866025,0.5);
        \coordinate (v3) at (0.866025,-0.5);
        \draw[] (v1) -- (v2)  (v3) -- (v1);
        \draw[out = -60, in = 60] (v2) to (v3);
        \draw[out = -120, in = 120] (v2) to (v3);
        \draw[] (v1) -- ++(-180:0.7);
        \draw[] (v3) -- ++(-35:0.7);
        \draw[] (v3) -- ++(-55:0.7);
        \draw[] (v2) -- ++(35:0.7);
        \draw[] (v2) -- ++(55:0.7);
        \node[] at ($(v2)+(1.1,0.7)$) {$p_1$};
        \node[] at ($(v3)+(1.1,-0.7)$) {$p_2$};
    \end{tikzpicture}
    \end{gathered}
\end{equation}
The parametrized integral is given by
\begin{equation}
    \mathcal{I}_{\icecream} = e^{2 \gamma_E \epsilon} \mu^{2 (4-D)} \, \Gamma(4-\D) \int_0^{\infty} \frac{\rd^4 \alpha}{\GL (1)} \frac{\U_{\icecream}^{4-3\D/2}}{(\F_{\icecream} - i \varepsilon)^{4-\D} } \,,
    \label{eq:Iicecream-sym}
\end{equation}
with 
\begin{equation}
    \F_{\icecream} = - \alpha _1 \alpha _2 \alpha _3 p_1^2- \alpha _2 \alpha _3 \alpha _4 p_2^2 \,, \qquad \U_{\icecream} = \alpha _1 \alpha _2{+}\alpha _3 \alpha _2{+}\alpha _4 \alpha _2{+}\alpha _1 \alpha _3{+}\alpha _3 \alpha _4
\end{equation}
and has its Landau singularities at
\begin{equation}
    p_1^2 = 0\,, \quad p_2^2 = 0 \,, \quad p_1^2-p_2^2 = 0\,.
\label{eq:Landau_sym}
\end{equation}
The first two are physical singularities while the last one is on the second sheet.

Let us form the dimensionless ratio $x = \frac{p_1^2}{p_2^2}$. The letters potentially appearing in the symbol $\mathcal{S}(\mathcal{I}_{\icecream})$ are then
\begin{equation}
    L_1 = x, \qquad L_2 = x-1\,.
\end{equation}
We also know that the function $\mathcal{I}_{\icecream}$ also has to be symmetric under the exchange $x \leftrightarrow \frac{1}{x}$, which corresponds to $p_1^2 \leftrightarrow p_2^2$. This requirement turns out to place powerful constraints on the symbol as we will see below. Let us start with using the trick that was proposed in Sec.~\ref{sec:dimless-ratios}.  If we factor out the scale $p_2^2$ from the integral, we can write
\begin{equation}
    \I_{\icecream} = \left( \frac{
    \mu^4}{p_1^2 p_2^2} \right)^{\epsilon} f_{\epsilon}(x) \,,
\end{equation}
where the function $f_{\epsilon}(x)$ depends only on $x$. Due to the symmetry of the prefactor, $f_{\epsilon}(x)$ must remain symmetric under $x \to \frac{1}{x}$, which will turn out to be crucial for placing constraints.

\begin{figure}[t]
    \centering
    \includegraphics[width=0.3\linewidth]{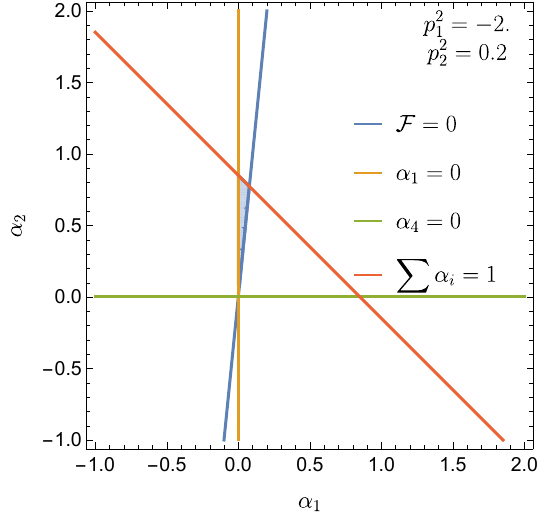}
    \hspace{1.5cm}
    \includegraphics[width=0.3\linewidth]{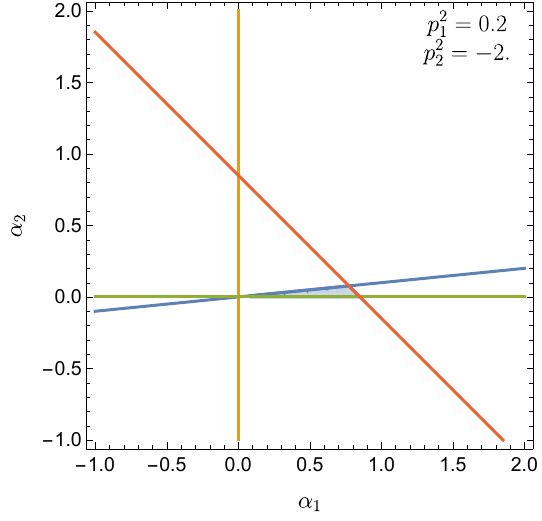}
    \caption{Tracking the integration contours of a symmetric ice cream cone. To fit the plot into two dimensions, we pick a slice through the integration space where $\sum_{i=1}^4 \alpha_i = 1$, and we arbitrarily put $\alpha_2+\alpha_3 = 0.15$, while plotting $\alpha_1$ and $\alpha_4$. \textbf{Left:} When encircling $p_2^2 = 0$, the monodromy contour is bounded by $\alpha_1=0$, $\F=0$ and the surface $\sum_{i=1}^4 \alpha_i =0$; \textbf{Right:} When encircling $p_1^2 = 0$, the monodromy contour is bounded by $\alpha_2=0$, $\F=0$ and the surface $\sum_{i=1}^4 \alpha_i =0$.}
    \label{fig:sym-ice-cream}
\end{figure}

Let us next analyze the sequential discontinuities of this integral. We plot the contours close to $p_1^2=0$ and $p_2^2=0$ in Fig.~\ref{fig:sym-ice-cream}, which correspond to $x=0$ and $x \to \infty$. The plots show that the integration contour obtained taking a discontinuity around $p_1^2=0$ is not singular when $p_2^2=0$, since $\alpha_2=0$ is not a boundary of the former discontinuity contour, but it is a boundary of the discontinuity contour corresponding to $p_2^2=0$. Thus, a sequential discontinuity in $p_2^2=0$ following $p_1^2=0$ and vice versa are absent in the full answer for $\I_{\icecream}$.

We now have stringent constraints on the symbol and on the form of $f_{\epsilon}(x)$. We know that $\I_{\icecream}$ is symmetric in $x \to \frac{1}{x}$, and also that there cannot be a sequential discontinuity in $x$ followed by $x$. Moreover, $x-1$ cannot be a physical singularity. These conditions forbid any symbol entries of the form $x \otimes x \otimes \ldots$ since they would lead to a sequential discontinuity in $p_2^2$ followed by $p_1^2$, and it also forbids symbol entries of the form $x \otimes x-1 \otimes \ldots$ since they cannot be made to obey the symmetries. What saves the function from being non-trivial is the prefactor, proportional to $(p_1^2 p_2^2)^{-\epsilon}$. Using that dependence, we can form the ansatz:
\begin{equation}
    \I_{\icecream} = g (p_1^2, p_2^2) \; (p_1^2/\mu^2)^{-2 \epsilon} + g (p_2^2, p_1^2) \; (p_2^2/\mu^2)^{-2 \epsilon} \,,
\end{equation}
where $g$ is a rational function in $p_1^2$ and $p_2^2$. Note that the function $g$ can potentially have poles in $p_1^2-p_2^2$ but not branch cuts, since that would lead to a physical singularity at $1-x=0$.

To fully bootstrap the integral we need to fix the prefactor and the power of $p_1^2-p_2^2$ by computing the leading singularity. We can get it by computing the next-to-maximal cut, which is the same as a discontinuity in, say, the $p_1^2$ channel.\footnote{Note that the maximal cut of this integral is zero. This can either be verified by direct computation, or by noting that the next-to-maximal cut is the discontinuity in $p_1^2$, and then, the maximal cut is obtained by taking a further discontinuity in $p_2^2$. However, we already saw that this sequential discontinuity is zero, which is consistent with the maximal cut being zero.}
Using the prefactor and the symmetry in $p_1^2 \leftrightarrow p_2^2$, we get: 
\begin{equation}
    \I_{\icecream} = e^{\gamma_E \epsilon}
    \left( \frac{
    \mu^2}{-p_2^2} \right)^{2 \epsilon} \frac{1}{p_1^2-p_2^2} \frac{\Gamma (1-\epsilon )^2 \Gamma(-1+2 \epsilon) \Gamma (-\epsilon )}{\Gamma (2-3 \epsilon )}
    \left(p_2^2 - x^{-2 \epsilon} p_1^2 \right) \,.
\end{equation}
To verify this result, we can directly compute the integral from~\eqref{eq:Iicecream-sym}.

\subsection{Asymmetric ice cream cone}
\label{sec:ice-asymmetric}
Next, we look at an asymmetric 2-scale configuration of the ice cream cone,
\begin{equation}
\I_{\icecream} =
    \begin{gathered}
    \begin{tikzpicture}[line width=1,scale=0.75]
    	\coordinate (v1) at (0,0);
        \coordinate (v2) at (0.866025,0.5);
        \coordinate (v3) at (0.866025,-0.5);
        \draw[] (v1) -- (v2)  (v3) -- (v1);
        \draw[out = -60, in = 60] (v2) to (v3);
        \draw[out = -120, in = 120] (v2) to (v3);
        \draw[] (v1) -- ++(-190:0.7);
         \draw[] (v1) -- ++(-170:0.7);
        \draw[] (v3) -- ++(-35:0.7);
        \draw[] (v3) -- ++(-55:0.7);
        \draw[] (v2) -- ++(45:0.7);
        \node[] at ($(v1)-(1.1,0.2)$) {$p_3$};
        \node[] at ($(v3)+(1.0,-0.7)$) {$p_2$};
        \node[color=white] at ($(v1)+(2,1.3)$) {$p_1^2$};
    \end{tikzpicture}
    \end{gathered}
\end{equation}
whose Feynman integral is given by
\begin{equation}
    \mathcal{I}_{\icecreamA} = e^{2 \gamma_E \epsilon} \mu^{2(4-D)} \Gamma(4-\D) \int \frac{\rd^4 \alpha}{\GL (1)} \frac{1}{\F_{\icecreamA}^{4-\D} }
    \frac{1}{\U_{\icecreamA}^{-4+3\D/2}}
    \,,
    \label{eq:Iicecream-asym}
\end{equation}
where
\begin{align}
   \F_{\icecreamA} & =- \alpha _1 \alpha _2 \alpha _4 p_3^2-\alpha_1 \alpha_3 \alpha_4 p_3^2-\alpha _2 \alpha _3 \alpha _4 p_2^2 \,,
   \\
   \U_{\icecreamA} & = \alpha _1 \alpha _2+\alpha _3 \alpha _2+\alpha _4 \alpha _2+\alpha _1 \alpha _3+\alpha _3 \alpha_4 \,.
\end{align}
In this case, the maximal cut computed using the Baikov representation gives the following result for the leading singularity,
\begin{equation}
    \Cut \; \I_{\icecreamA} \propto \frac{\mu^{4 \epsilon} (p_3^2)^{- \epsilon } \left(p_3^2-p_2^2\right){}^{-\epsilon } \Gamma (1-\epsilon )}{(2 \epsilon -1) \Gamma (2-3 \epsilon )} \,.
\end{equation}

\begin{figure}[t]
    \centering
    \includegraphics[width=0.3\linewidth]{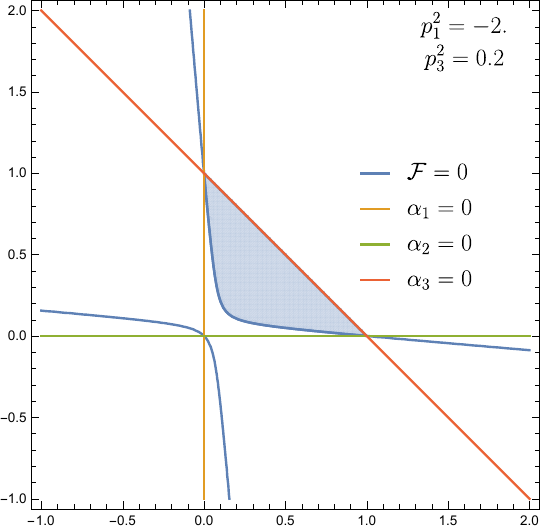}
    \hspace{1.5cm}
    \includegraphics[width=0.3\linewidth]{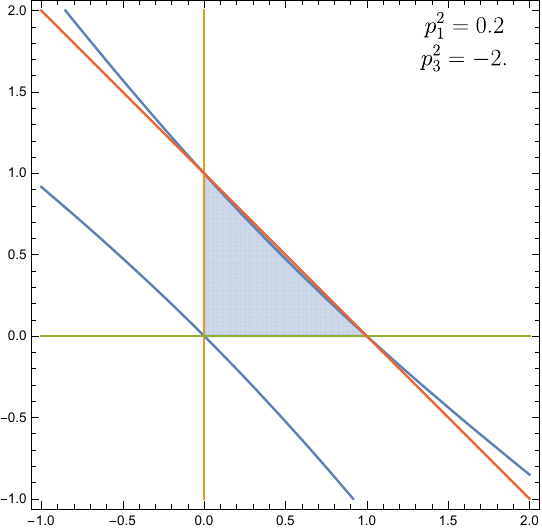}
    \caption{Tracking the integration contours of an asymmetric ice cream cone. Since $\alpha_4$ is a factor of $\F$ while the other factor is independent of  $\alpha_4$, we can conveniently restrict the plot of the parametric regions to a cross section in any plane of a constant positive value of $\alpha_4$. The GL(1) gauge fixing can be implemented by setting $\sum_{i=1}^3 \alpha_i = 1$.
   \textbf{Left:} When encircling $p_2^2 = 0$, the monodromy contour is bounded by $\alpha_3=0$, $\F=0$ and the plane $\alpha_4=0$ which is parallel to the plot in the third dimension; \textbf{Right:} When encircling $p_1^2 = 0$, the monodromy contour is bounded by $\alpha_1=0$, $\alpha_2=0$, $\F=0$ and the same parallel plane $\alpha_4=0$.}
    \label{fig:asym-ice-cream}
\end{figure}

The sequential-discontinuity analysis shows that $p_2^2$ cannot follow $p_3^2$ in the full answer. 
See Figure~\ref{fig:asym-ice-cream}. This condition is connected to the fact that $p_2^2$ is not a factor of the discriminant of $\F$, while $p_3^2$ is and leads to the factorization that is approached in the figure on the left.
We form the dimensionless ratio $x = \frac{p_3^2}{p_2^2}$. The singularities of the integral are then at $p_2^2=0$, $p_3^2=0$ and $p_3^2-p_2^2 = 0$, so we form the letters
\begin{equation}
    L_1 = \frac{p_2^2}{\mu^2} \, \qquad L_2 = \frac{p_3^2}{\mu^2}\,, 
    \qquad L_3 = \frac{p_3^2-p_2^2}{\mu^2}\,.
\end{equation}

We note that $L_3$ contains the unphysical singularity at $p_3^2-p_2^2$, so it cannot appear in the first entry. We also cannot have an entry of the form $L_2 \otimes L_1$, since it would indicate a sequential discontinuity in $p_2^2$ after $p_3^2$.

\begin{table}
    \centering
    \begin{tabular}{l|c}
    Constraints imposed & $\substack{\text{Number of}\\ \text{coeff's for } \mu^{2\epsilon}p_2^{-2\epsilon}\cJ_{\text{w2}}}$ \\ 
    \hline
    Sequences of three letters \ \ & \ \ 9 \ \ \\
    Integrability \ \  & \ \ 7 \ \   \\
    Physical branch cuts  & 4 \\
    Sequential-discontinuities  & 3 \\
    Dim-reg consistency & 2
    \end{tabular}
    \caption{The number of free parameters $c_{ij}$ in the weight 2 part of the symbol $\mathcal{S} \big( \I_{\icecreamA} \big)$ for the asymmetric ice cream cone, from imposing different constraints.}
\end{table}

We use the same approach as for the hook triangle integral from Sec.~\ref{sec:3scaletri}, factoring out the leading singularity and writing the symbol of the function as
\begin{equation} \label{eq:I-tri-splitting}
    \left( \frac{\mu^2}{p_3^2} \right)^{-2 \epsilon} \I_{\icecreamA}
    =
    \frac{1}{\epsilon}
    +
    \cJ_{\text{w1}}
    +
    \epsilon \,
    \cJ_{\text{w2}}
    +
    \mathcal{O}(\epsilon^2) \,.
\end{equation}
The right-hand side is then independent of $\mu^2$, but we cannot directly apply the sequential-discontinuity constraints, since we have multiplied the Feynman integral by $(p_3^2)^{-2 \epsilon}$. We will start with bootstrapping the right-hand side, followed by including the $\mu^2$ dependence in a controlled way.
We will only bootstrap the terms of leading weight, but we will discuss below how to get the subleading terms. 

We first note that if $\mathcal{J}_{\text{w}1}$ contains a factor of $\log(p_2^2)$, a cross term between the $\frac{1}{\epsilon}$ and $(p_3^2)^{2 \epsilon}$ would generally lead to a sequential discontinuity in $p_2^2$ after $p_3^2$. $\mathcal{J}_{\text{w}1}$ also cannot contain $\log(p_3^2-p_2^2)$, since it would lead to an unphysical singularity on the first sheet. Thus, we must have that at leading weight, either $\mathcal{J}_{\text{w}1}=0$ or $\mathcal{J}_{\text{w}1}=2 \log\left(\frac{p_2^2}{p_3^2}\right)$, which are the only possibilities consistent with the sequential discontinuities. A comparison with the leading singularity shows that the former choice must be correct.

Next, we bootstrap the $\mathcal{O}(\epsilon)$ term. By imposing integrability, physical branch cuts and the constraints on sequential discontinuities, we go from 9 undetermined coefficients to 3. Finally, by imposing that the term proportional to $\log(\mu^2)$ is consistent with the expansion in~\ref{eq:I-tri-splitting}, we get down to two undetermined coefficients. Denoting them as $c_1$ and $c_2$, the symbol is
\begin{align}
\label{eq:ice-asym-terms}
    \mathcal{S}(\I_{\icecreamA}) & = (c_1+c_2) p_2^2\otimes (p_2^2-p_3^2)-(c_1+c_2) (p_2^2\otimes p_3^2+p_3^2\otimes (p_2^2-p_3^2))\\ & \hspace{2cm} +c_1 p_2^2\otimes p_2^2+c_2 p_3^2\otimes p_3^2 \,.
    \nonumber
\end{align}
We can use the leading singularity to fix the overall scale, e.g., by setting $c_1+c_2=1$.

The full answer for the integral is 
\begin{align}
    \mathcal{I}_{\icecreamA} & = \frac{(p_3^2)^{-2 \epsilon } \Gamma (1-\epsilon )}{(1-2 \epsilon ) \Gamma (2-3 \epsilon )}
    \Bigg[\Gamma (1-\epsilon ) \Gamma (-\epsilon ) \, _2F_1\left(1,2 \epsilon ;1+\epsilon;\frac{p_2^2}{p_3^2}\right)
    \\ & \hspace{6cm} +2 \pi  \cot (\pi  \epsilon ) \Gamma (1-2 \epsilon ) \left(\frac{p_2^2 (p_3^2-p_2^2)}{(p_3^2)^2}\right)^{-\epsilon }\Bigg] \,,
    \nonumber
\end{align}
which expands to
\begin{align}
\mathcal{I}_{\icecreamA} & =
\frac{1}{\epsilon } +(5-2 \log (p_2^2)) + 
    \epsilon  \Bigg[2 \text{Li}_2\left(\frac{p_2^2}{p_3^2}\right)+2 \log (p_2^2) (\log (p_3^2-p_2^2)-5)
    \\ & \hspace{2cm} -2 \log (p_3^2) \log (p_3^2-p_2^2)+\log ^2(p_2^2)+\log ^2(p_3^2) -\frac{5 \pi ^2}{6}+19\Bigg]
+ {\mathcal O}(\epsilon^2)
\nonumber
\end{align}
for $p_2^2>0$ and $p_3^2>0$. Note that there is actually no singularity at $p_2^2=p_3^2$, since the naive divergence cancels between the dilogarithm and the logarithms. The weight-two part of the symbol is
\begin{equation}
2 \epsilon^{-1} {\mathcal S}\left(\mathcal{J}_{\text{w2}}\right) = p_2^2 \otimes \frac{(p_2^2)^2 (p_2^2-p_3^2)}{p_3^2} - p_3^2 \otimes p_3^2(p_2^2-p_3^2) \,.
\end{equation}
Going back to Eq.~\eqref{eq:ice-asym-terms}, we see that the correct answer is obtained by setting $c_2=-1$, assuming we have fixed the scale by $c_1+c_2=1$.

\section{Summary and discussion}
\label{sec:conclusions}

In this paper, we have demonstrated that constraints on sequential discontinuities can be obtained by tracking integration contours of Feynman integrals in parametric space while varying the external kinematics. The main idea is that sequences of discontinuities reveal their presence in singular behavior of the shape of the integration contour. In several examples, we have found that these new constraints can be used to fix symbols uniquely up to normalization.

We have presented an initial exploration of this main idea, and there are many avenues for further study. Most obviously, we have focused on constraints on sequences of length 2. The case featured in Fig.~\ref{fig:G1p} hints that we might be able to probe longer sequences through a more detailed understanding of the nature of singularities of contours. In Sec.~\ref{sec:3scaletri}, we saw an example of how such sequences can directly lead to excluding terms from the symbol. It might be possible to develop a set of complementary constraints from the point of view of a diagrammatic coaction, which leads naturally to a recursive construction of symbols \cite{Abreu:2017enx,Abreu:2017mtm,Abreu:2021vhb}. In fact, the concept of coaction \cite{Brown:2015ylf,Panzer:2016snt,Brown:2015fyf} might be a preferable framework for this study, because it is a map to pairs of objects, whose iteration leads to the symbol.

We have also focused on sequences of single symbol letters, and hence on the contours associated to momentum thresholds and mass singularities. In the example of Sec.~\ref{sec:monodromy_theory}, we discussed how the maximal cut of that triangle corresponds to an integration contour whose singularities are revealed in the third entry of the symbol. Exploring the full set of possible cuts of an integral would be expected to provide additional constraints. We note some overlap between these and genealogical constraints~\cite{Hannesdottir:2024cnn}, which are the ones obtained when the cut integration contours are no longer bounded by some boundaries $\alpha_i=0$ needed for other Landau singularities; once a boundary is lost through cutting, it can play no further role as we track the region. 

A thorough understanding of the homology of Feynman integrals would help tremendously towards addressing questions such as these. The foundations laid in \cite{boyling1968homological} deserve to be updated and extended in light of IR/UV divergences and dimensional regularization. For the purpose of tracking integration contours, we would like to understand better how analytic continuation proceeds across singular loci, for example as in the case shown in Fig.~\ref{fig:G12}, where we expect to see smooth behavior in the fully complexified kinematic phase space.

\acknowledgments
It is a pleasure to thank Samuel Abreu, Souvik Bera, Sudeepan Datta, Alexander Farren, Einan Gardi, Craig Larkin, Andrew McLeod, Sebastian Mizera, Eliza Somerville, and Mikey Whelan for enlightening and stimulating conversations. 
Computations were performed with the assistance of the packages {\tt PolyLogTools} \cite{Duhr:2019tlz}, {\tt MultiHypExp} \cite{Bera:2023pyz} and {\tt SOFIA} \cite{Correia:2025yao}.
R.B. wishes to acknowledge the support of the Munich Institute
for {Astro-,} Particle and BioPhysics (MIAPbP) which is
funded by the Deutsche Forschungsgemeinschaft (DFG,
German Research Foundation) under Germany’s Excellence Strategy—EXC-2094—390783311; the Galileo Galilei Institute for Theoretical Physics and the INFN; and the Mainz Institute for Theoretical Physics (MITP) of the Cluster of Excellence PRISMA+ (project ID 39083149).
R.B. was supported by a J. Robert Oppenheimer Visiting Professorship at the Institute for Advanced Study in the initial stages of this work. 
H.S.H. gratefully acknowledges funding provided by the J. Robert Oppenheimer Endowed Fund of the Institute for Advanced Study. This material is based upon work supported by the U.S. Department of Energy, Office of Science, Office of High Energy Physics under Award Number DE-SC0009988.
This work is funded by the European Union (ERC, MaScAmp, 101167287). Views and opinions expressed are however those of the author(s) only and do not necessarily reflect those of the European Union or the European Research Council Executive Agency. Neither the European Union nor the granting authority can be held responsible for them.

\appendix

\bibliographystyle{JHEP}
\bibliography{refs}

\end{document}